\documentclass[pra,superscriptaddress,preprintnumbers,preprint,showpacs,amsmath,amssymb]{revtex4-1}
\usepackage{morefloats}
\usepackage{psfrag}
\usepackage{graphicx}
\usepackage{subfigure}
\usepackage{array}
\usepackage{natbib} 
\def\be{\begin{equation}}       \def\ee{\end{equation}}
\def\bea{\begin{eqnarray}}      \def\eea{\end{eqnarray}}

\begin{document}
\title{Entanglement robustness and spin relaxation in thermal states of two-dimensional dissipative spin system in an inhomogeneous magnetic field}
\author{Gehad Sadiek}
\email{Corresponding author: gsadiek@sharjah.ac.ae}
\affiliation{Department of Applied Physics and Astronomy, University of Sharjah, Sharjah 27272, UAE}
\affiliation{Department of Physics, Ain Shams University, Cairo 11566, Egypt}
\author{Samaher Almalki}
\affiliation{Department of Physics, King Saud University, Riyadh 11451, Saudi Arabia}
\date{\today}
\begin{abstract}
{We consider a finite two-dimensional Heisenberg triangular spin lattice, with a central spin surrounded by equally distant spins, at different degrees of anisotropy, coupled to a dissipative Lindblad environment obeying the Born-Markovian constrain at finite temperature. We show how applying an inhomogeneous magnetic field to the system may significantly affect the entanglement distribution and properties among the spins in the asymptotic steady state of the system. Particularly, applying an inhomogeneous field with an inward (growing) gradient toward the central spin is found to considerably enhance the nearest neighbor entanglement and its robustness to the thermal dissipative decay effect in the completely anisotropic (Ising) system, whereas all the beyond nearest neighbor entanglements vanish entirely.
Applying the same field to a partially anisotropic (XYZ) system, does not only enhance the nearest neighbor entanglements and their robustness but also all the beyond nearest neighbor ones. Nevertheless, the inhomogeneity of the field shows no effect on the asymptotic behavior of the entanglement in the isotropic (XXX) system, which vanishes under any system configuration. Moreover, the same inhomogeneous field exhibits the most influential impact, compared with the other ones, on the the spin dynamics as well. Although in the isotropic system the spins relax to a separable (disentangled) steady state with all the spins reaching a common spin state regardless of the field inhomogeneity, the spins in the steady state of the completely anisotropic system reach different distinguished spin states depending on their positions in the lattice. However, in the XYZ system, though the anisotropy is lower, the spin states become even more distinguished, accompanying the long range quantum correlation across the system, which is a sign of a critical behavior taking place at this combination of system anisotropy and field inhomogeneity.
}

\end{abstract}
\maketitle
\section{Introduction}

Quantum entanglement is considered as the physical resource responsible for manipulating the linear superposition of the quantum states in many body quantum systems \cite{Peres1993}. Entanglement, and its derivatives, show scaling behavior as the quantum system crosses a quantum phase transition critical point \cite{Sachdev2001}. Particularly, it is crucial in quantum information processing fields such as quantum teleportation, cryptography, and quantum computation  \cite{Nielsen2000}. However, quantum entanglement is very fragile due to the induced decoherence caused by the inevitable coupling of the quantum system to its surrounding environment \cite{Zurek1991, Bacon2000}. The main effect of decoherence is to randomize the relative coherent phases of the possible states of the quantum system diminishing its quantum aspects. As a result, it is considered as one of the main obstacles toward realizing an effective quantum computing system. The decoherence in the system sweepes out entanglement between the different parties of the system. Therefore, creating, quantifying, transferring and protecting entanglement in quantum states of many body systems is in the focus of interest of both theoretical and experimental research.

Many of the newly engineered quantum systems that are considered promising candidates for the underlying technology of quantum information processing, such as cold atoms in optical lattices, ultracold atoms, optical microcavities, trapped ions and superconducting circuits \cite{Diehl2010, Torre2010,Ritsch2013,Carusotto2013,Markus2012,Heule2011, Houck2012,Fitzpatrick2017}, represent great experimental framework for studying dissipative effects in driven many-body quantum systems. On the other hand, the Heisenberg interacting spin systems have been in focus of interest for their own sake as they describe the novel physics of localized spins in magnetic systems as well as for their successful role in modeling many of theses new customized physical systems. Moreover, many of these new systems can be used to simulate Heisenberg spin systems in a highly controllable manner.

Entanglement properties and dynamics in Heisenberg spin chains in absence of dissipative environments have been studied intensively \cite{Sadiek2010, Barouch1970, Sadiek2008, Sen(De)2004, HuangZ2006, Lieb1961, Lashin2014, XuQ2010, XuQ2011, Sadiek2013, Sadiek2019}. The dynamics of a system of interacting qubits, represented by Heisenberg spin model, coupled to a dissipative environment has been studied in several works as well. Particularly, the problem of two interacting qubits coupled to dissipative environments has been treated both analytically and numerically \cite{Wang2006,ABLIZ2006,Dubi2009}. The one-dimensional interacting spin chains, $N > 2$, coupled to dissipative environments were investigated as well at different degrees of anisotropy, magnetic field strength and temperatures \cite{Hein2005, Tsomokos2007, Buric2008, Hu2009, Hu2009a, Buric2009, Pumulo2011, Zhang2013}. The dynamics of entanglement in the Ising and isotropic ($XXX$) one-dimensional spin chains has been investigated \cite{Buric2008} using a numerical stochastic approach by applying the quantum state diffusion theory \cite{Lakshminarayan2001} to reduce the needed storage space from $2^{2N}$ to $2^N$ for $N$ interacting spins. An Ising one-dimensional spin system in an external magnetic field with two non-vanishing components in the $x$ and $z$ direction and coupled to a Markovian environment was investigated using stochastic calculations too \cite{Buric2009}. One particular work of special interest considered a one-dimensional chain of superconducting Josephson qubits with experimentally realistic conditions \cite{Tsomokos2007}. The effect of the environmental disorder and noise on the entanglement in the chain was tested. The influence of the noise was introduced as a set of bosonic baths such that each one of them is coupled to a single qubit. It was shown that this noise environment causes significant change to the entanglement dynamics of the Josephson qubits. In the limiting case when the internal degrees of freedom of the bath's were traced out the system behaved as an Ising spin chain coupled to a Born-Markovian environment with an asymptotic steady state entanglement. Other recent works have investigated the entanglement dynamics in spin systems under different environmental and external effects and focused on the entanglement and information transfer through the system \cite{Petrosyan2010, Ronke2011, Alkurtass2013,Wu2014, Ciliberti2013, Sadiek2013}.
In a previous work we have studied the entanglement dynamics in a one-dimensional spin chain in an external homogeneous magnetic field coupled a Markovian dissipative environment. We performed an exact numerical treatment for the time evolution of the system and showed how the interplay of the different system parameters, such as interaction anisotropy,  magnetic field strength and temperature can control the asymptotic steady state of the system \cite{Sadiek2016}. 

Recently there has been great interest in studying unconventional magnetism in spin systems in the absence and presence of dissipative effects, where new non-traditional magnetic phases emerged as a result of varying the system key parameters such as the system anisotropy and the inhomogeneity of the external magnetic field \cite{Lee2013, Sun2003, Asoudeh2005, Zhang2005, Hu2007, Hassan2010, Guo2011, Zhang2011, Albayrak2010, Guo2010, Hu2014, Rios2017}. While the ground state properties of the system were found to dictate its behavior in the equilibrium critical phenomena at zero temperature, the steady state density matrix was the major player in the presence of dissipative effects.
In a pioneering work, Lee et al. studied an anisotropic XYZ Heisenberg system of localized spins on a d-dimensional lattice at zero temperature under dissipative spin-flip process, associated with optical pumping. They showed how the asymptotic behavior of the system can exhibit new novel magnetic phases, as the degree of anisotropy of the spin-spin interaction is varied in the absence of external magnetic fields \cite{Lee2013}.  
The impact of an inhomogeneous magnetic field on entanglement and coherence in a closed system of a pair of XXZ interacting {\it s} spins was investigated at zero and finite temperature \cite{Rios2017}. The critical behavior of the system and its different phases at different values of the spin ($s\ge 1/2$), field gradient, spin interaction anisotropy and temperature was investigated. It was demonstrated how the inhomogeneity of the magnetic field can be utilized to control the system energy eigenlevels and enhance its entanglement content. Also, it was shown that the limiting temperature of entanglement in this system is mainly decided by the magnetic field gradient, where at a certain temperature, the system becomes entangled above a threshold value of that gradient. 
A finite XXZ systems of arbitrary spin under inhomogeneous fields were studied too  \cite{Cerezo2017}, where it was shown that highly degenerate exactly separable symmetry breaking ground states can be obtained for a wide range of inhomogeneous field configurations of zero sum in arrays of any dimension. 
Recently, the dissipative phase transition of an anisotropic XYZ Heisenberg spin-1/2 system in two-dimensional lattice was investigated by applying the corner-space renormalization method \cite{Rota2017}. The linear response of the system was studied under the effect of 
an applied polarizing magnetic field in the {\it xy-}plane and subject to a dissipative incoherent spin relaxation process. The finite size scaling of the susceptibility of the system was carried out, which exhibited a critical behavior, where its peak value increases as a power law of the system size. The finite size scaling of the quantum fisher information indicated a critical behavior of the entanglement at the transition point.
Very recently, spin-{\it s} chains with ferromagnetic XXZ coupling in the presence of sparse alternating magnetic fields have been studied. The exact ground state of the system was investigated and was found to exhibit a non-trivial magnetic behavior, where it shows significant magnetization plateaus sustainable at large system size \cite{Cerezo2019}.  

In this paper, we study the time evolution and the asymptotic steady state of the bipartite quantum entanglement and spin relaxation in a finite two-dimensional Heisenberg spin-1/2 triangular lattice, where a single central spin is surrounded by equally distant spins, with nearest-neighbor spin interaction under the influence of dissipative Lindblad environment at zero and finite temperature. We investigate the impact of an external inhomogeneous magnetic field on the entanglement sharing, dynamics, asymptotic behavior and robustness against the thermal dissipative effect of the environment. We show how a particular inhomogeneous magnetic field setup, where its gradient is directed toward the central spin, can significantly enhance the bipartite and global bipartite entanglement among the nearest neighbor spins and boost their thermal robustness in the completely anisotropic (Ising) system and even the beyond nearest neighbors in the partially anisotropic system, which indicates that a long range quantum correlation is taking place across the lattice at this combination of inhomogeneity of the magnetic field and anisotropy of spin-spin coupling. Also, we explore the associated spin dynamics and relaxation process as we vary the inhomogeneity of the field, the anisotropy of the spin interaction and the environment temperature. We demonstrate how the same particular inhomogeneous magnetic field setup has the strongest influence, compared with all other setups, on the steady state of the spins in the system except at zero anisotropy. We show how in the steady state of the system the spins reach different states that are most distinguished from each other in the partially anisotropic system accompanying the long rang quantum correlation. 

This paper is organized as follows. In the next section, we present our model and calculations. In secs. III and IV, we study the time evolution of entanglement and the spin relaxation, respectively, in the spin system at different degrees of anisotropy. We conclude in sec. V.

\section{The Model}
\begin{figure}[htbp]
\begin{minipage}[c]{\textwidth}
 \centering
   \includegraphics[width= 9 cm]{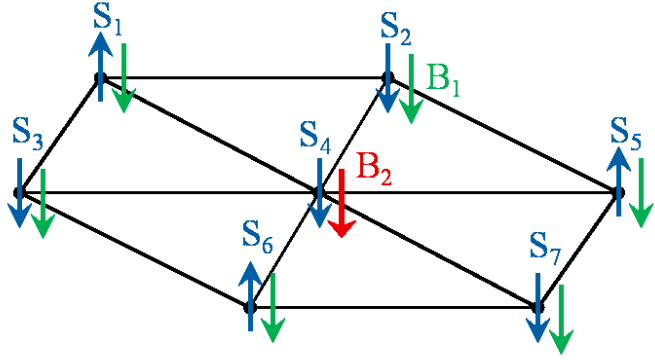}
  \caption{{\protect\footnotesize A two dimensional triangular spin lattice in the presence of an external inhomogeneous magnetic field with strengths $B_1$ at the border sites and $B_2$ at the central one.}} 
 \label{fig1}
 \end{minipage}
 \end{figure}

We consider a set of 7 localized spin-$\frac{1}{2}$ particles in a two dimensional triangular lattice coupled through nearest neighbor exchange interaction $J$ and subject to an external inhomogeneous magnetic field, as shown in Fig~\ref{fig1}. The Hamiltonian of the system is given by 
\begin{equation}
H=\frac{(1+\gamma)}{2} J \sum_{i=1}^{N} S_{i}^{x} S_{i+1}^{x} 	+  \frac{(1-\gamma)}{2} J \sum_{i=1}^{N} S_{i}^{y} S_{i+1}^{y} 	+   \delta J \sum_{i=1}^{N} S_{i}^{z} S_{i+1}^{z}	 +  \sum_{i=1}^{N} h_{i}^{z} S_{i}^{z}\;,
\label{eqn:H}
\end{equation}
where: $S^{\alpha}_{i}$ = $\frac{1}{2}$ $\sigma^{\alpha}_{i}$ ($\alpha$ = $\textit{x}$, $\textit{y}$ or $\textit{z}$) and $\sigma^{\alpha}_{i}$ are the local spin-$\frac{1}{2}$ operators and Pauli operators, respectively, for convenience we set $\hbar = k = 1$. $\gamma$ and $\delta$ are the anisotropy parameters which determines the relative strength of the spin-spin coupling in the $x$, $y$ and $z$-directions. We study different classes of the Heisenberg spin system, by changing the values of the anisotropic parameters $\gamma$ and $\delta$, such as the Ising ($\gamma=1$ and $\delta=0$), $XXX$ ($\gamma=0$ and $\delta=0.5$), $XYZ$ ($\gamma=0.5$ and $\delta=1$), etc. The system is subject to an external inhomogeneous static magnetic field applied in the z-direction such that the magnetic field strength at the border sites is $h_i=B_1 \;(i=1,2,3,5,6,7)$, whereas the strength at the central site is $h_4=B_2$. We assume the maximum external magnetic field strength is $\omega$.

The dynamics of an isolated quantum system is described by the time evolution of its density matrix $\rho(t)$ according to the quantum Liouville equation $\dot{\rho}\left(t\right) = -i \left[\textit{H},\rho\right]$. For an open quantum system interacting with its environment such that the system and the environment satisfy the Born-Markovian approximation, the time evolution of the system is best described by the Lindblad Master equation \cite{Lindblad1976, Breuer2002}, which is defined as  
\begin{equation}
\dot{\rho}\left(t\right) = -i \left[\textit{H},\rho\right] + \mathcal{D}_{\rho}\;,
\label{eqn:Lindblad}
\end{equation}
where $\mathcal{D}_{\rho}$ is the extra term that describes the dissipative dynamics and is represented in the Lindblad form as
\begin{equation}
\mathcal{D}_{\rho} = -\frac{1}{2} \sum_{j=1}^M \sum_{k=1}^N \left\{ [L_{k}^{(j)} \rho, L_{k}^{(j)\dagger}] + [L_{k}^{(j)}, \rho L_{k}^{(j) \dagger}]\right\} \;,
\label{eqn:dissipative}
\end{equation}
where the Lindblad operator $L_{k}^{(j)}$ represents the effect of the considered environment on the system site $k$, where the environment is assumed to couple to each site independently of the other sites, {\it M} is the total number of Lindblad operators and {\it N} is the total number of sites. It is more convenient to work in the Liouville space, where the density operator is represented as a vector. As a result, Eq.~(\ref{eqn:Lindblad}) can be recasted into the matrix equation form 
\begin{equation}
\vec{\dot{\rho}}(t) = (\hat{\mathcal{L}}^H+\hat{\mathcal{L}}^D)\vec{\rho} = \hat{\mathcal{L}}\vec{\rho}\;,
\label{eqn:matrixform}
\end{equation}
where $\hat{\mathcal{L}}^H$ and $\hat{\mathcal{L}}^D$ are superoperators acting on the vector $\rho$ in the Liouville space, where the first one represents the unitary evolution due to the free Hamiltonian while the second represents the dissipation process. The solution of Eq.~(\ref{eqn:matrixform}) yields the density vector as
\begin{equation}
\vec{\rho}(t)= \sum_{i} A_i \; \vec{\eta}_i \; e^{\lambda_i \; t} \;,
\end{equation}
where the coefficients $A_i$ are determined from the initial conditions of the evolution process, $\left\{\lambda_i \right\}$ and $\left\{\vec{\eta}_i\right\}$ are the sets of all eigenvalues and eigenvectors of the tetrahedral matrix $\mathcal{L}$ \cite{Sadiek2016}.
For a two-dimensional system with $N$ spin-1/2 particles, the dimension of the Hilbert space is $2^N$ and the dimension of the tetrahedral matrices is $2^{2N}$ which, even for a small number of spins, is extremely large. 

For the considered spin system, the effects of thermal relaxing and exciting environment, respectively, are represented by the two operators
\be
L_{k}^{(1)} =  \sqrt{\Gamma} (\sqrt{(\bar{n} + 1)} \; S_{i}^{-}), \;\;\;  L_{k}^{(2)} = \sqrt{\Gamma}  (\sqrt{\bar{n}} \; S_{i}^{+} ) \;\;\;\;\;\;\;\;\;\;\;\;\;\;\;\;\;\;\;\;\;\;\;\;\;\;\;\; (k=1,2,\dots, N ) \;,
\label{L_operators}
\ee
where $S^{+}$ and $S^{-}$ are the spin raising and lowering operators, $S^{\pm}$=$S^{x}$ $\pm$ $iS^{y}$.
The quantity $\bar{n}$ accounts for the thermal influence of the environment and is proportional to its temperature, whereas $\Gamma$ is a phenomenological parameter that represents the interaction between the quantum system and the environment \cite{Breuer2002, Tsomokos2007, Mintert2005}. 

When a single spin-$1/2$ particle is inserted, at rest, into a homogeneous magnetic field, it precesses around the magnetic field direction with a constant angle that depends on the initial state of the spin, and with a (Larmor) frequency that is determined by the strength of the applied field.
For a spin system with an XYZ nearest neighbor interaction, as described by the Hamiltonian (\ref{eqn:H}), in absence of an external magnetic field $(h_{i}^{z}=0)$, every spin in the system experiences an effective net magnetic field due to interaction with all its neighboring spins. This magnetic field forces the spin to precess about the field direction, where the precession strength of every spin depends on its location in the system and the degree of anisotropy, the higher the anisotropy in the system is, the stronger is the precession.
When this system couples to the Lindblad environment at zero temperature, only the first term , $L_{k}^{(1)}$, in Eq.~(\ref{L_operators}) is active with a decay effect on the precessing spin that acts to align it into the negative $z$-direction, $|\downarrow\rangle$. 
Turning on the temperature activates the second term, $L_{k}^{(2)}$, which acts to align the spin into the positive $z$-direction, $|\uparrow\rangle$, however, its effect is much smaller than the first term, as can be noticed from Eq.~(\ref{L_operators}), especially at very low temperatures when the quantum character of the system is preserved. 
The asymptotic steady state of every spin and its entanglement to the other spins is determined by the interplay between the spin-spin interaction effective field, responsible for precession, and the dissipative environment decay effect. While the initial state of the system may affect the initial and intermediate dynamics of the system, the asymptotic behavior is independent of it, as we will show in our results.
In the extreme case of a completely isotropic system, the spins don not precess at all around the effective field and the dissipative decay effect dominates, at zero temperature, forcing all the spins to point downwards, parallel to each other, leading to an asymptotic separable steady state with zero entanglement. At finite temperature, the spins in the steady state stay parallel but slightly deviates from the downward direction.  
Introducing anisotropy to the system enhances the precession process, which competes the dissipative decay effect and makes the system evolve to a steady state with a finite entanglement, where each spin may end up in a different state from the others depending on its location in the lattice and its symmetry. 
Applying an external homogeneous magnetic field to the spin system adds up to the effective magnetic field and impacts the precession process in a way that depends on its magnitude and direction. 
Applying an inhomogeneous magnetic field causes a big variance in the asymptotic behavior and entanglement of each spin compared with the others depending on the field gradient magnitude and direction, which changes the entanglement distribution and sharing among the spins depending on their locations.

As mentioned before, for Eq.~(\ref{eqn:Lindblad}) to represent a good approximation for the time evolution of the system, certain restrictions have to apply to the system parameters, the coupling parameter between the system and the environment $\Gamma$ as well as the relaxation time scale of the environment dynamics should be small compared to that of the system dynamics manifested by the parameter $\omega$ representing the spin precession frequency around the $z$-axis. As a result, we consider values of $\Gamma$ and $J$ such that $\Gamma$ and $J << \omega$, where we set $\Gamma = J = 0.05\;\omega$, $\omega = 1$, and the temperature parameter $0 \leq \bar{n} \leq 0.1 \; (\sim 41 mK)$.

We adopt the concurrence as a measure of the bipartite entanglement in the system, where Wootters \cite{Wootters1998} has shown that for a pair of two-state systems $i$ and $j$, the concurrence $C_{i, j}$, which varies between $0$ to $1$, can be used to quantify the entanglement between them and is defined by $C_{i, j}(\rho_{i, j})=max\{0,\epsilon_1-\epsilon_2-\epsilon_3-\epsilon_4\}$,
where $\rho_{i, j}$ is the reduced density matrix of the two spins under consideration, $\epsilon_i$'s are the eigenvalues of the Hermitian matrix
$R\equiv\sqrt{\sqrt{\rho_{i, j}}\tilde{\rho_{i, j}}\sqrt{\rho_{i, j}}}$ with
$\tilde{\rho_{i, j}}=(\sigma^y \otimes
\sigma^y)\rho_{i, j}^*(\sigma^y\otimes\sigma^y)$ and $\sigma^y$ is the
Pauli matrix of the spin in the $y$-direction. In addition to the concurrence, which quantifies the entanglement between any two spins in the system in a pure or mixed state, it is very insightful to evaluate another  quantity, $\tau_2$, which is a measure of the global pairwise entanglement in the system and is defined as the sum of the squared pairwise concurrences, between a single spin, for instance i, and every other spin in the system, where $\tau_2 = \sum_{j \neq i} C_{i,j}^2$ \cite{Amico2004, Roscilde2004}. We study the time evolution of the system using the standard basis $\left\{\left|\uparrow\uparrow\cdots\uparrow\right\rangle,\left|\uparrow\uparrow\cdots\downarrow\right\rangle,\cdots,\left|\uparrow\downarrow\cdots\downarrow\right\rangle,\cdots,\left|\downarrow\downarrow\cdots\downarrow\right\rangle\right\}$ and starting from different initial typical states: a separable (disentangled) state, $\left|\psi_s\right\rangle=\left|\uparrow\uparrow\cdots\uparrow\right\rangle$; a partially entangled ($W$-state), $\left|\psi_w\right\rangle$=$\frac{1}{\sqrt{N}}\left(\left|\uparrow\downarrow\cdots\downarrow\right\rangle+\left|\downarrow\uparrow\cdots\downarrow\right\rangle+\cdots+\left|\downarrow\downarrow\cdots\uparrow\right\rangle\right)$ and a maximally entangled state,  $\left|\psi_m\right\rangle=\frac{1}{\sqrt{2}}\left( \left| \uparrow\downarrow\right\rangle +\left| \downarrow\uparrow\right\rangle \right) \left|\downarrow\downarrow\cdots\downarrow\right\rangle$.

\section{Dynamics of entanglement}
\subsection{Anisotropic spin system (Ising Model)}
We start by studying the two-dimensional completely anisotropic (Ising) spin system coupled to the dissipative environment. For convenience we consider the time evolution of the system in terms of the dimensionless time $T=\omega \; t$. For the rest of the paper we consider three different combinations of the magnetic fields $B_1$ and $B_2$, namely ($\omega$,$\omega$), ($\omega$,0.1$ \omega$) and (0.1$\omega$,1), which are represented in panels (a), (b) and (c) respectively in every figure in the paper, unless otherwise is stated explicitly. Also, we adopt a color code for the temperature parameter in all figures in this paper, where we use a blue (solid) line for $\bar{n}=0$, a green (dashed) line for $\bar{n}=0.001$, a red (dash-dotted) line for $\bar{n}=0.005$, a violet (dotted) line for $\bar{n}=0.01$, a black (solid with x marks) line for $\bar{n}=0.05$ and brown (dashed with x marks) line for $\bar{n}=0.1$, unless otherwise is stated explicitly.
\begin{figure}[htbp]
 \centering
 \subfigure{\includegraphics[width=8cm]{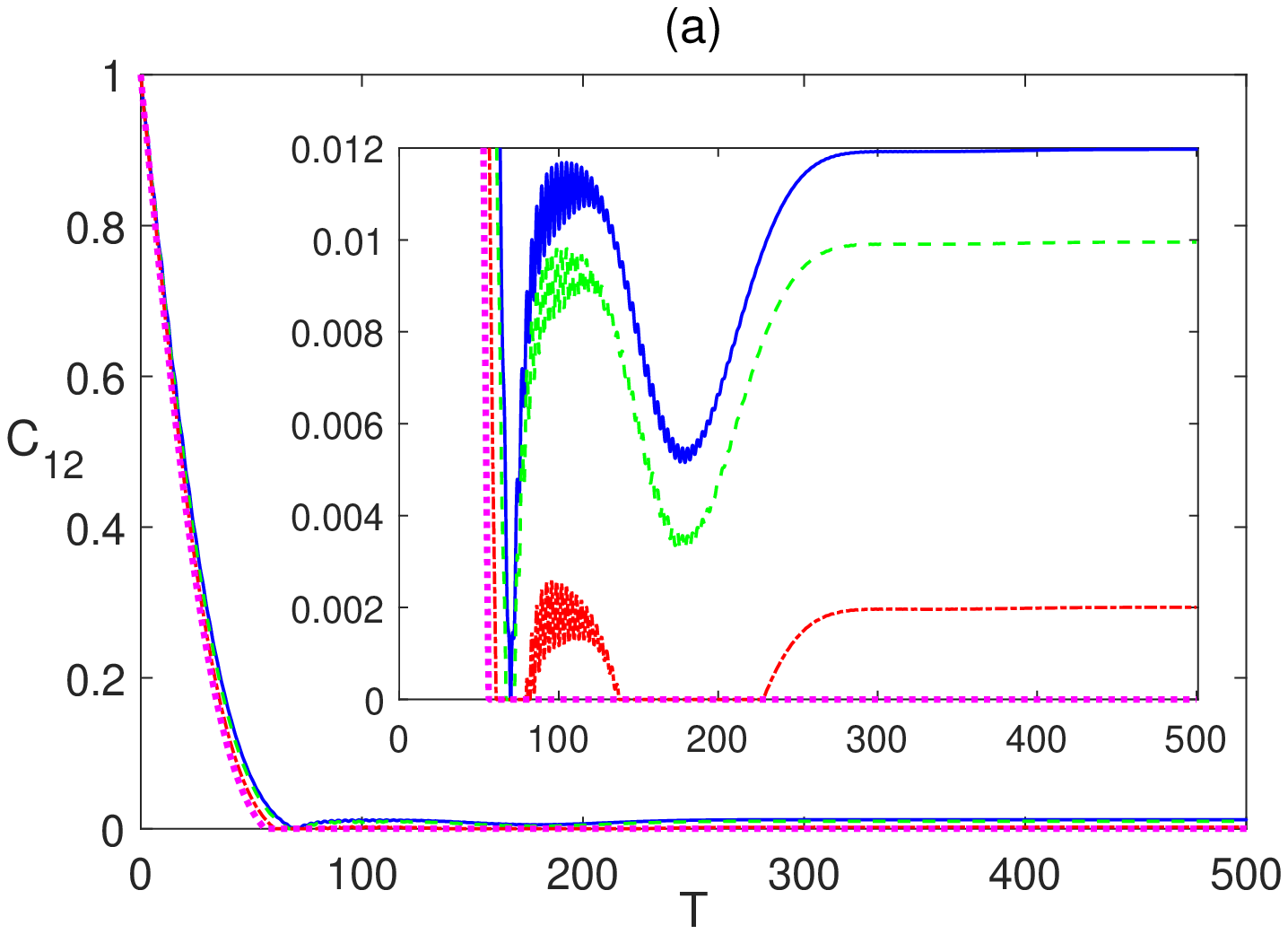}}\quad
 \subfigure{\includegraphics[width=8cm]{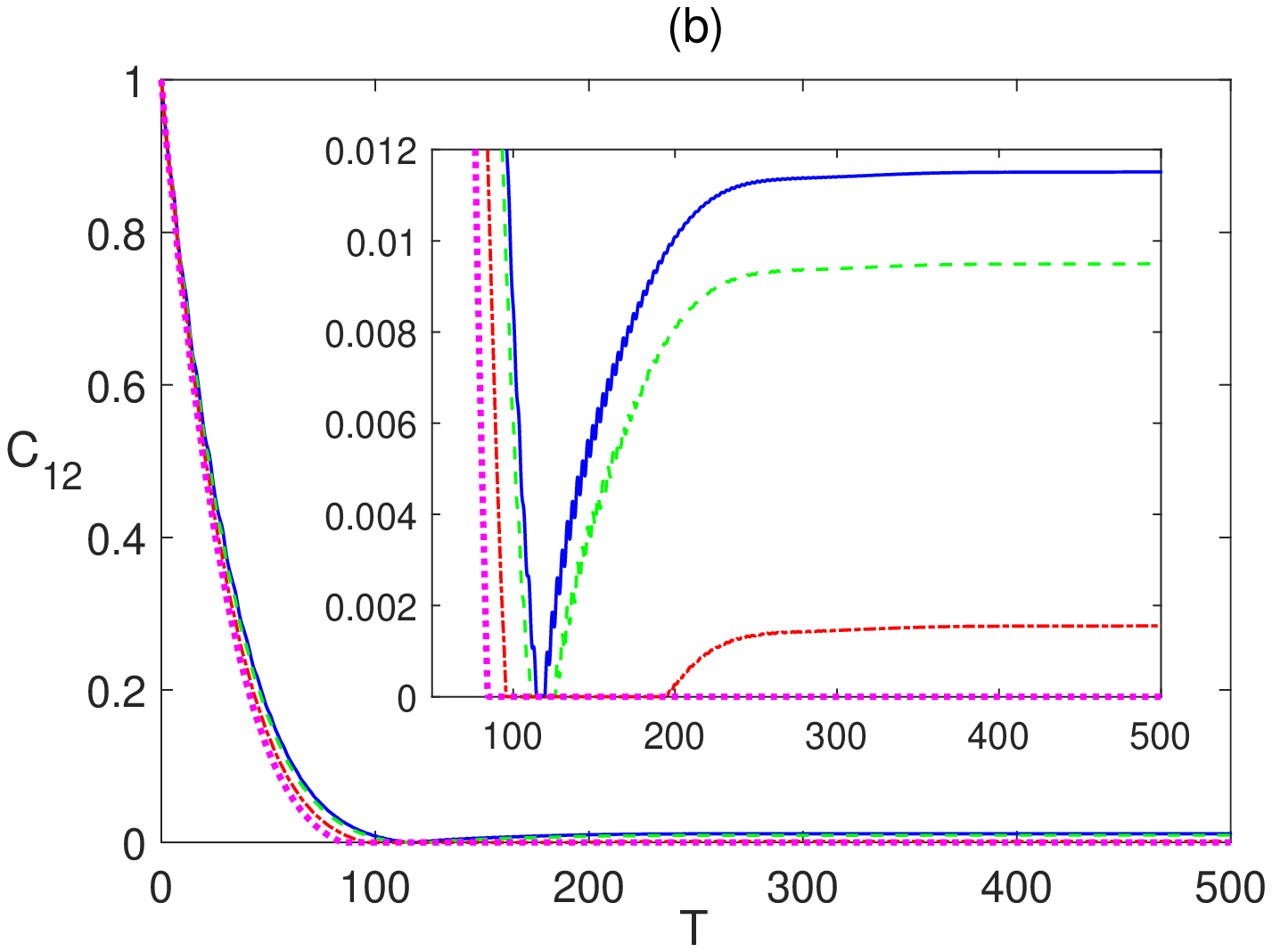}}\\
 \subfigure{\includegraphics[width=8cm]{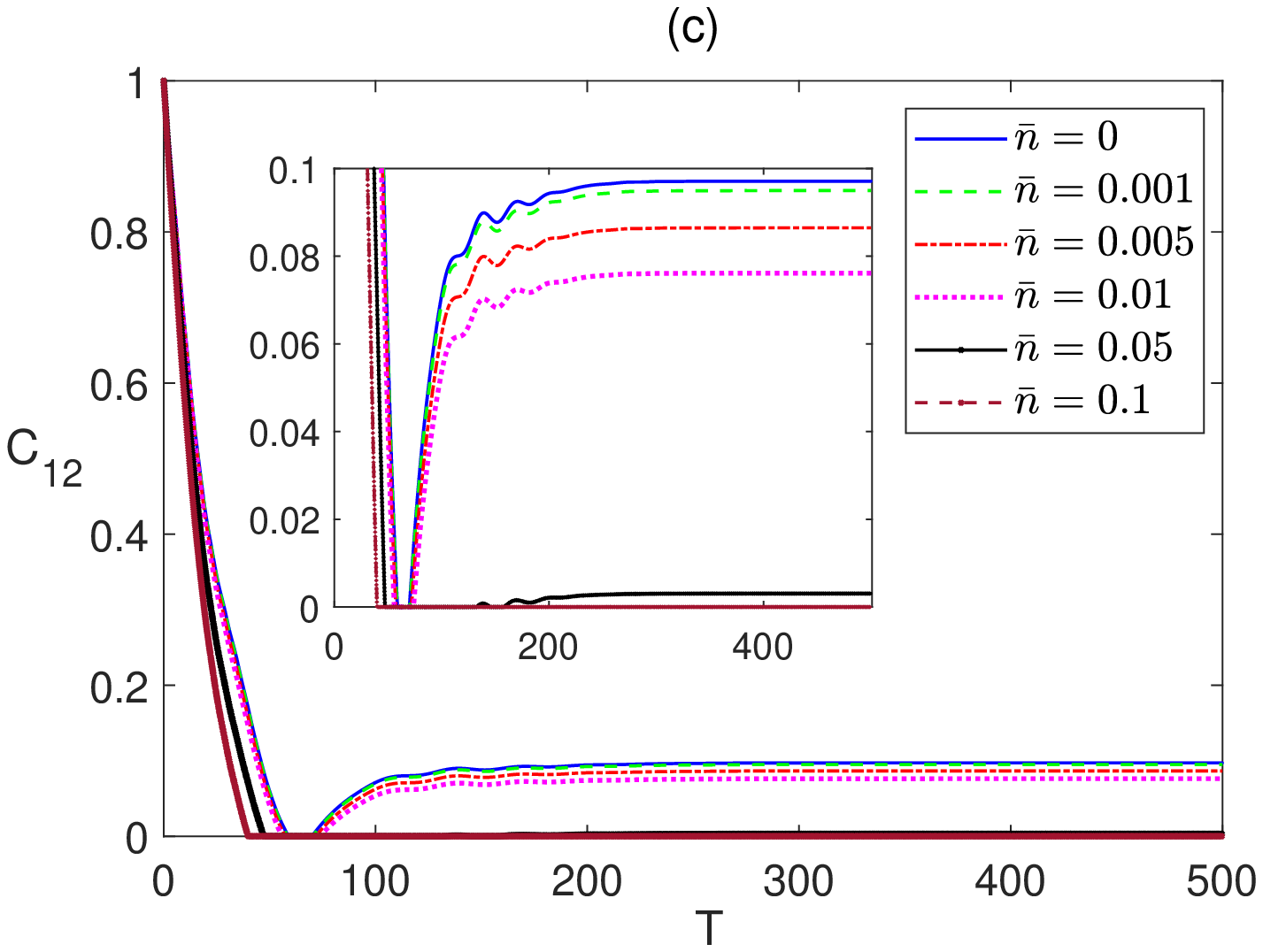}}\quad
 \caption{{\protect\footnotesize Time evolution of $C_{12}$ in the Ising system in the presence of the environment $(\Gamma=0.05)$ starting from an initial maximally entangled state at different temperatures ($0 \leq \bar{n} \leq 0.1$), and different magnetic field strengths (a) $B_1=1$ and $B_2=1$, (b) $B_1=1$ and $B_2=0.1$ and (c) $B_1=0.1$ and $B_2=1$. The legend for all panels is as shown in panel (c).}}
\label{fig2}
\end{figure}
In Fig.~\ref{fig2}, we depict the dynamical behavior of the nearest neighbor (nn) bipartite entanglement between the two border spins 1 and 2, $C_{12}$, starting from an initial maximally entangled state at different temperatures, where spins 1 and 2 are in a Bell state while all the other spins are in a separable state as described by $|\psi_m\rangle$. The inner inset plots in all panels in this paper provide a magnifying look at the asymptotic behavior of the entanglement. As can be noticed, in all the three magnetic field cases, $C_{12}$ starts at the maximum value and decays to zero before reviving again and  asymptotically reaching a final steady state. The steady state entanglement value depends on both of inhomogeneity of the magnetic field and the environment temperature. Applying a non-homogeneous magnetic field where the border magnetic field is higher than the central one, $B_1>B_2$, slightly reduces the entanglement asymptotic value as shown in Fig.~\ref{fig2}(b) compared with that of the homogeneous case presented in Fig.~\ref{fig2}(a) at all temperatures. Clearly, raising the temperature is devastating to entanglement, where the entanglement value decreases significantly as the temperature increases until completely vanishing at around $\bar{n}=0.005$, as shown in Fig.~\ref{fig2}(a). Fig.~\ref{fig2}(b) illustrates that in this inhomogeneous magnetic field case entanglement is more fragile under the thermal effect. Interestingly, applying an inhomogeneous field with higher value at the center, $B_1 < B_2$, as illustrated in Fig.~\ref{fig2}(c), enhances the steady state entanglement significantly and makes it much more robust to the thermal excitation, where it persists up to $\bar{n}=0.1$, which is about 20 times higher than that of the other two cases.
\begin{figure}[ht]
 \centering
 \subfigure{\includegraphics[width=8cm]{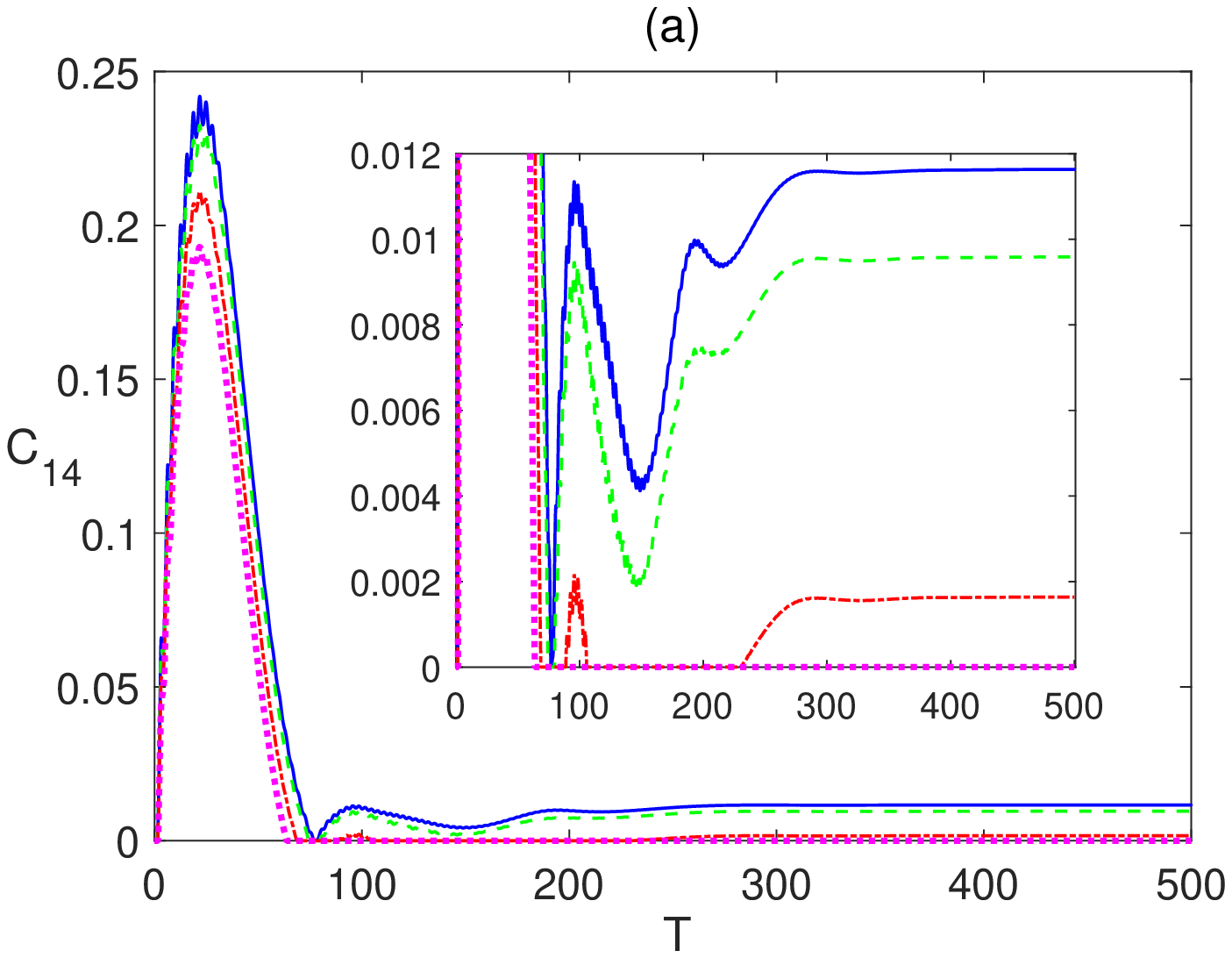}}\quad 
 \subfigure{\includegraphics[width=8cm]{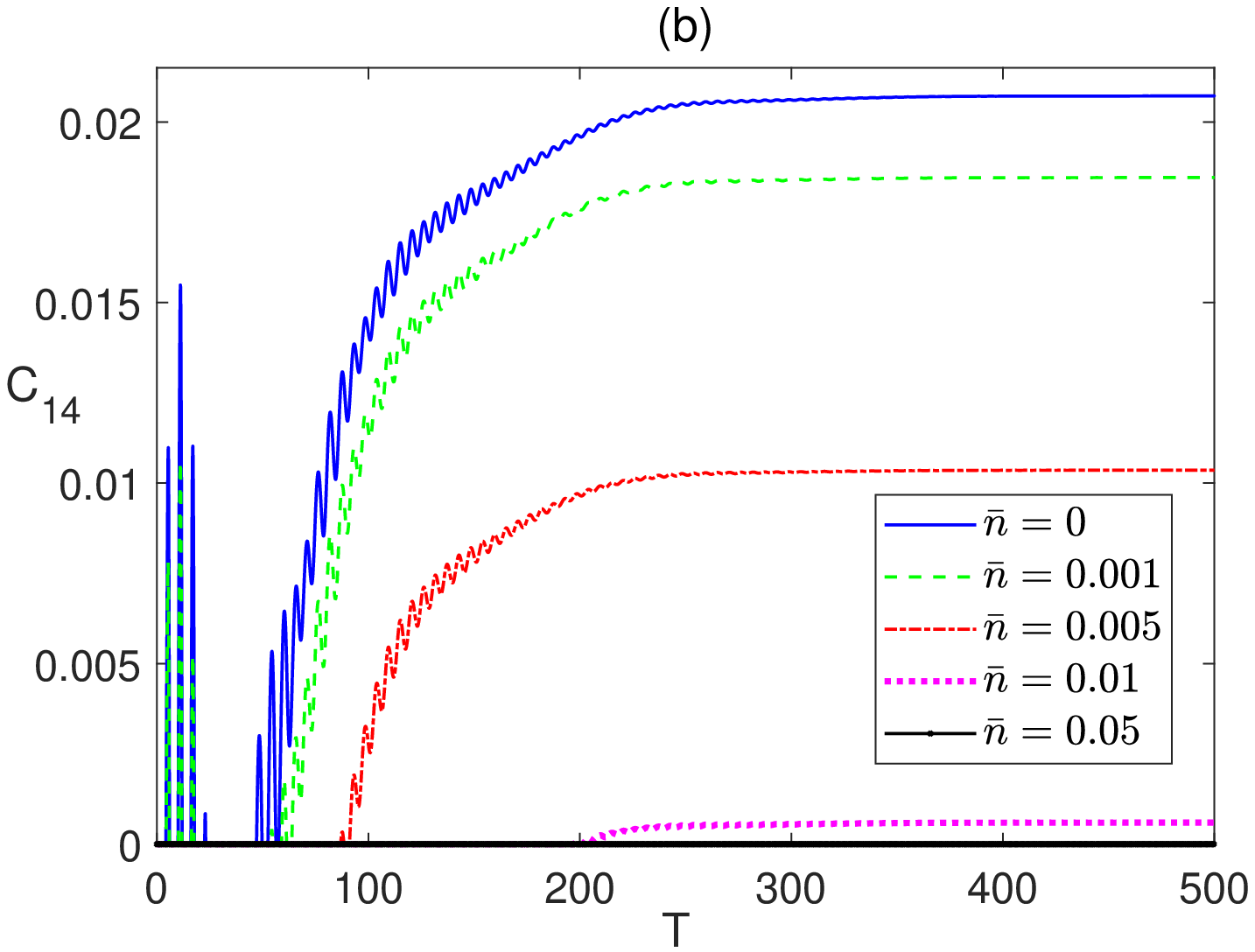}}\\
 \subfigure{\includegraphics[width=8cm]{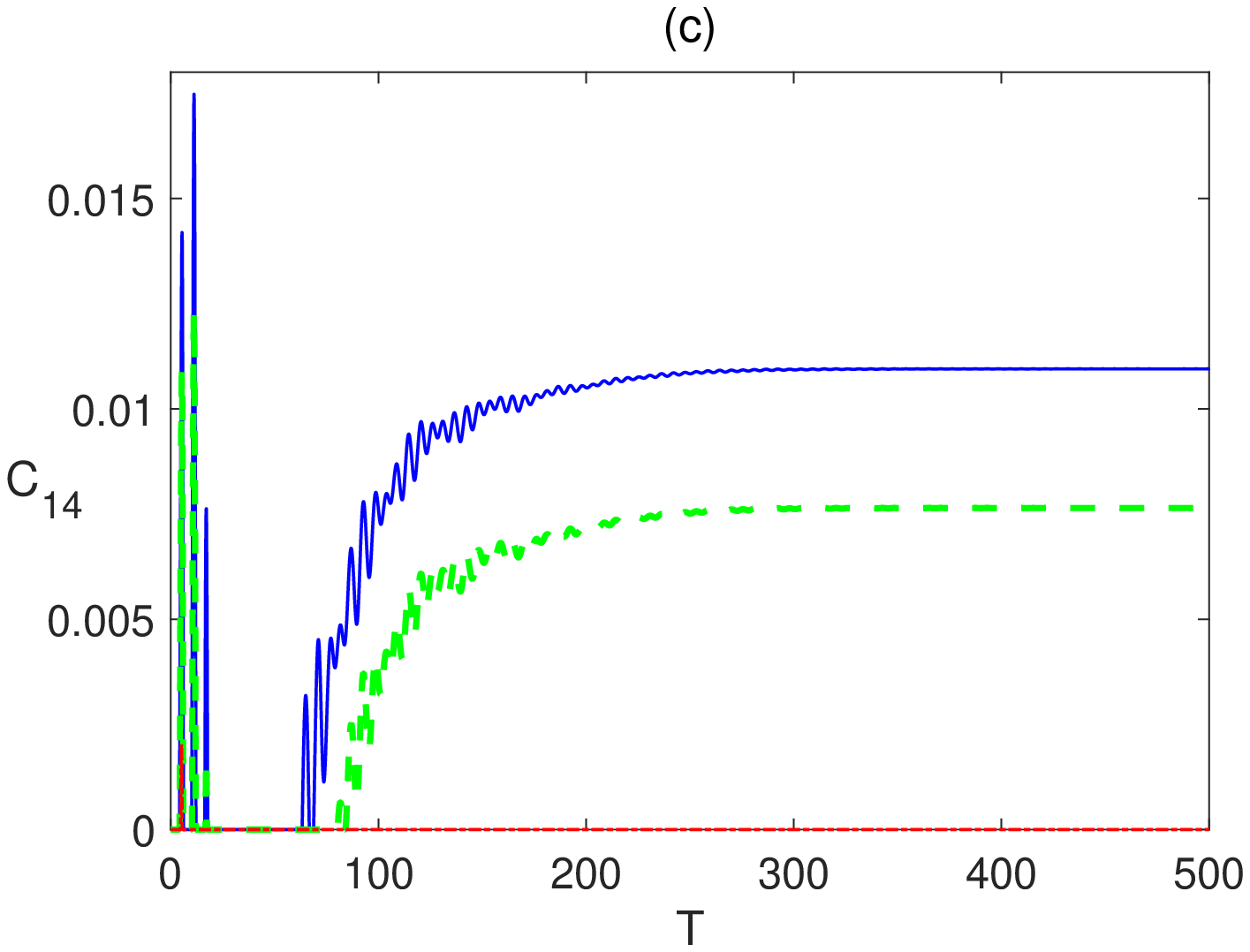}}\quad
 \caption{{\protect\footnotesize Time evolution of $C_{14}$ in the completely anisotropic, Ising, system in the presence of the environment $(\Gamma=0.05)$ starting from an initial maximally entangled state at different temperatures ($0 \leq \bar{n} \leq 0.05$), and different magnetic field strengths (a) $B_1=1$ and $B_2=1$, (b) $B_1=1$ and $B_2=0.1$ and (c) $B_1=0.1$ and $B_2=1$. The legend for all panels is as shown in panel (b).}}
\label{fig3}
\end{figure}

The time evolution of the bipartite entanglement between the border spin 1 and the central spin 4, $C_{14}$, is explored in Fig.~\ref{fig3}. The (nn) entanglement $C_{14}$ starts at zero before rising up to a maximum value then decaying again, vanishing for a period of time that increases with increasing temperature before reviving again to maintain an asymptotic steady state value. 
While the asymptotic value of $C_{14}$ is slightly lower than that of $C_{12}$ at all temperatures in the homogeneous case, as shown in Fig.~\ref{fig3}(a), which is expected as a result of the entanglement sharing of the central spin with more nearest neighbor spins compared with spin 1. However, it almost doubles when a weaker magnetic field is applied at the central spin compared with the border one with a higher robustness to temperature as illustrated in Fig.~\ref{fig3}(b). Fig.~\ref{fig3}(c) shows that the asymptotic values of $C_{14}$ are lower than that of the homogeneous case when the stronger magnetic field is applied at the central spin, $B_2>B_1$. 

\begin{figure}[htbp]
 \centering
 \subfigure{\includegraphics[width=8cm]{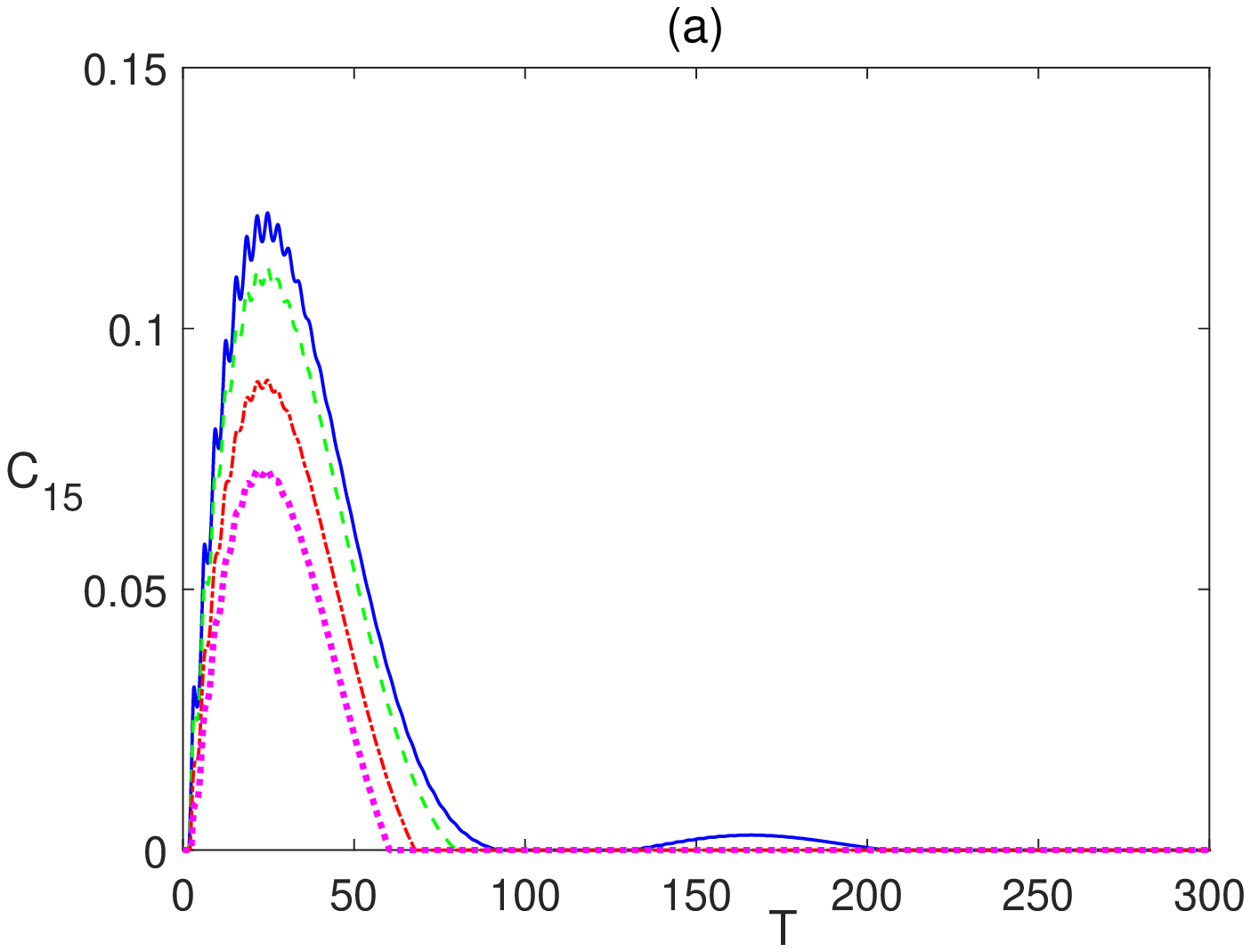}}\quad 
 \subfigure{\includegraphics[width=8cm]{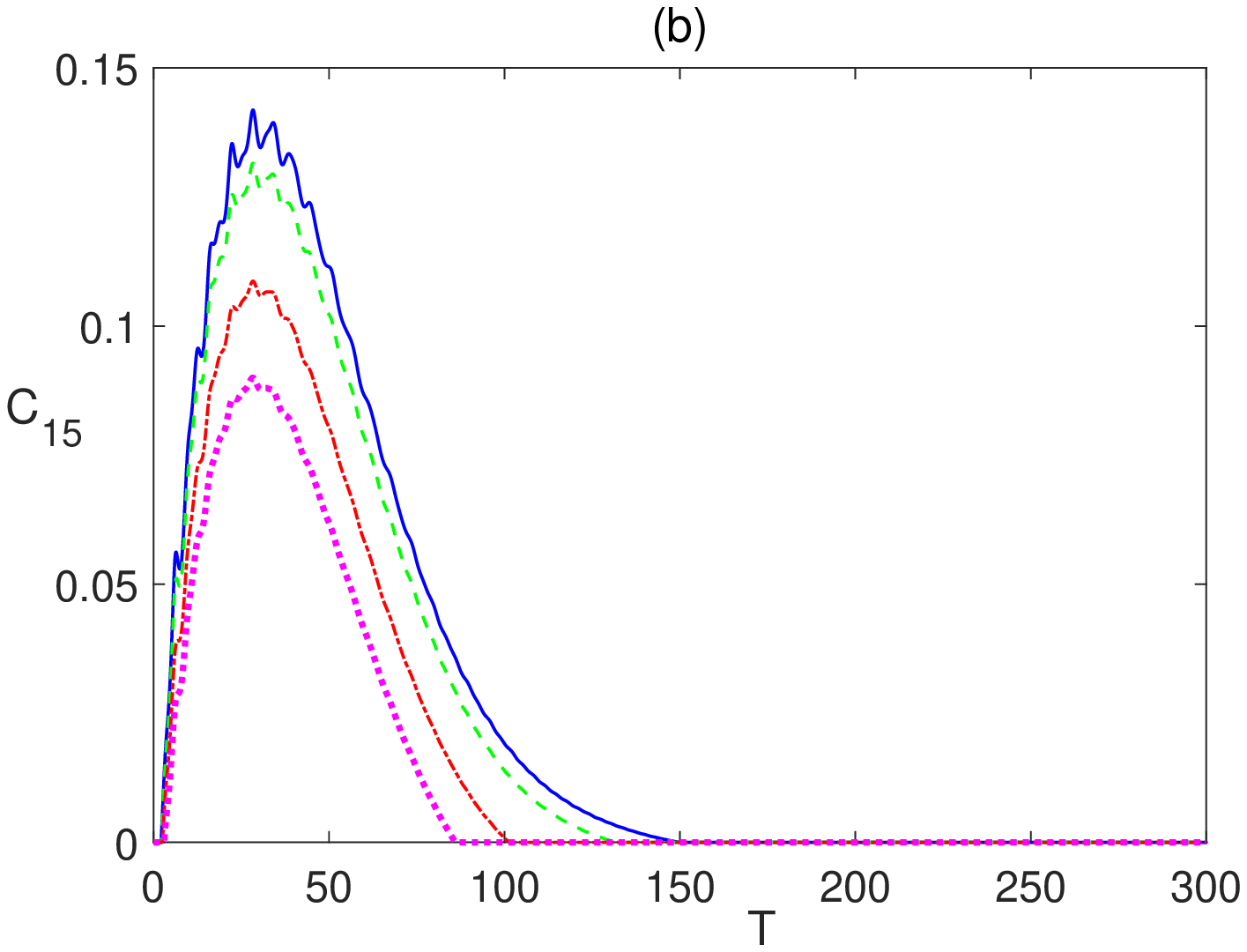}}\\
 \subfigure{\includegraphics[width=8cm]{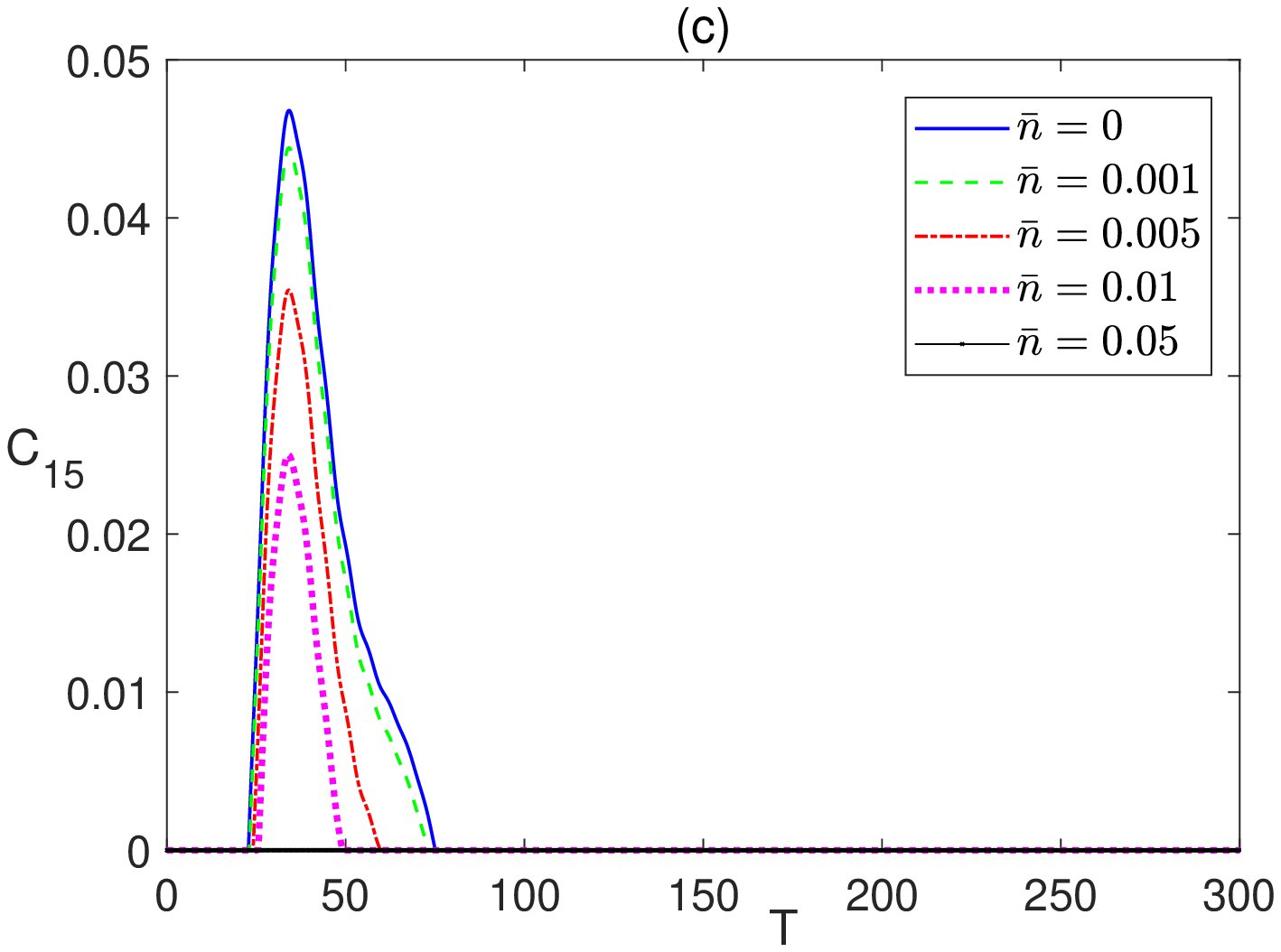}}\quad
 \caption{{\protect\footnotesize Time evolution of $C_{15}$ in the Ising system in the presence the of the environment $(\Gamma=0.05)$ starting from an initial maximally entangled state at different temperatures ($0 \leq \bar{n} \leq 0.05$), and different magnetic field strengths (a) $B_1=1$ and $B_2=1$, (b) $B_1=1$ and $B_2=0.1$ and (c) $B_1=0.1$ and $B_2=1$. The legend for all panels is as shown in panel (c).}}
\label{fig4}
\end{figure}

The dynamical behavior of the next to nearest neighbor (nnn) entanglement $C_{15}$ is depicted in Fig.~\ref{fig4}. In Fig.~\ref{fig4}(a), $C_{15}$ starts at a zero value which is maintained for a very short period of time that increases as the temperature is raised, then it increases reaching a peak value that decreases with temperature before decaying and maintaining a zero value at all temperatures except zero, where it revives again making a much smaller peak before completely vanishing. The inhomogeneous magnetic field, $B_2>B_1$, case shows a very similar behavior but with a slightly longer zero period at the beginning and higher peak value, as shown in Fig.~\ref{fig4}(b). Applying an inhomogeneous magnetic field with a weaker strength at the central spin leads to a similar behavior as the previous cases but with much lower peak values and much longer zero entanglement period at the beginning as can be seen in Fig.~\ref{fig4}(c). The entanglement $C_{17}$ was found to maintain a zero value at all times at all magnetic field combinations.

\begin{figure}[htbp]
 \centering
 \subfigure{\includegraphics[width=8cm]{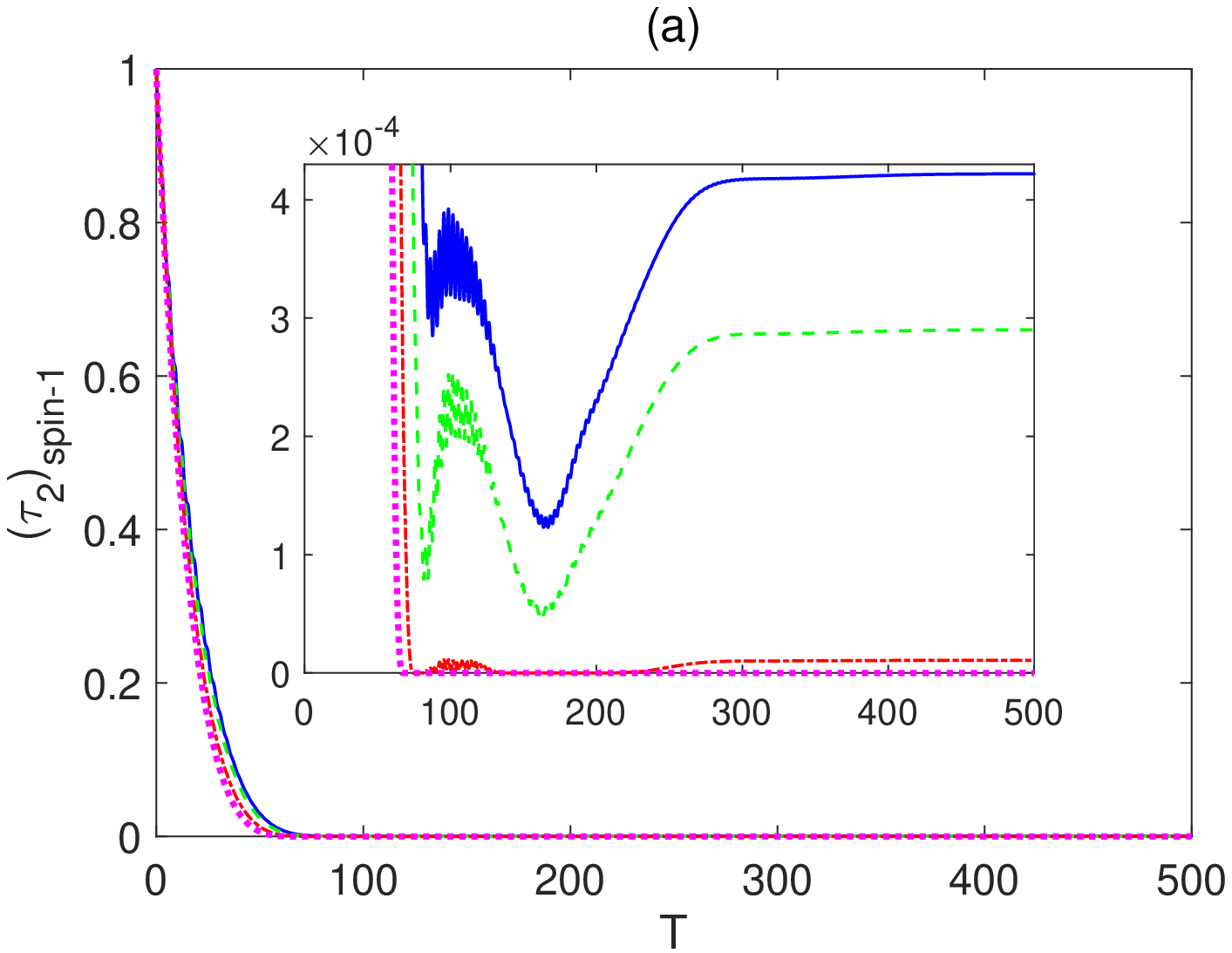}}\quad 
 \subfigure{\includegraphics[width=8cm]{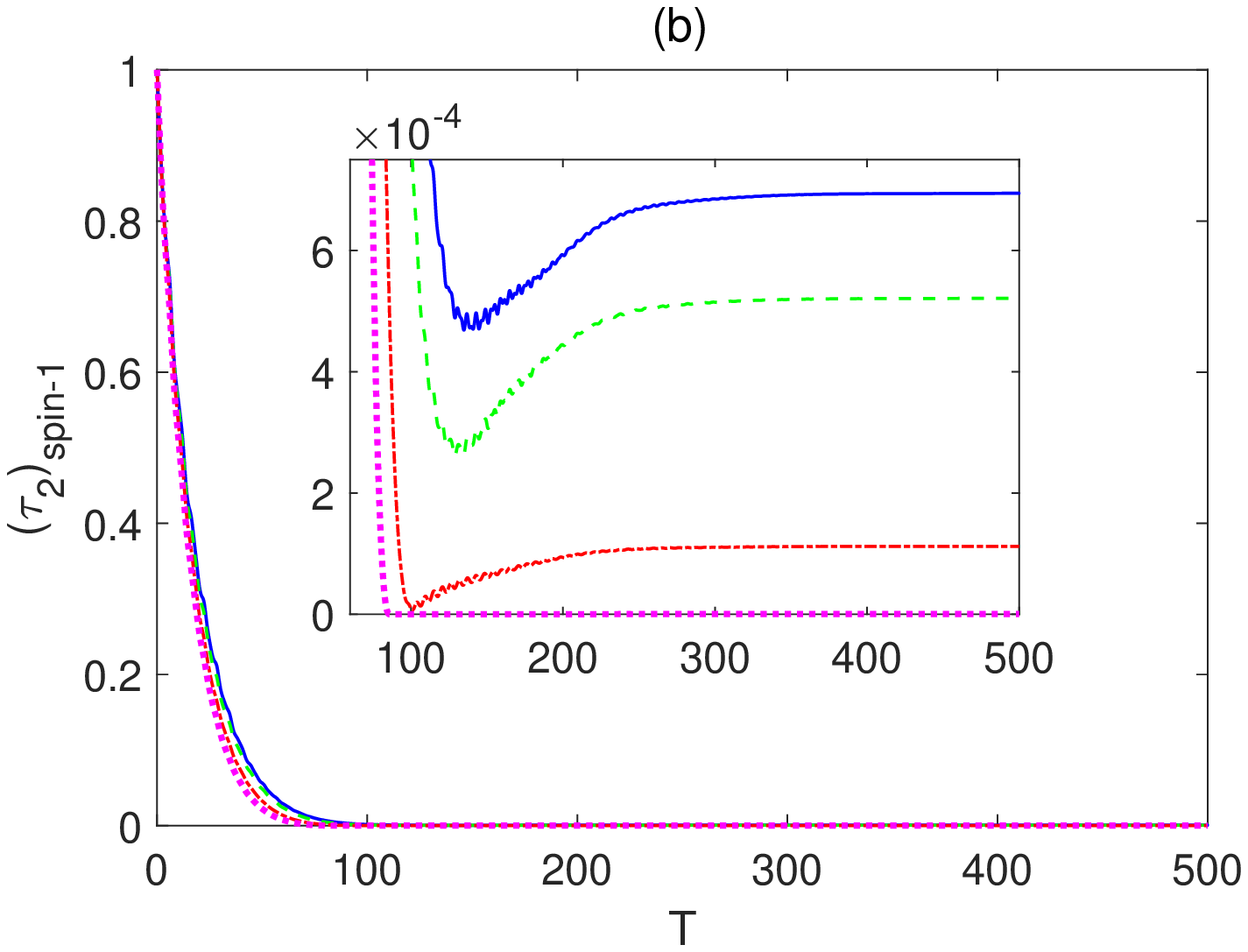}}\\
 \subfigure{\includegraphics[width=8cm]{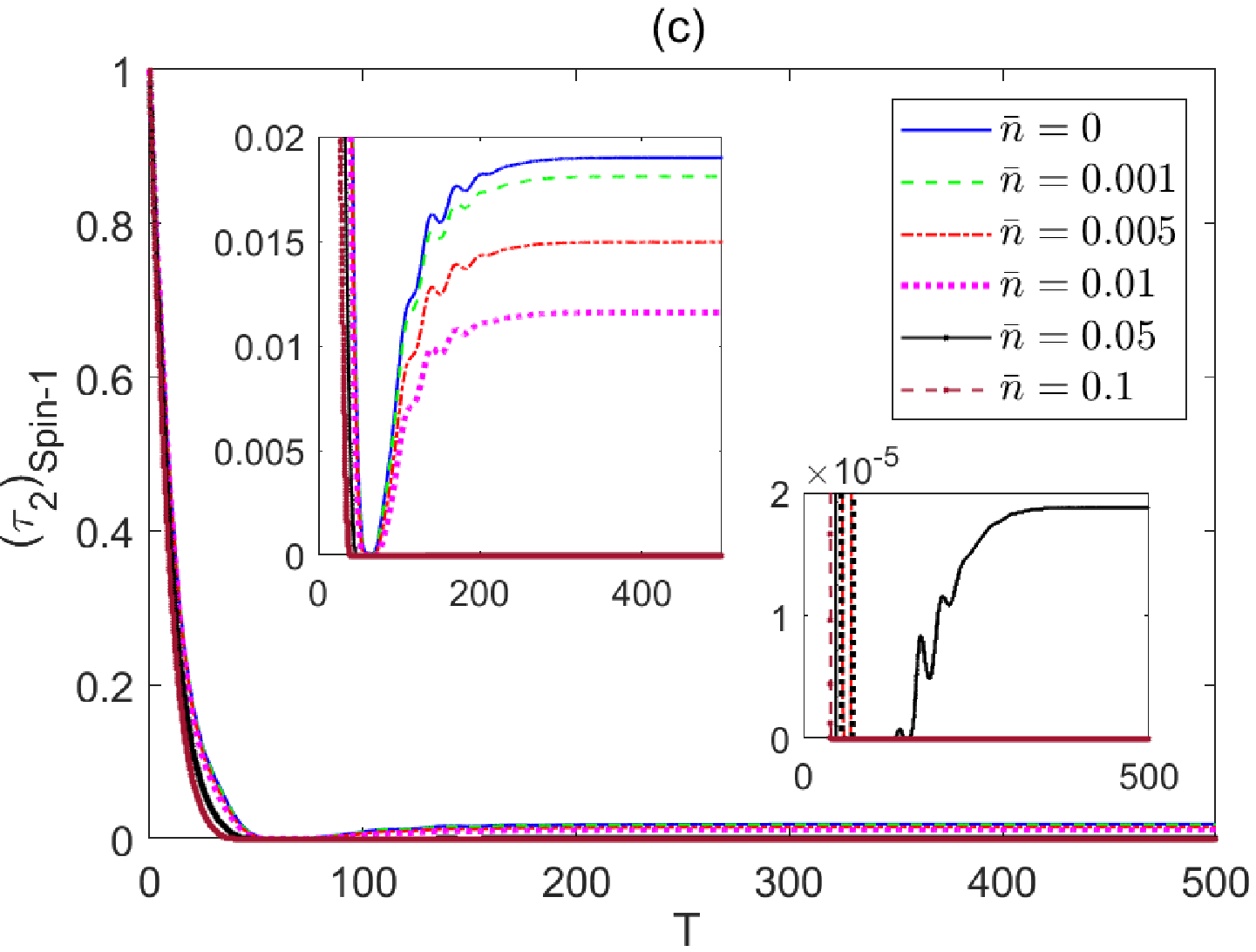}}\quad
 \caption{{\protect\footnotesize Time evolution of $\tau_2$ for spin 1 in the Ising system in the presence of the environment $(\Gamma=0.05)$ starting from an initial maximally entangled state at different temperatures ($0 \leq \bar{n} \leq 0.1$), and different magnetic field strengths (a) $B_1=1$ and $B_2=1$, (b) $B_1=1$ and $B_2=0.1$ and (c) $B_1=0.1$ and $B_2=1$. The legend for all panels is as shown in panel (c).}}
\label{fig5}
\end{figure}

\begin{figure}[htbp]
 \centering
 \subfigure{\includegraphics[width=8cm]{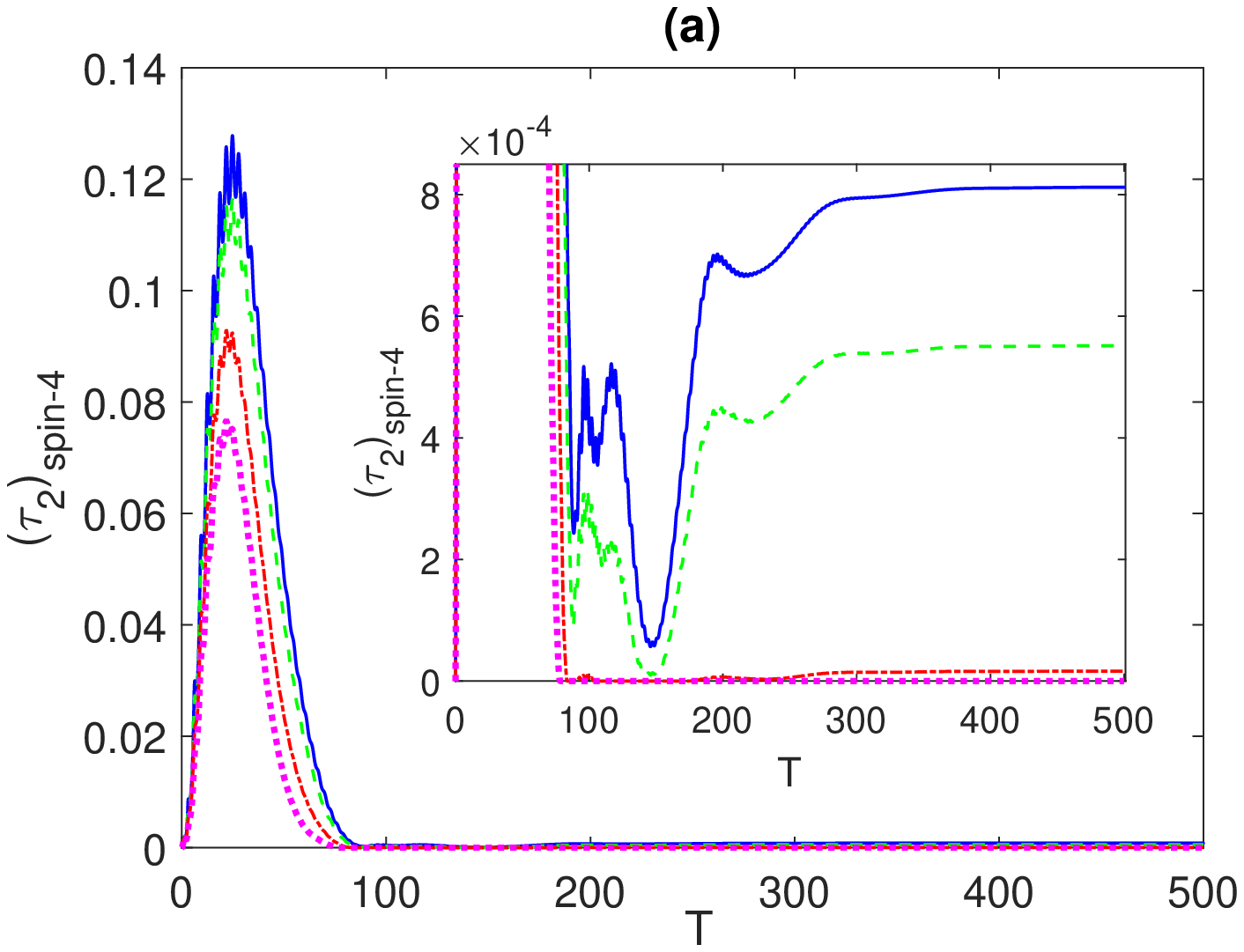}}\quad
 \subfigure{\includegraphics[width=8cm]{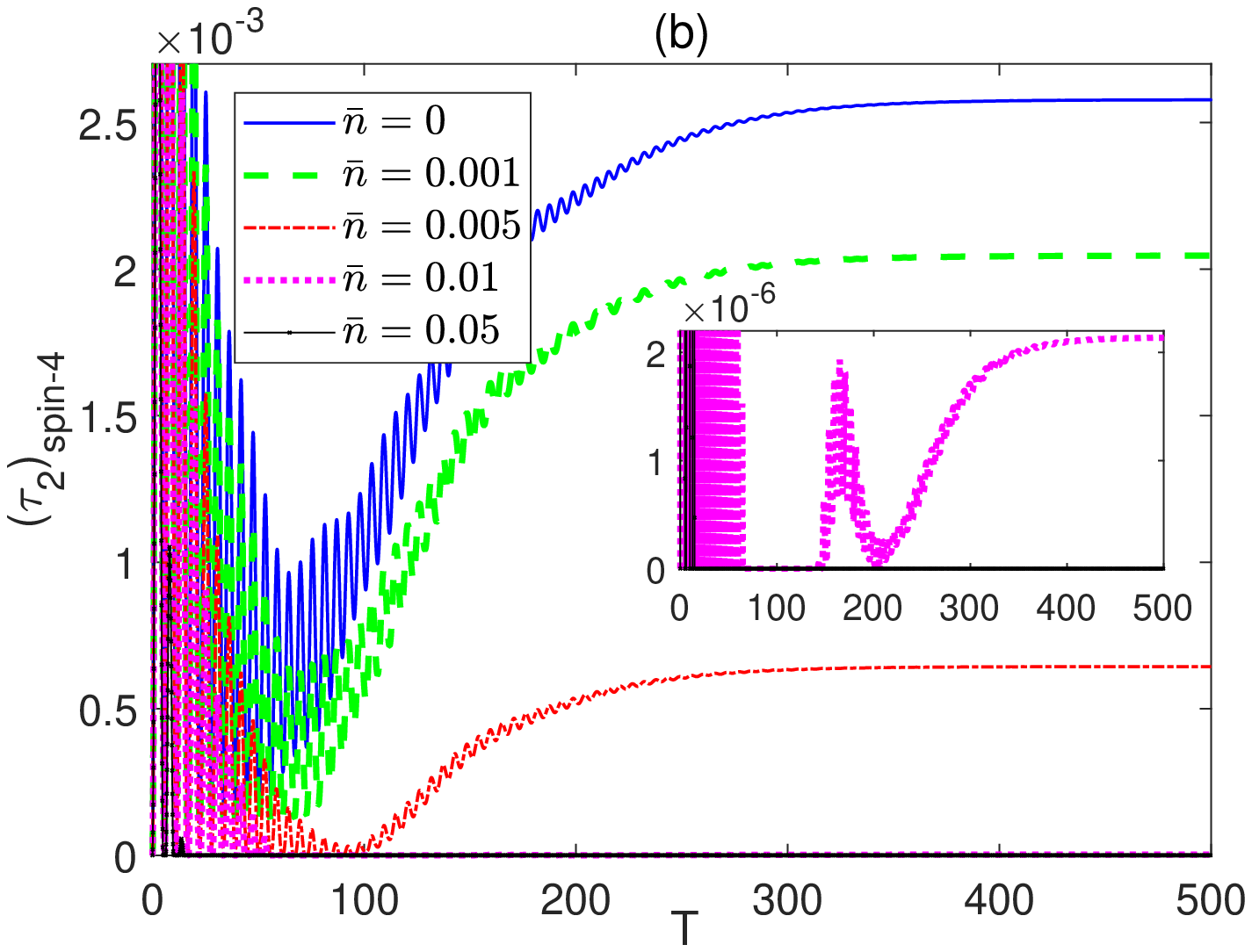}}\\
 \subfigure{\includegraphics[width=8cm]{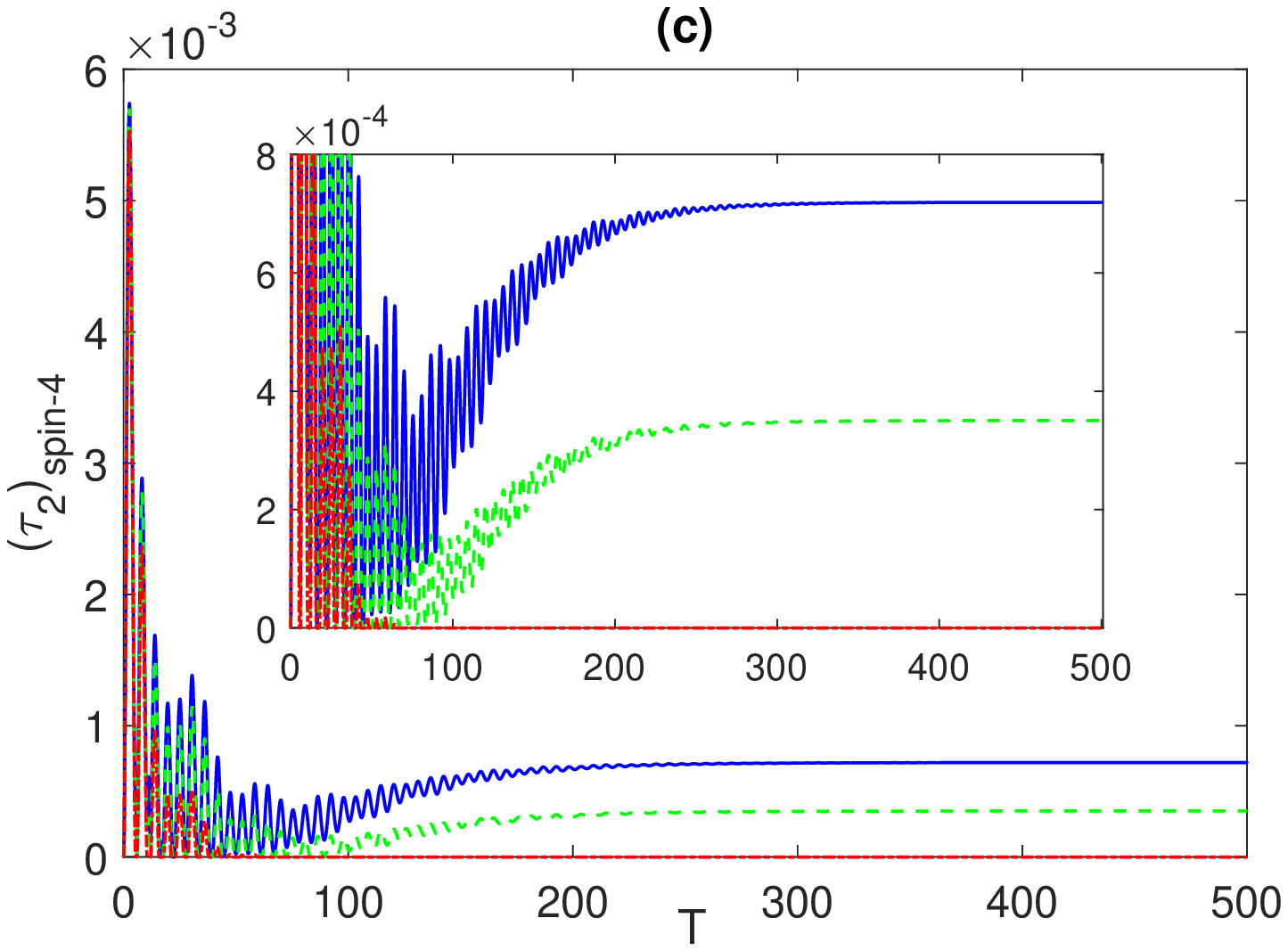}}\quad
 \caption{{\protect\footnotesize Time evolution of $\tau_2$ for spin 4 in the Ising system in the presence of the environment $(\Gamma=0.05)$ starting from an initial maximally entangled state at different temperatures ($0 \leq \bar{n} \leq 0.1$), and different magnetic field strengths (a) $B_1=1$ and $B_2=1$, (b) $B_1=1$ and $B_2=0.1$ and (c) $B_1=0.1$ and $B_2=1$. The legend for all panels is as shown in panel (c).}}
\label{fig6}
\end{figure}

The dynamics of the global bipartite entanglement $\tau_2$ for spin 1 is depicted in Fig.~\ref{fig5}. As can be noticed, the inhomogeneous magnetic field, where $B_1>B_2$, slightly increases the entanglement asymptotic value, as shown in \ref{fig5}(b), compared with the homogeneous case presented in \ref{fig5}(a), while reversing the gradient of the field, enhances the asymptotic value considerably and boosts its robustness against temperature from a critical value of about $\bar{n}=0.01$ to $0.1$. This indicates that applying an inhomogeneous magnetic field enhances the global bipartite entanglement in general for the border spins, which is not the case for the central spin, where it depends on the field gradient direction, as can be noticed in Fig.~\ref{fig6}.
The asymptotic value of the global bipartite entanglement of the central spin 4 comes out to be higher than that of spin 1 under the effect of the homogeneous magnetic field as shown in Fig.~\ref{fig6}(a), but increases significantly when the inhomogeneous magnetic field is applied, $B_1>B_2$, as illustrated in Fig.~\ref{fig6}(b), with much higher robustness against temperature. Reversing the gradient of the magnetic field such that a strong field is action on spin 4, reduces the asymptotic value and thermal robustness considerably as shown in \ref{fig6}(c). Comparing Figs.~\ref{fig5} and \ref{fig6} shows that the asymptotic values of $\tau_2$ for spin 1 is in general lower than the corresponding ones of $\tau_2$ for spin 4, except for the case of inhomogeneous magnetic field $B_1 < B_2$, i.e. when the higher magnetic field strength acts on spin 4. Therefore, in the first two magnetic field cases, the global bipartite entanglement of spin 4 is higher than that of spin 1, which is expected as spin 4 has much more neighboring spins and the only way to reverse this is to apply an inhomogeneous magnetic field with an inward increasing gradient.

\begin{figure}[htbp]
 \centering
 \subfigure{\includegraphics[width=8cm]{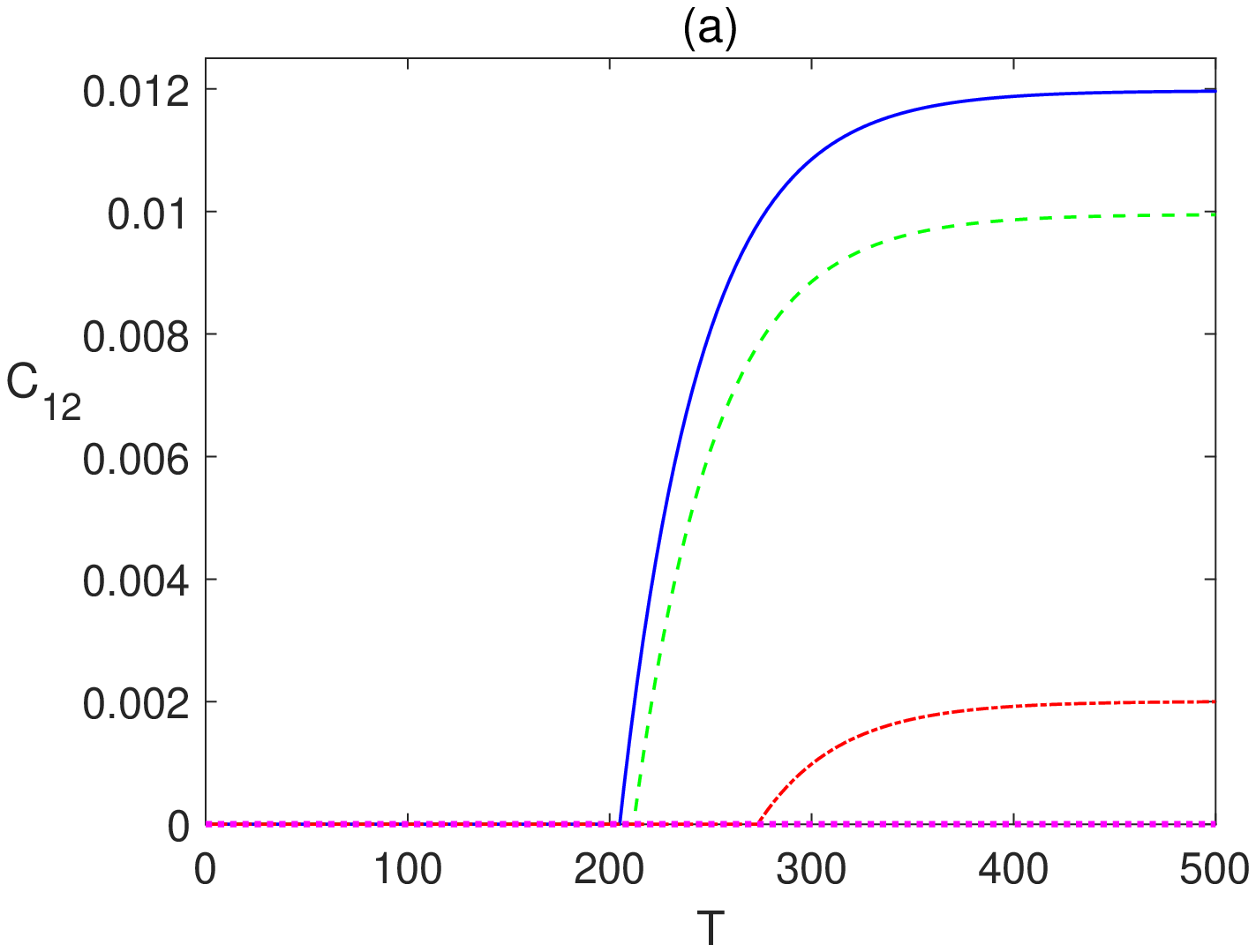}}\quad 
 \subfigure{\includegraphics[width=8cm]{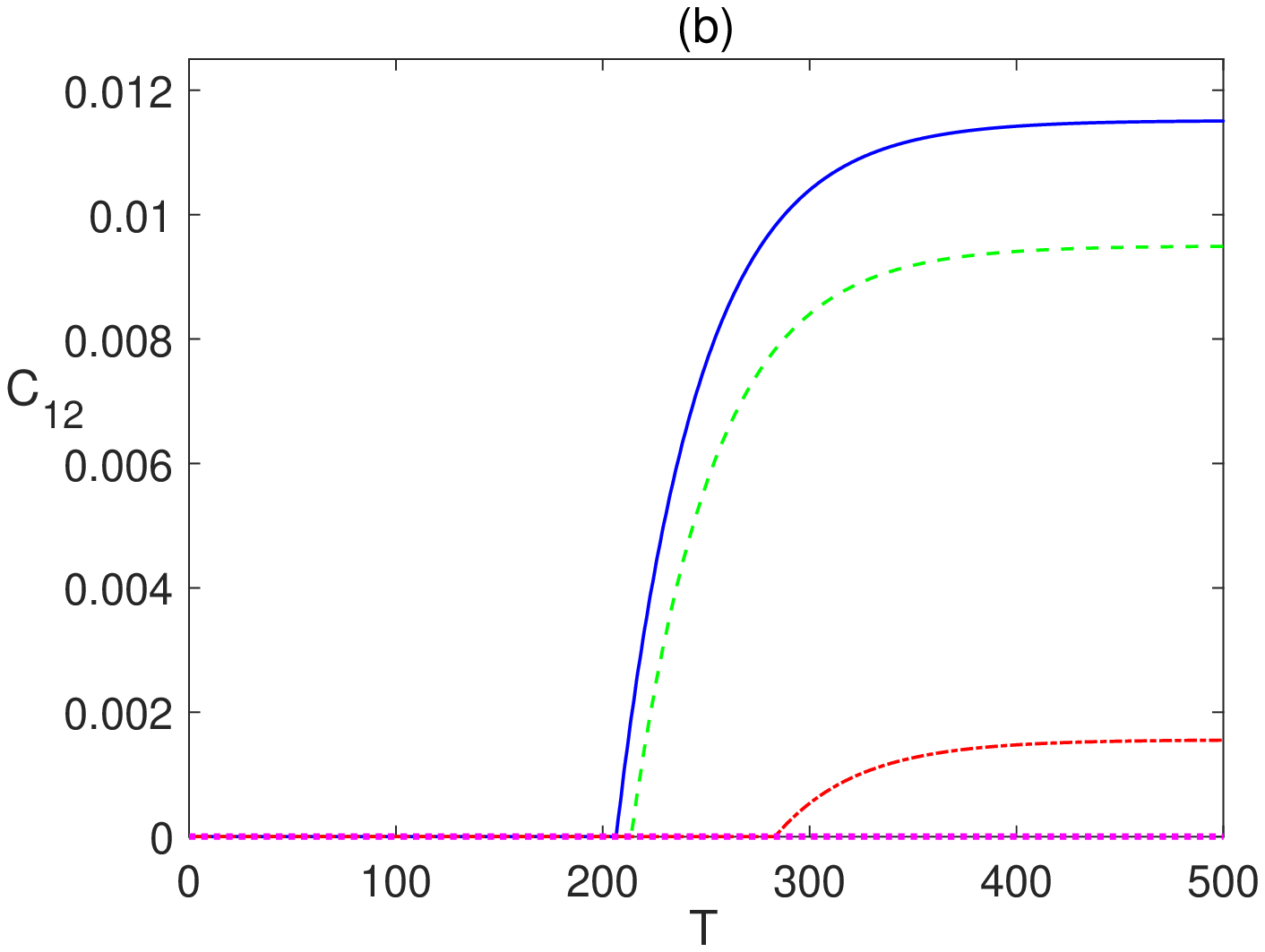}}\\
 \subfigure{\includegraphics[width=8cm]{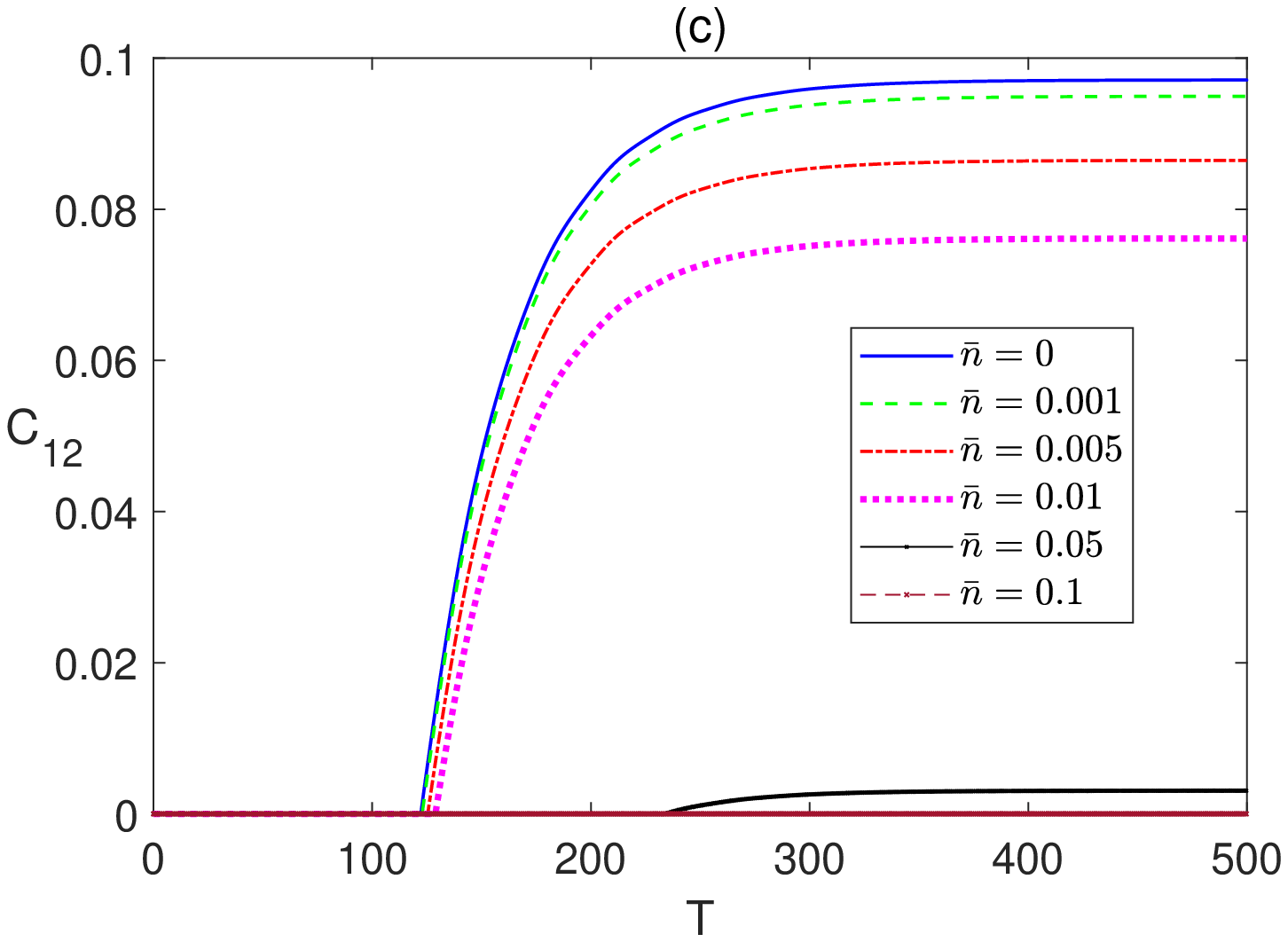}}\quad
 \caption{{\protect\footnotesize Time evolution of $C_{12}$ in the Ising system in the presence of the environment $(\Gamma=0.05)$ starting from an initial disentangled state at different temperatures ($0 \leq \bar{n} \leq 0.1$), and different magnetic field strengths (a) $B_1=1$ and $B_2=1$, (b) $B_1=1$ and $B_2=0.1$, and (c) $B_1=0.1$ and $B_2=1$. The legend for all panels is as shown in panel (c).}}
\label{fig7}
\end{figure}

\begin{figure}[htbp]
 \centering
 \subfigure{\includegraphics[width=8cm]{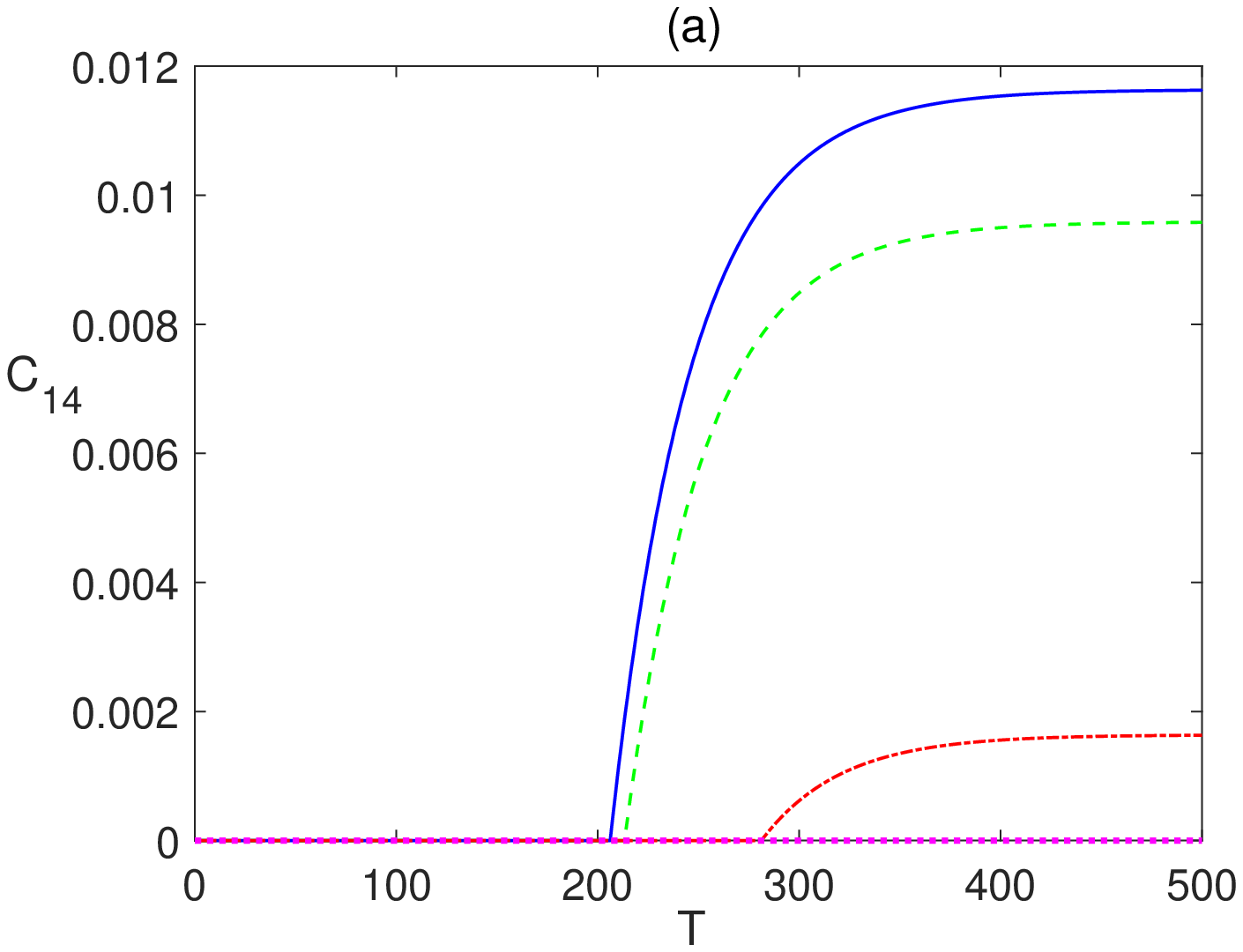}}\quad 
 \subfigure{\includegraphics[width=8cm]{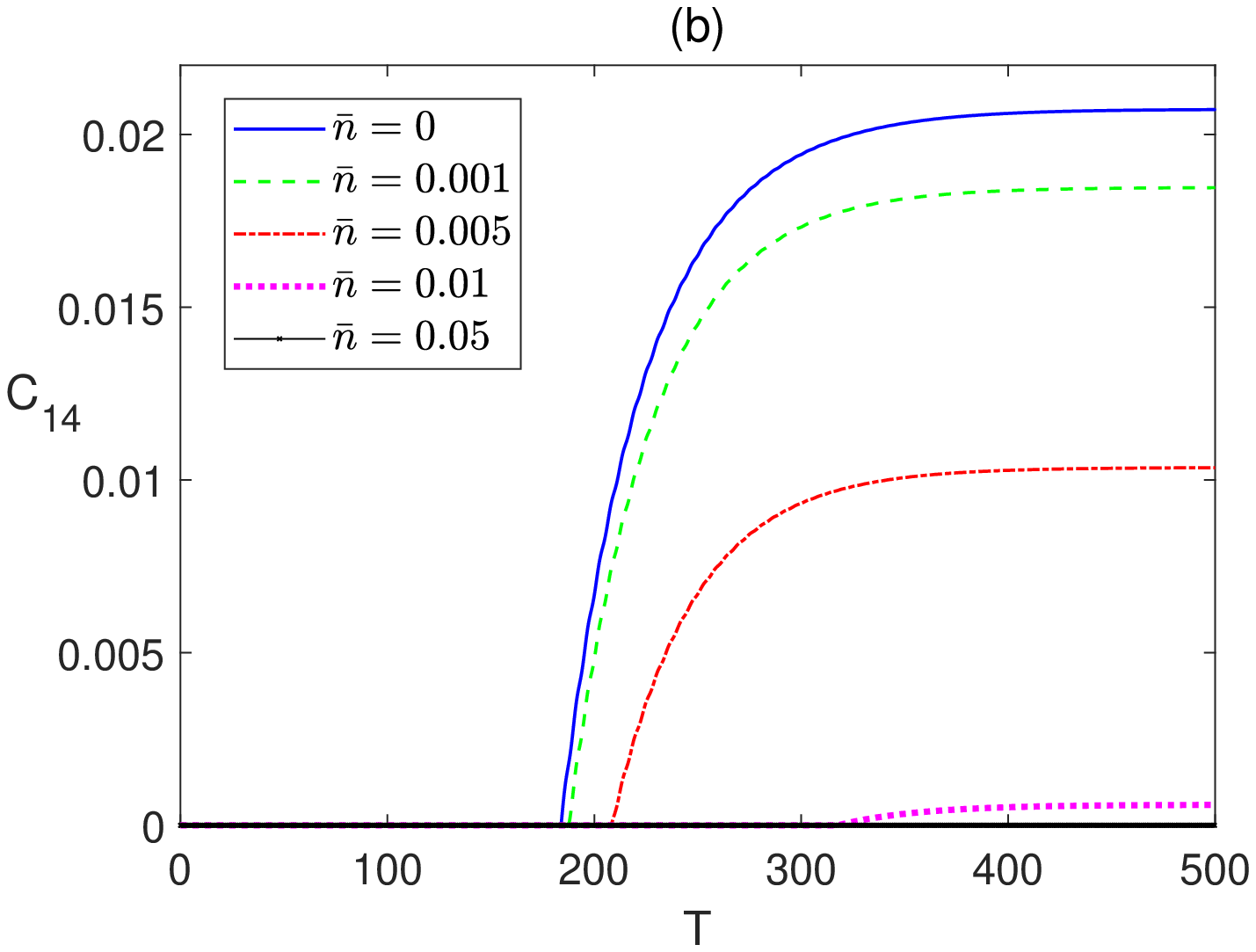}}\\
 \subfigure{\includegraphics[width=8cm]{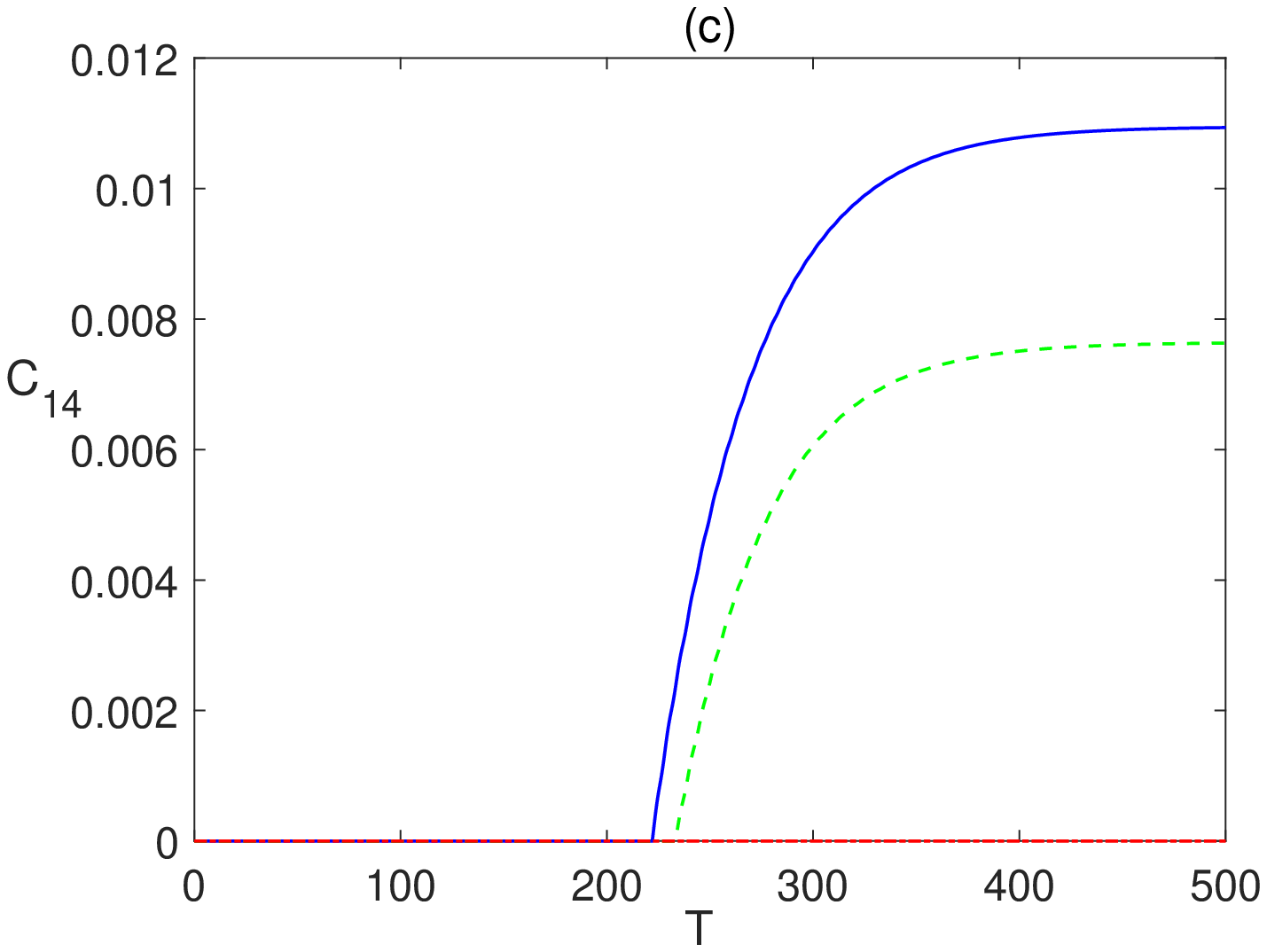}}\quad
 \subfigure{\includegraphics[width=8cm]{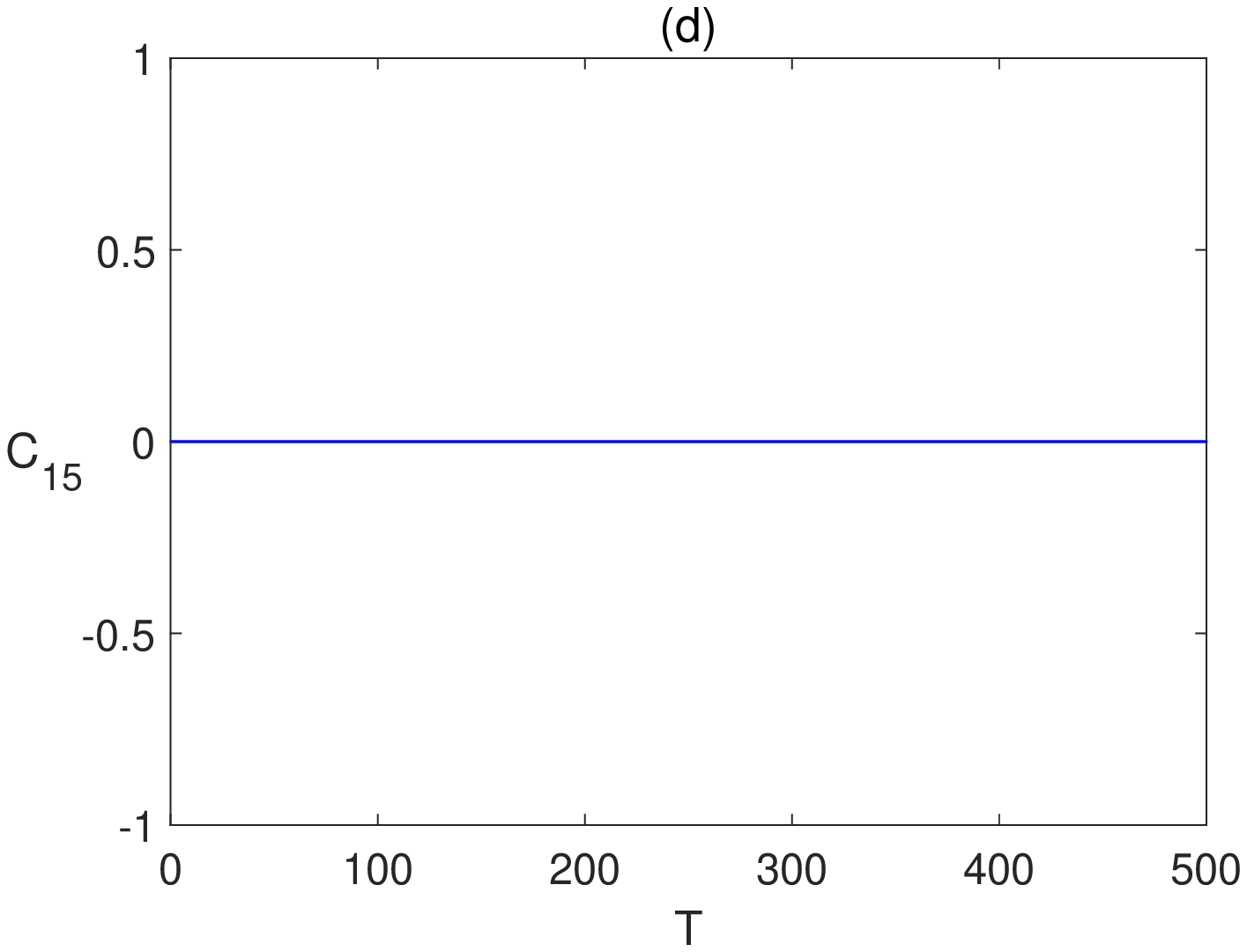}}\quad
 \caption{{\protect\footnotesize Time evolution of $C_{14}$ and $C_{15}$ (in fig (d)) in the Ising system in the presence of the environment $(\Gamma=0.05)$ starting from an initial disentangled state at different temperatures ($0 \leq \bar{n} \leq 0.05$), and different magnetic field strengths (a) $B_1=1$ and $B_2=1$, (b) $B_1=1$ and $B_2=0.1$, and (c) $B_1=0.1$ and $B_2=1$. The legend for all panels is as shown in panel (b).}}
\label{fig8}
\end{figure}

The time evolution of the Ising system starting form an initially disentangled, separable, state is presented in Figs.~\ref{fig7}-\ref{fig10}. The dynamics of $C_{12}$ is depicted in Fig.~\ref{fig7}, which shows that the entanglement starting at a zero value revives within a finite period of time to reach an asymptotic steady state value in all magnetic field arrangements. The difference between the homogeneous case, in Fig.~\ref{fig7}(a), and the inhomogeneous case, where $B_1>B_2$, in Fig.~\ref{fig7}(b) is quite small, where the steady state values are slightly reduced. But applying an inhomogeneous field, where $B_2>B_1$ raises the steady state value significantly and increases robustness against temperature as shown in Fig.~\ref{fig7}(c). Moreover, the zero entanglement period in Figs.~\ref{fig7}(a) and (b) are almost the same but much longer than the one in Fig.~\ref{fig7}(c). 

Comparing Figs.~\ref{fig7} and \ref{fig2}, one can conclude that the steady state values of $C_{12}$ are exactly the same regardless of the initial state of the system, i.e. the system evolves to the same final state independent of its initial setup, including the partially entangled state, $|\psi_w \rangle$, which we have tested as well, but is not presented here. The entanglement $C_{14}$, also, revives form zero to a steady state value that depends again on the inhomogeneity of the magnetic field, where as can be seen, the asymptotic value and thermal robustness are much higher in panel (b), where $B_2 < B_1$, compared with panels (a) and (c) in Fig.~\ref{fig8}, where particularly in Fig.~\ref{fig8}(c), $C_{14}$ becomes very fragile under the thermal effect. The nnn entanglement $C_{15}$ maintains a zero value at all times at zero and no-zero temperatures as shown in Fig.~\ref{fig8}(d). Again, comparing Figs.~\ref{fig8} and \ref{fig3}, shows that the asymptotic value of $C_{14}$ is independent of the initial state of the system. 

Studying the overall bipartite entanglements $\tau_2$ of spin 1 and 4 for an initially disentangled state shows that they start from zero value but revive and increase monotonically to reach asymptotically the same exact steady state values reported in the case of an initial maximally entangled state, which emphasizes the independence of the system steady state of the initial state of the system. The relevant figures are not shown here to save space.

\subsection{Partially anisotropic system (XYZ Model)}
\begin{figure}[htbp]
 \centering
 \subfigure{\includegraphics[width=8cm]{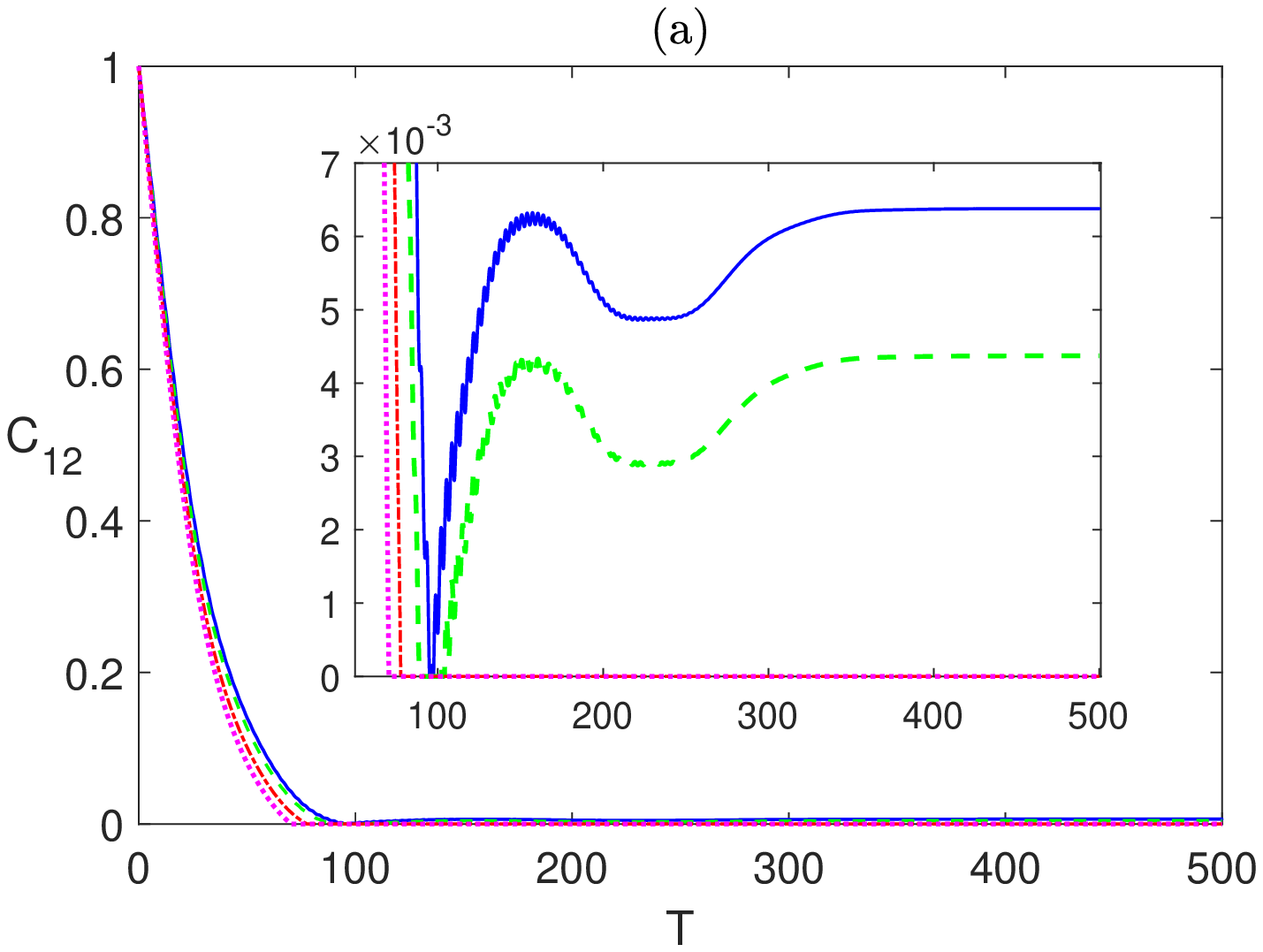}}\quad 
 \subfigure{\includegraphics[width=8cm]{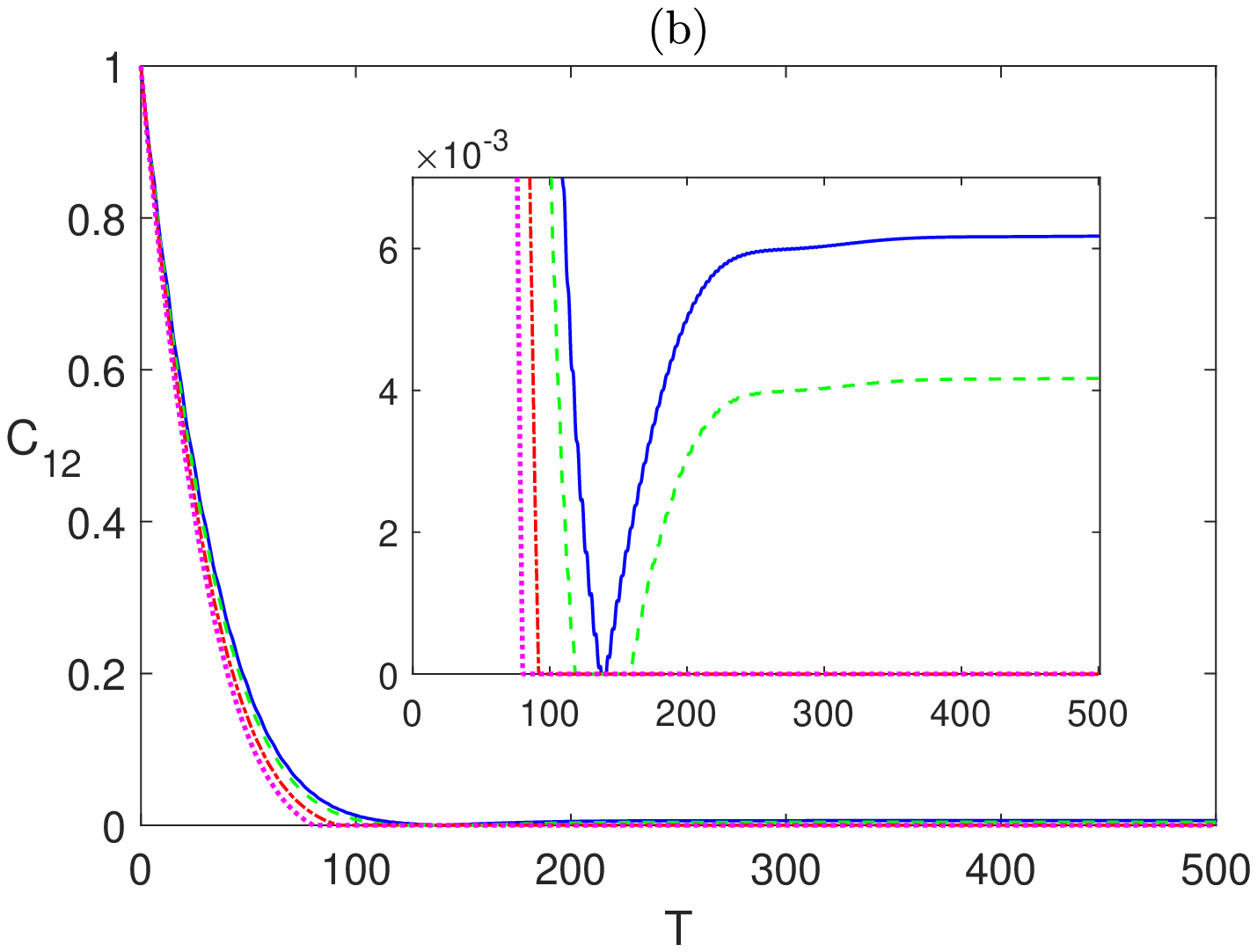}}\\
 \subfigure{\includegraphics[width=8 cm]{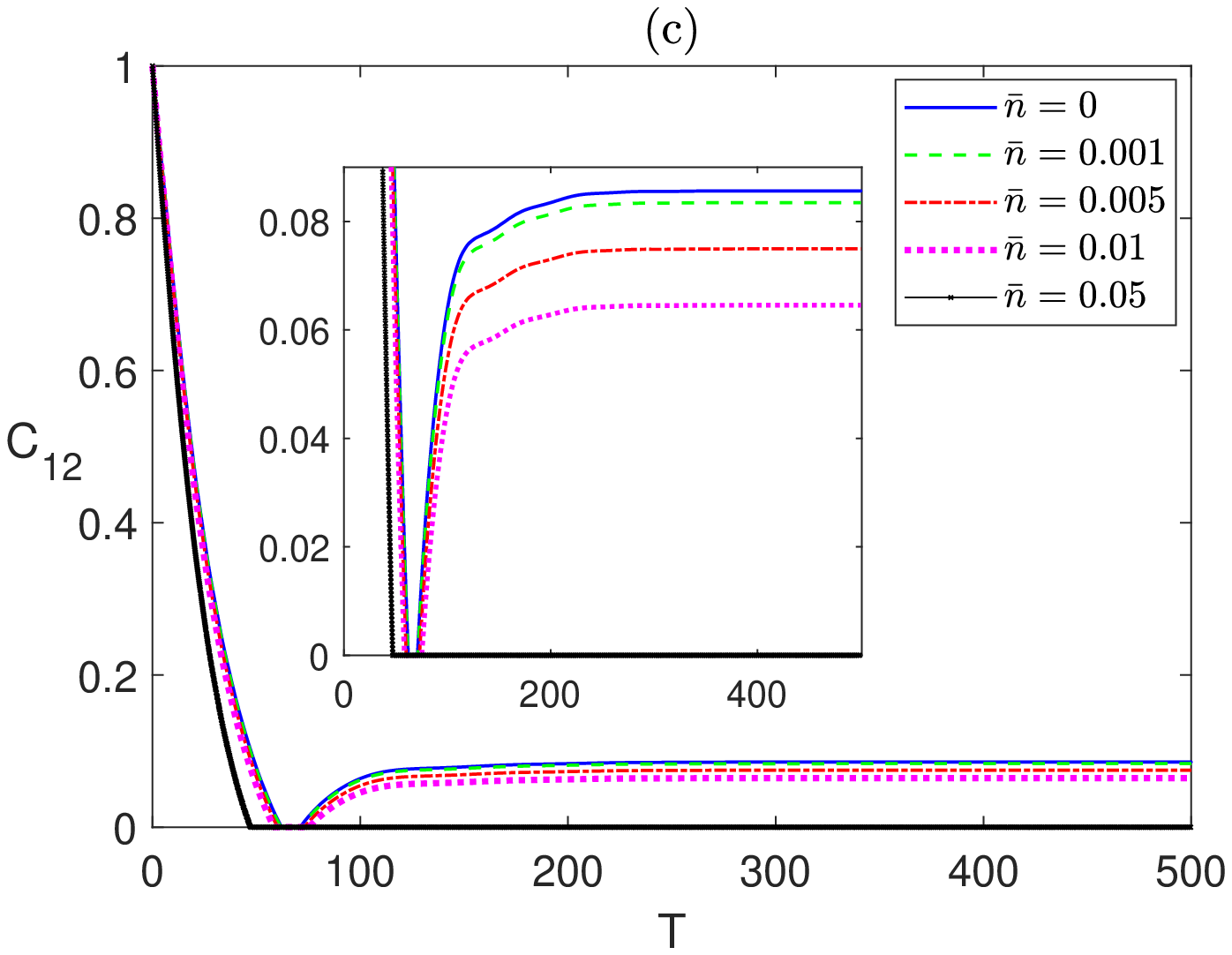}}\quad
 \caption{{\protect\footnotesize Time evolution of $C_{12}$ in the XYZ system in the presence of the environment $(\Gamma=0.05)$ starting from an initial maximally entangled state at different temperatures ($0 \leq \bar{n} \leq 0.05$), and different magnetic field strengths (a) $B_1=1$ and $B_2=1$, (b) $B_1=1$ and $B_2=0.1$, and (c) $B_1=0.1$ and $B_2=1$. The legend for all panels is as shown in panel (c).}}
\label{fig9}
\end{figure}

\begin{figure}[htbp]
 \centering
 \subfigure{\includegraphics[width=8cm]{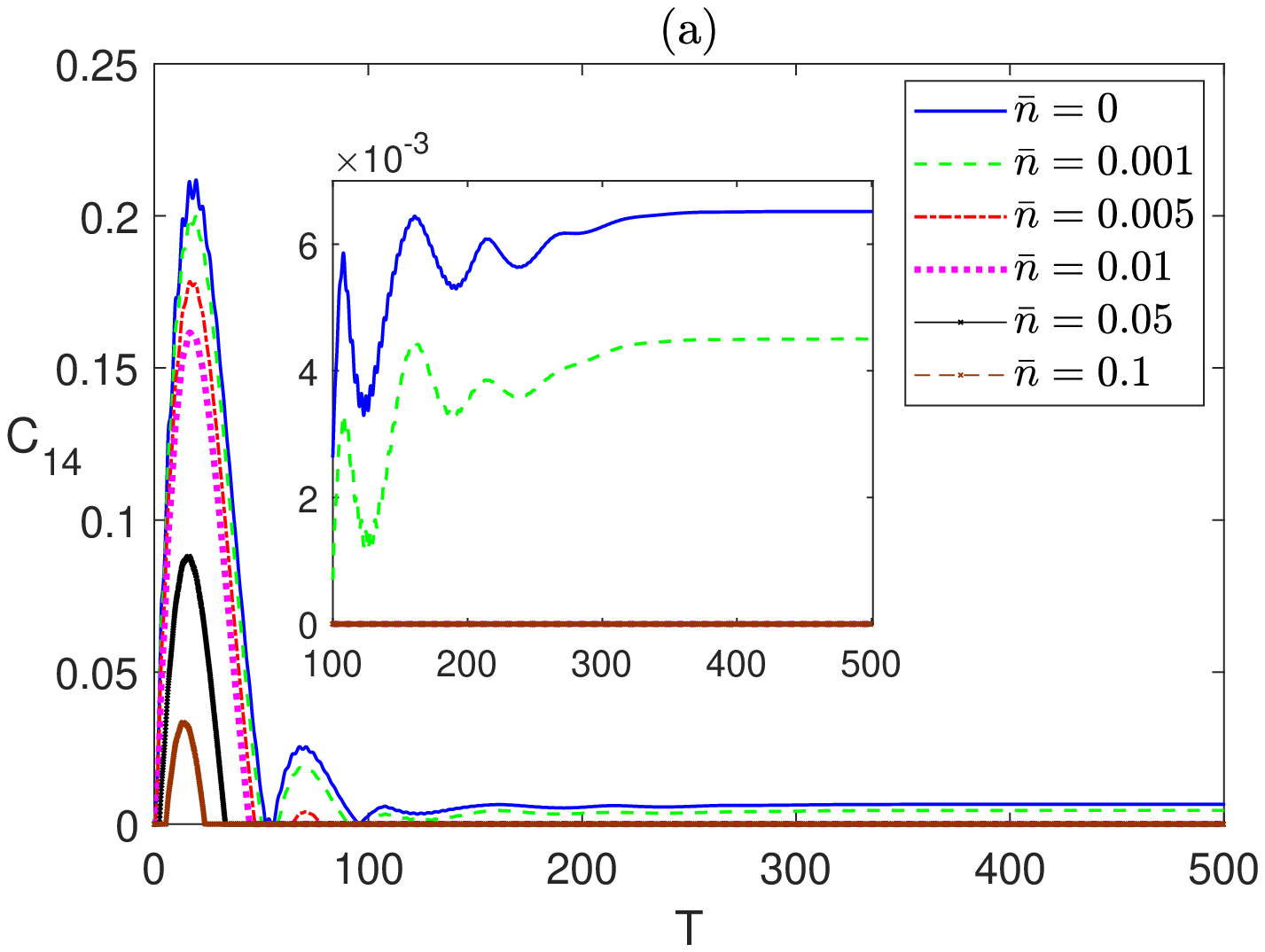}}\quad 
 \subfigure{\includegraphics[width=8cm]{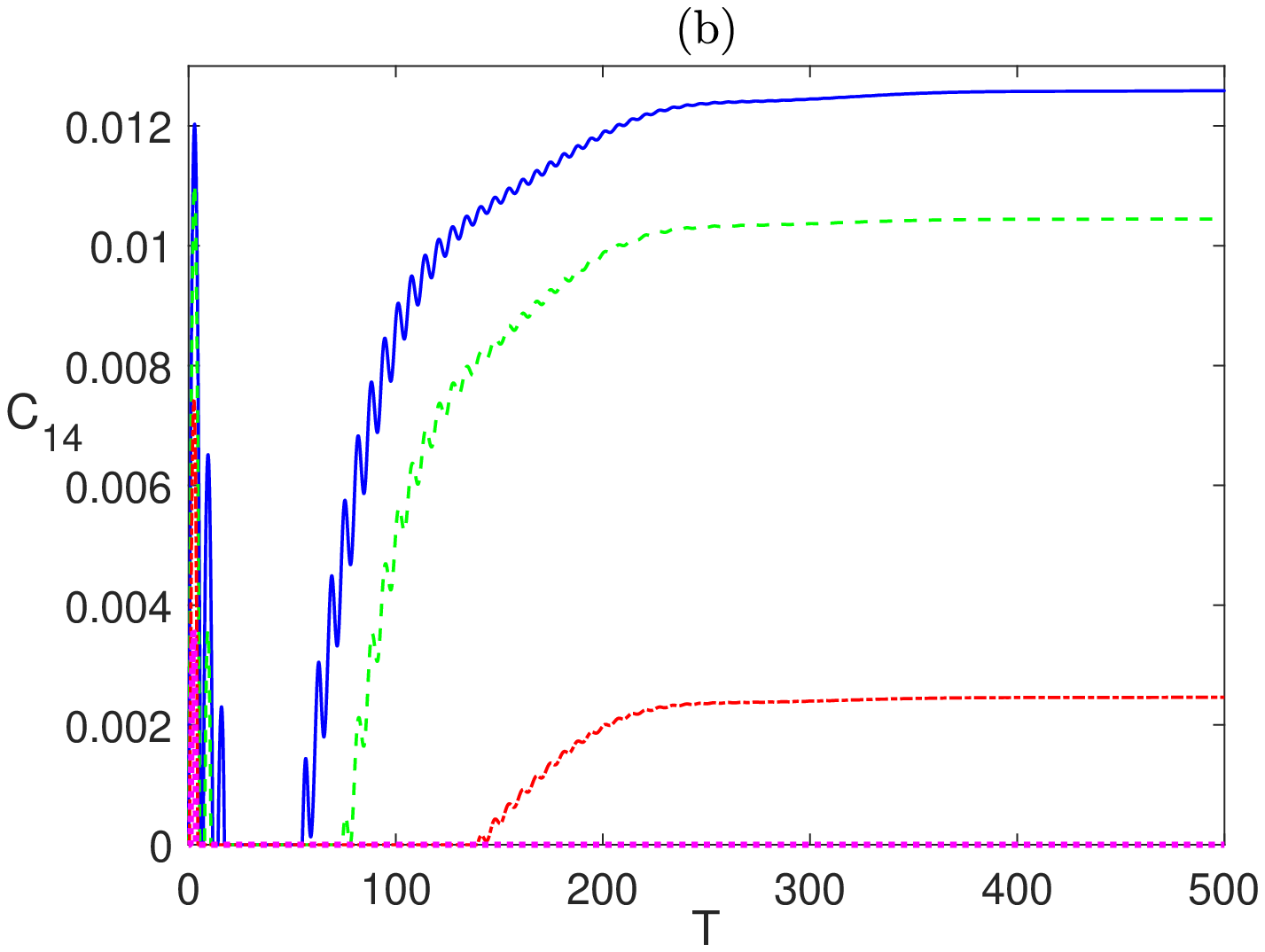}}\\
 \subfigure{\includegraphics[width=8 cm]{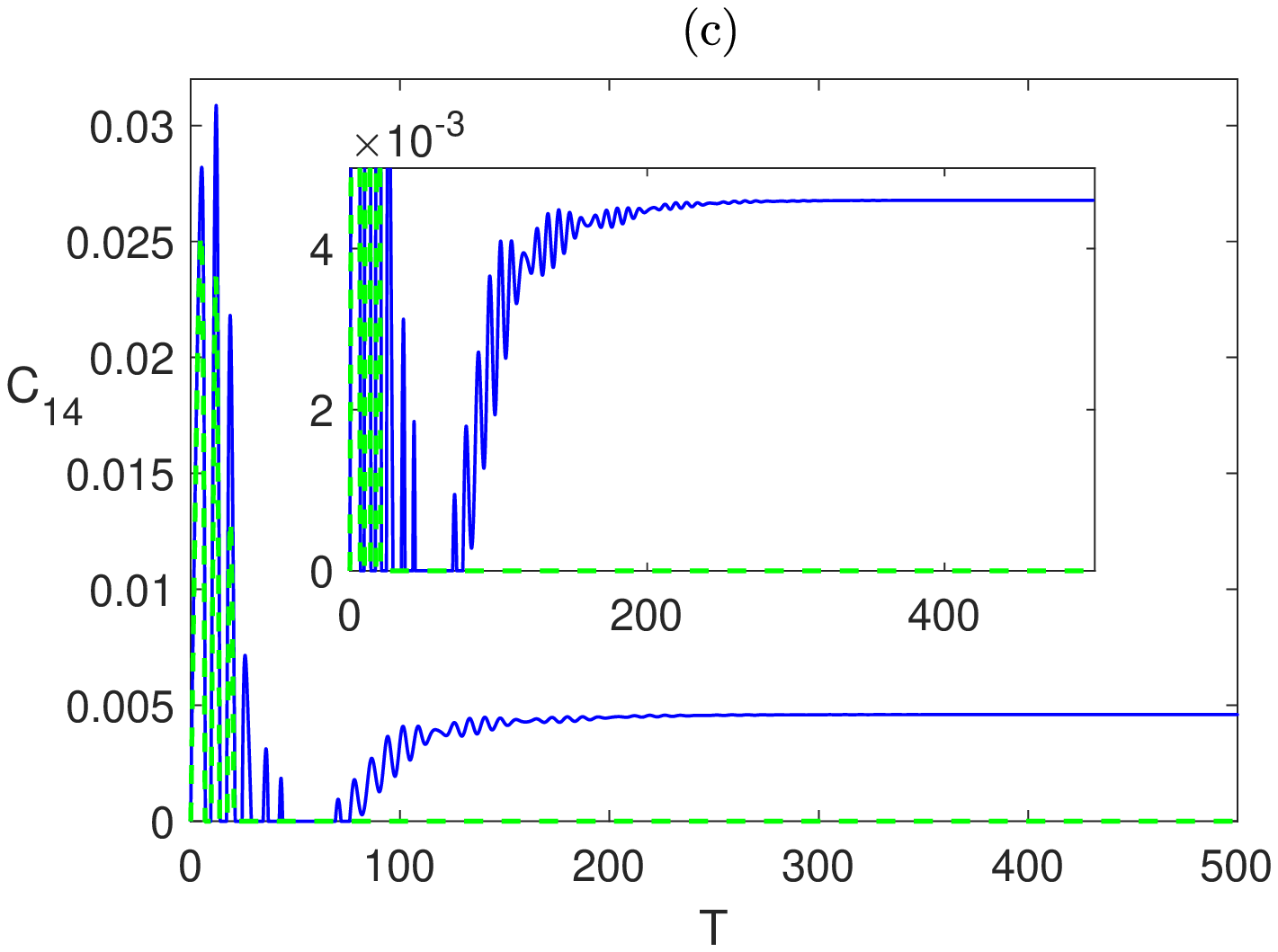}}\quad
 \caption{{\protect\footnotesize Time evolution of $C_{14}$ in the XYZ system in the presence of the environment $(\Gamma=0.05)$ starting from an initial maximally entangled state at different temperatures ($0 \leq \bar{n} \leq 0.1$), and different magnetic field strengths (a) $B_1=1$ and $B_2=1$, (b) $B_1=1$ and $B_2=0.1$, and (c) $B_1=0.1$ and $B_2=1$. The legend for all panels is as shown in panel (a).}}
\label{fig10}
\end{figure}

\begin{figure}[htbp]
 \centering
 \subfigure{\includegraphics[width=8cm]{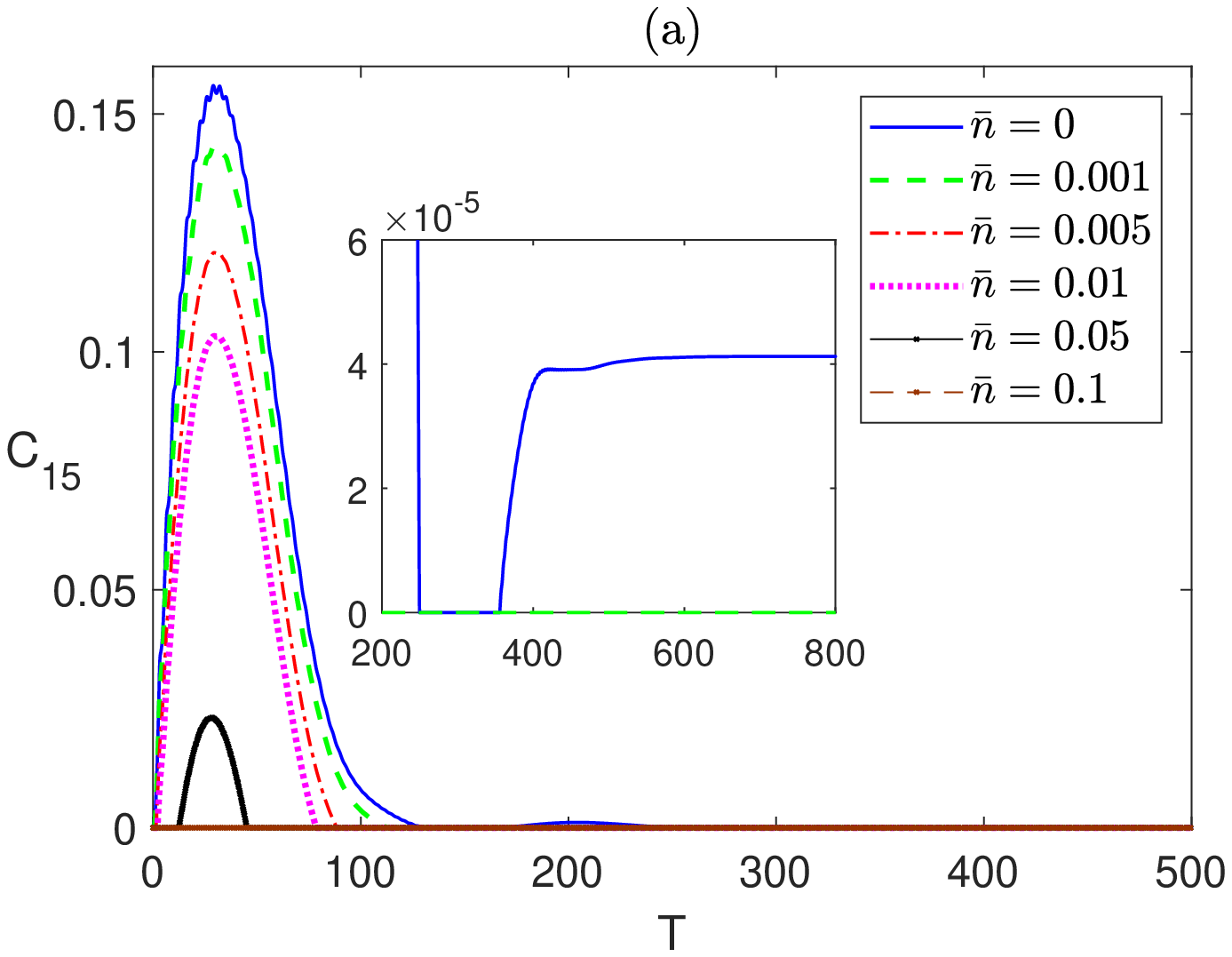}}\quad 
 \subfigure{\includegraphics[width=8cm]{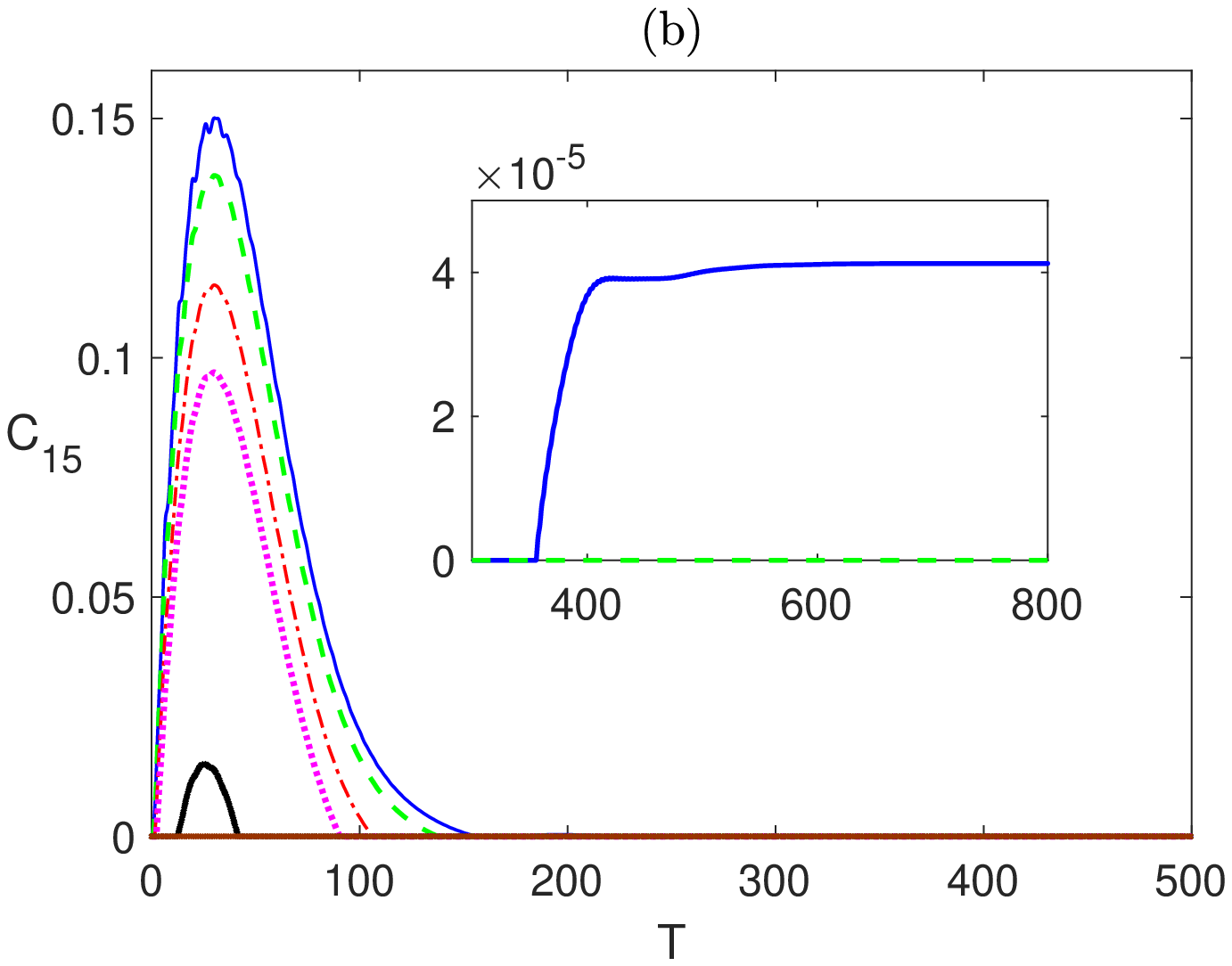}}\\
 \subfigure{\includegraphics[width=8 cm]{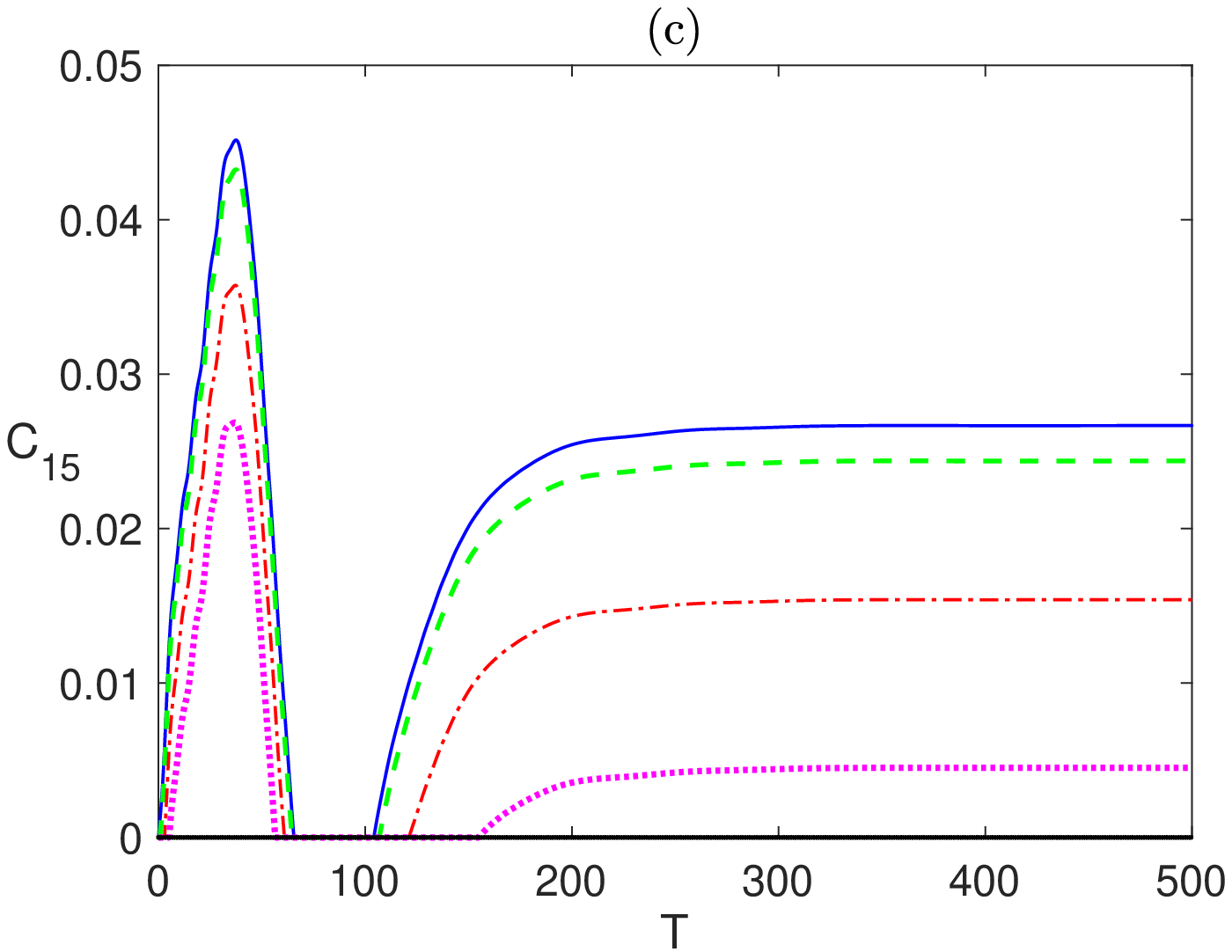}}\quad
 \caption{{\protect\footnotesize Time evolution of $C_{15}$ in the XYZ system in the presence of the environment $(\Gamma=0.05)$ starting from an initial maximally entangled state at different temperatures ($0 \leq \bar{n} \leq 0.1$), and different magnetic field strengths (a) $B_1=1$ and $B_2=1$, (b) $B_1=1$ and $B_2=0.1$, and (c) $B_1=0.1$ and $B_2=1$. The legend for all panels is as shown in panel (a).}}
 \label{fig11}
\end{figure}

Studying the same spin system at a partial degree of anisotropy of the spin-spin interaction shows some similarities to the completely anisotropic (Ising) system but manifests striking differences as well. In this section, we investigate the time evolution of the partially anisotropic (XYZ) Heisenberg system, where $\gamma =1/2$ and $\delta=1$. We present the dynamics of the system starting form a maximally entangled state in Figs.~\ref{fig9} - \ref{fig11}.

\begin{figure}[htbp]
 \centering
 \subfigure{\includegraphics[width=8cm]{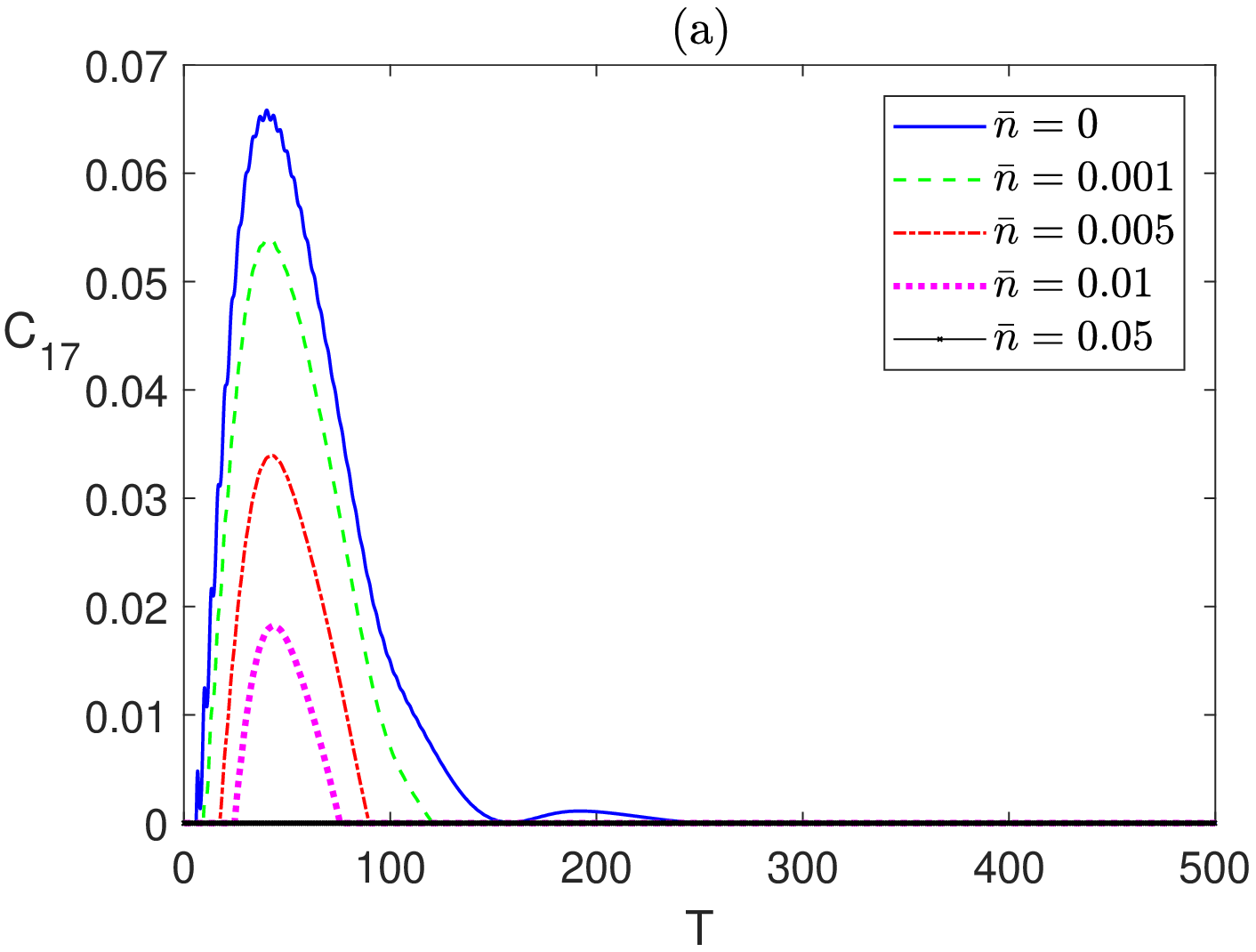}}\quad 
 \subfigure{\includegraphics[width=8cm]{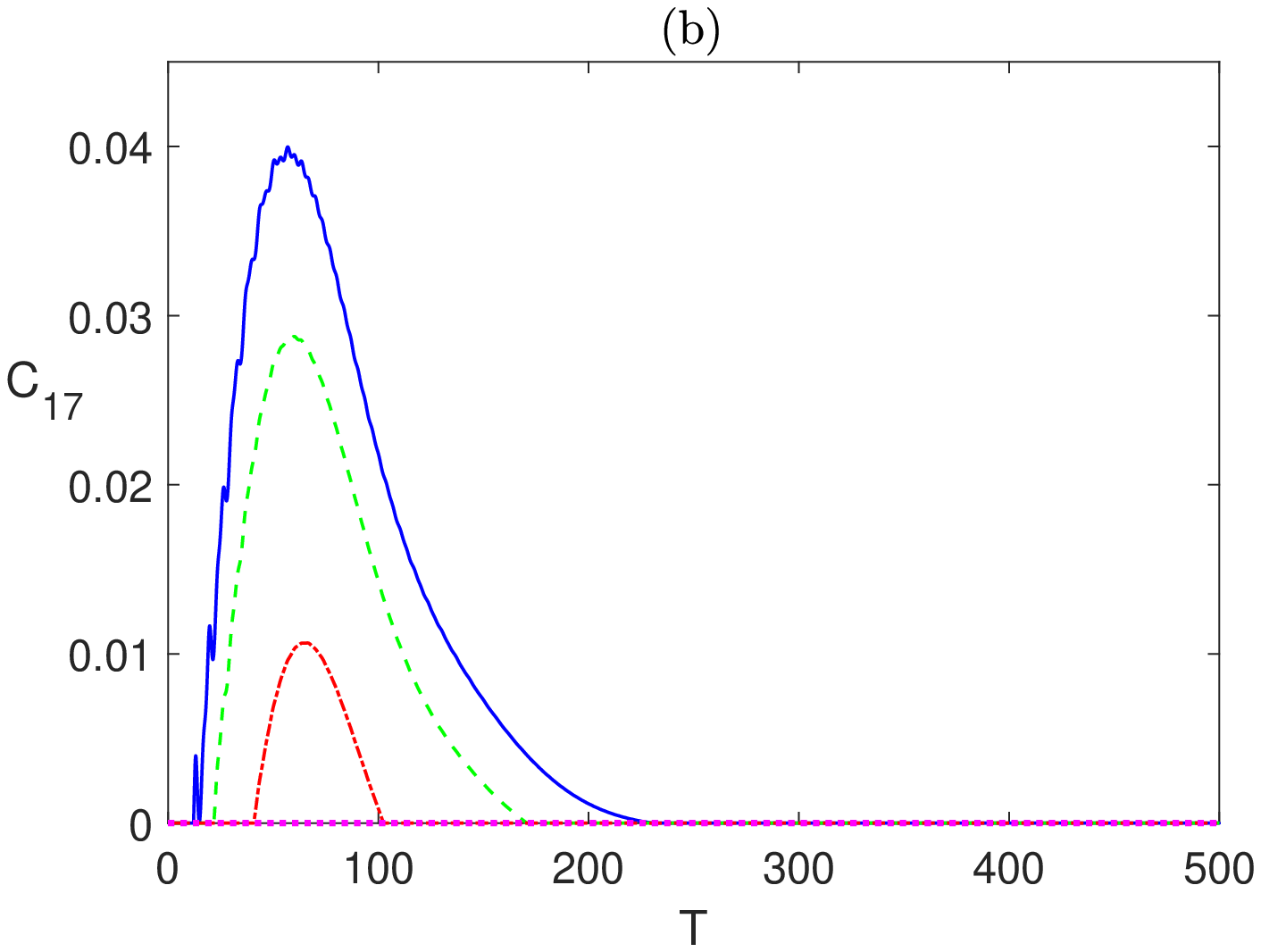}}\\
 \subfigure{\includegraphics[width=8 cm]{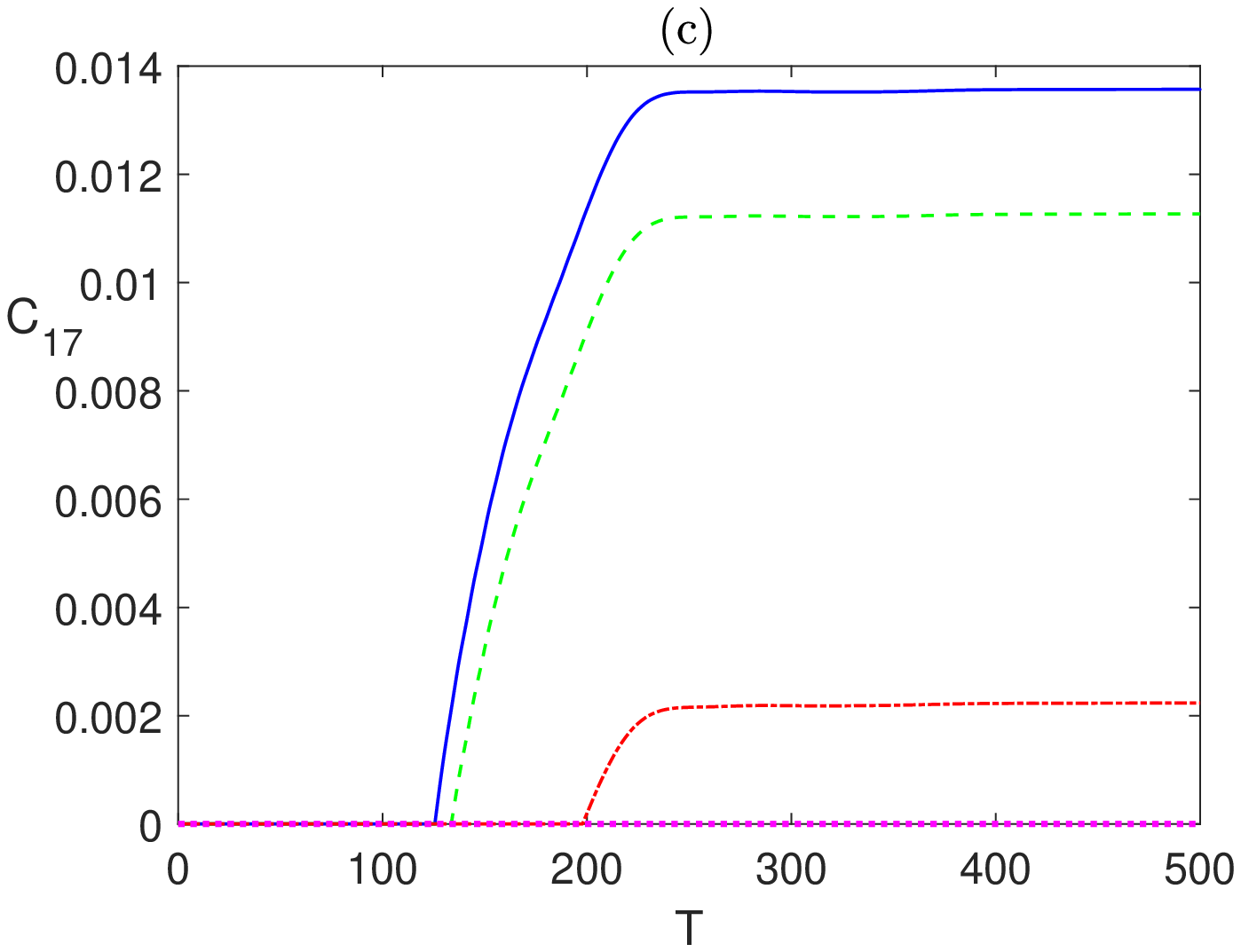}}\quad
 \caption{{\protect\footnotesize Time evolution of $C_{17}$ in the XYZ system in the presence of the environment $(\Gamma=0.05)$ starting from an initial maximally entangled state at different temperatures ($0 \leq \bar{n} \leq 0.05$), and different magnetic field strengths (a) $B_1=1$ and $B_2=1$, (b) $B_1=1$ and $B_2=0.1$, and (c) $B_1=0.1$ and $B_2=1$. The legend for all panels is as shown in panel (a).}}
 \label{fig12}
\end{figure}

The entanglement $C_{12}$ shows a very similar profile to the corresponding one in Ising case, presented in Fig.~\ref{fig2}, at all magnetic field setups, as shown in Fig.~\ref{fig9}. Nevertheless, There is a notable difference in the asymptotic values of $C_{12}$, lowering the anisotropy reduces these values to almost its half magnitude, except in Fig.~\ref{fig9}(c), where there is only a slight decrease in the steady state values. On the other hand, increasing anisotropy has another damaging effect as it reduces the robustness of the system to thermal excitation at all degrees of inhomogeneity of the field as can be concluded from Fig.~\ref{fig9}(a), (b) and (c) compared with Fig.~\ref{fig2}(a), (b) and (c) respectively. 
The dynamics of $C_{14}$ is plotted in Fig.~\ref{fig10}, which in a similar fashion to $C_{12}$, shows asymptotic values that are about half the magnitude of the corresponding ones in the Ising system for all cases of magnetic field including the case of $B_1 < B_2$ illustrated in Fig.~\ref{fig10}(c), in contrary to the behavior of $C_{12}$ in that particular case.
The most distinguished behavior of the XYZ system, compared with the Ising system, manifests itself in the nnn entanglement $C_{15}$ and the nnn entanglement $C_{17}$, which are depicted in Figs. \ref{fig11} and \ref{fig12} respectively. In contrary to the Ising system case, the nnn entanglement $C_{15}$ doesn't vanish asymptotically at zero temperature, although it does at non-zero temperatures, where it reaches a very small steady state value in both cases of homogeneous and inhomogeneous magnetic field $(B_1>B_2)$, shown in Fig. \ref{fig11}(a) and (b) respectively. Interestingly, in the other case of the inhomogeneous magnetic field $(B_1 < B_2)$, depicted in panel (c), $C_{15}$ shows high robustness against thermal effects and higher asymptotic values that even exceeds that of the nn entanglement $C_{14}$, illustrated in Fig.~\ref{fig10}(c). 
More interestingly, the nnnn entanglement $C_{17}$, shown in Fig.~\ref{fig12}(c), where $B_1<B_2$, evolves from zero before reviving and reaching asymptotically a non-zero steady state value at zero and non-zero temperatures, which are also higher than the corresponding $C_{14}$ values illustrated in Fig.~\ref{fig10}(c). The other two cases of magnetic field, shown in Fig.~\ref{fig12}(a) and (b) result in an asymptotically vanishing $C_{17}$. Therefore, applying an inhomogeneous magnetic field to the spin system, where the field gradient is directed inward enhances the entanglement among the border spins, even the nnnn neighbors, and increases its robustness against thermal excitation.

\begin{figure}[htbp]
 \centering
 \subfigure{\includegraphics[width=8cm]{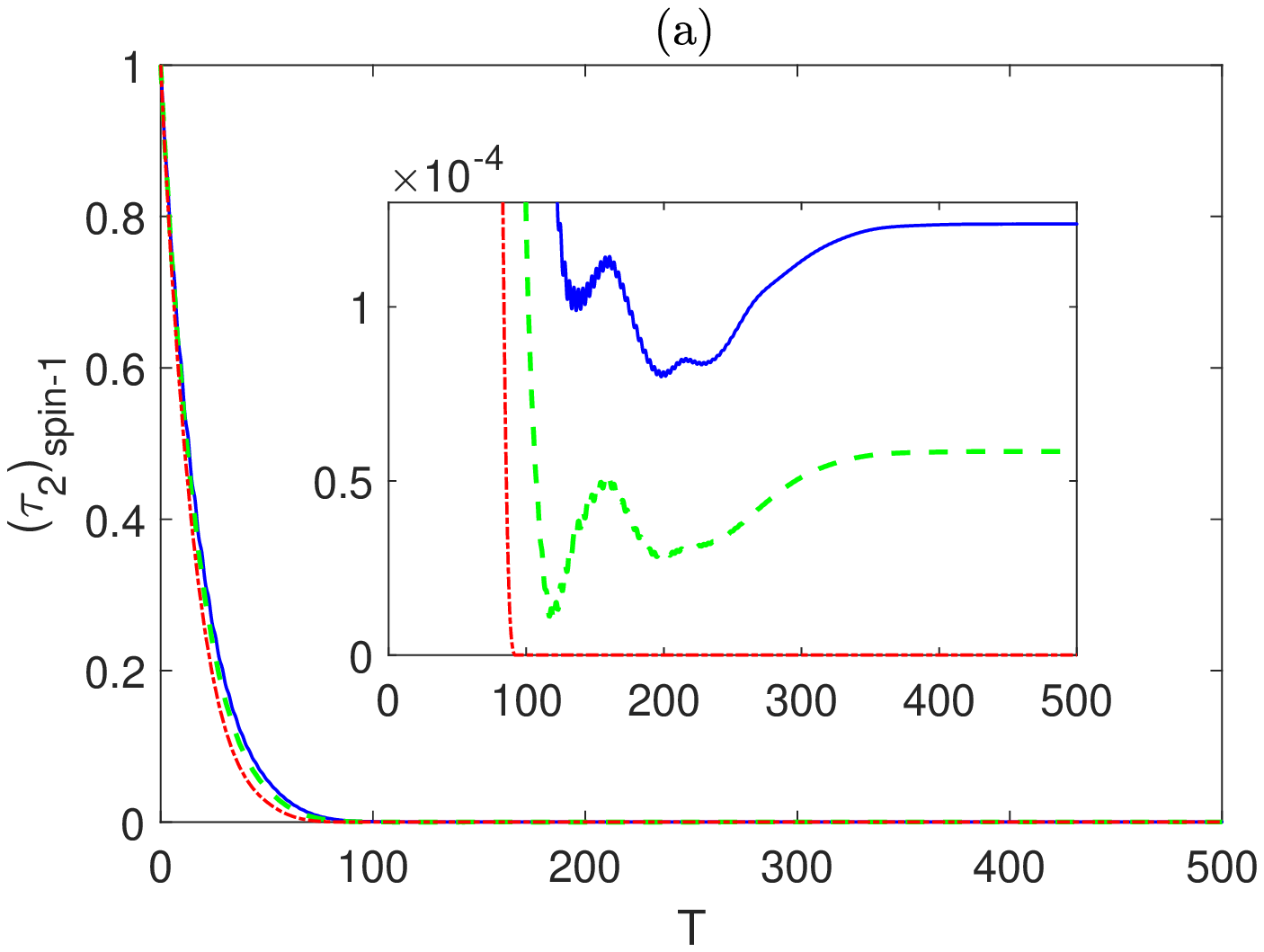}}\quad 
 \subfigure{\includegraphics[width=8cm]{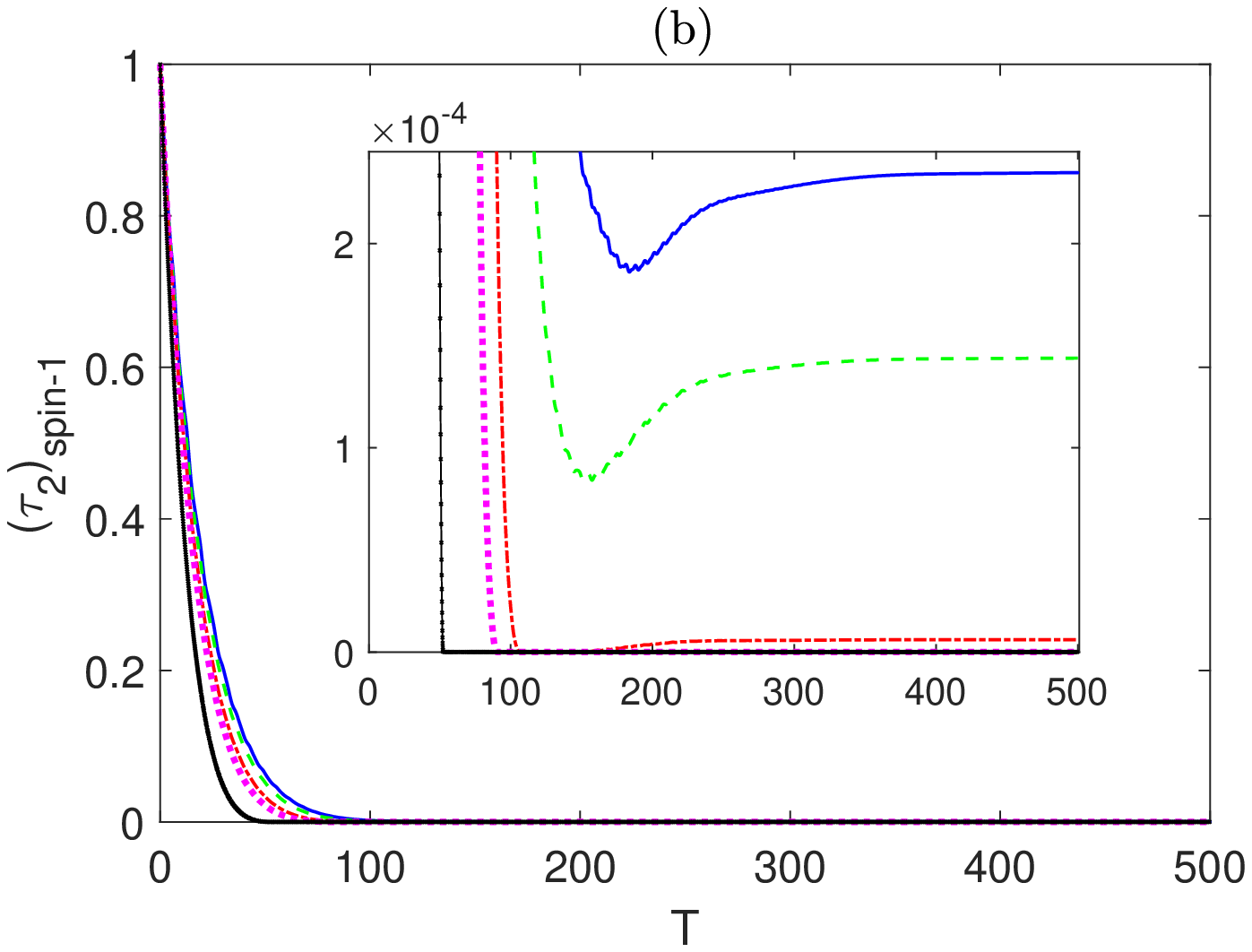}}\\
 \subfigure{\includegraphics[width=8cm]{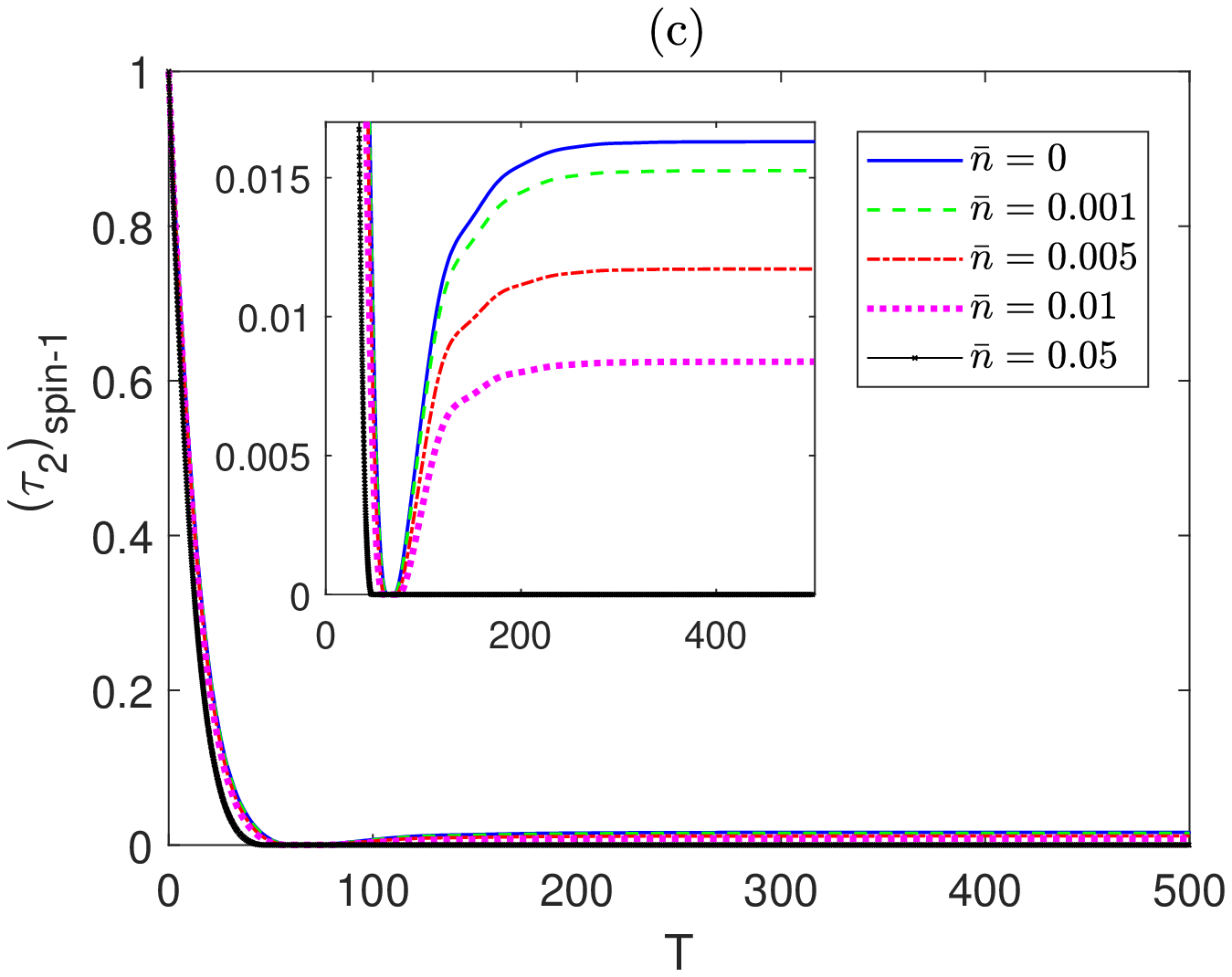}}\quad
 \caption{{\protect\footnotesize Time evolution of $\tau_2$ for spin 1 in the XYZ system in the presence of the environment $(\Gamma=0.05)$ starting from an initial maximally entangled state at different temperatures ($0 \leq \bar{n} \leq 0.05$), and different magnetic field strengths (a) $B_1=1$ and $B_2=1$, (b) $B_1=1$ and $B_2=0.1$, and (c) $B_1=0.1$ and $B_2=1$. The legend for all panels is as shown in panel (c).}}
\label{fig13}
\end{figure}
\begin{figure}[htbp]
 \centering
\subfigure{\includegraphics[width=8cm]{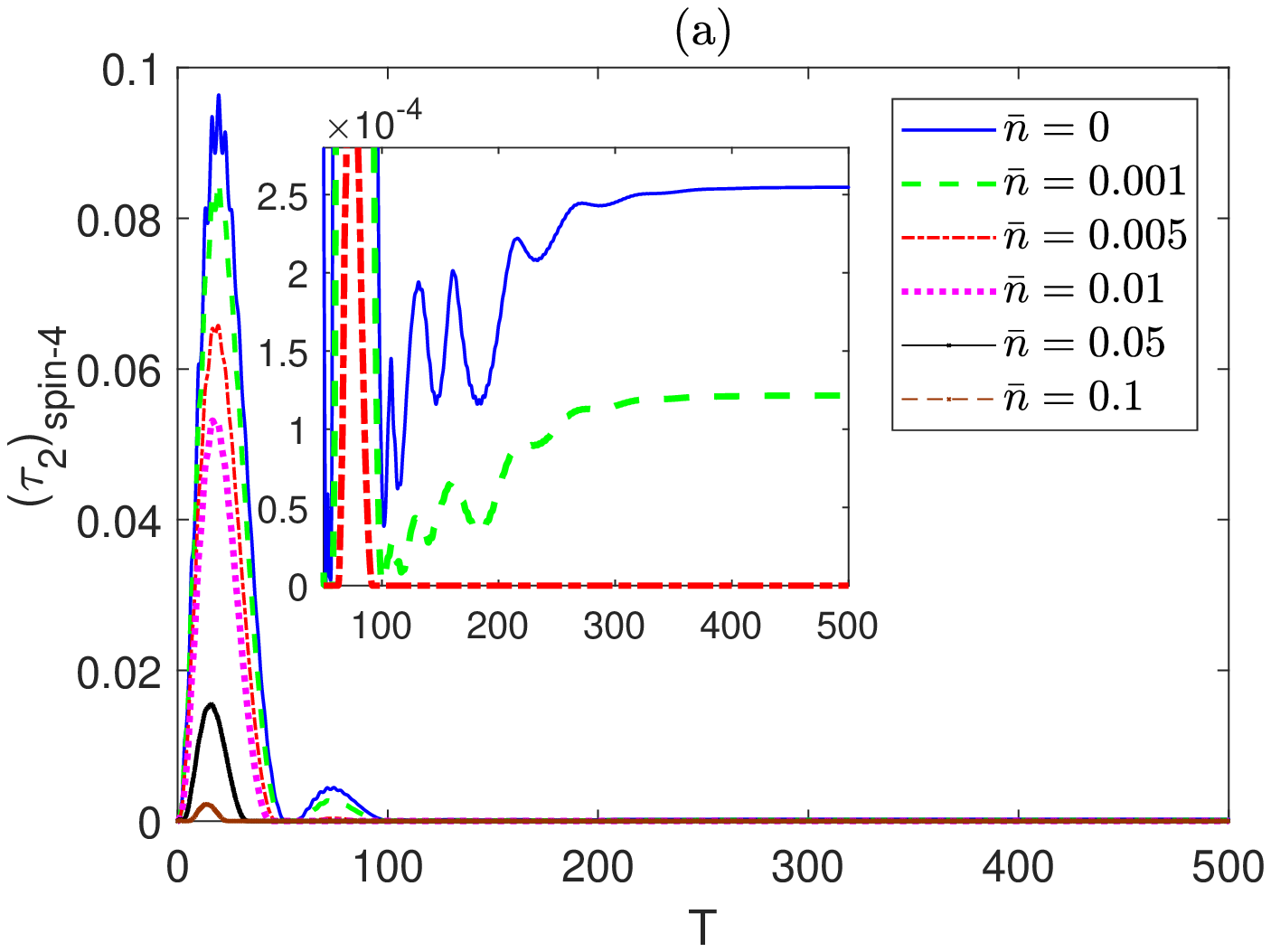}}\quad 
 \subfigure{\includegraphics[width=8cm]{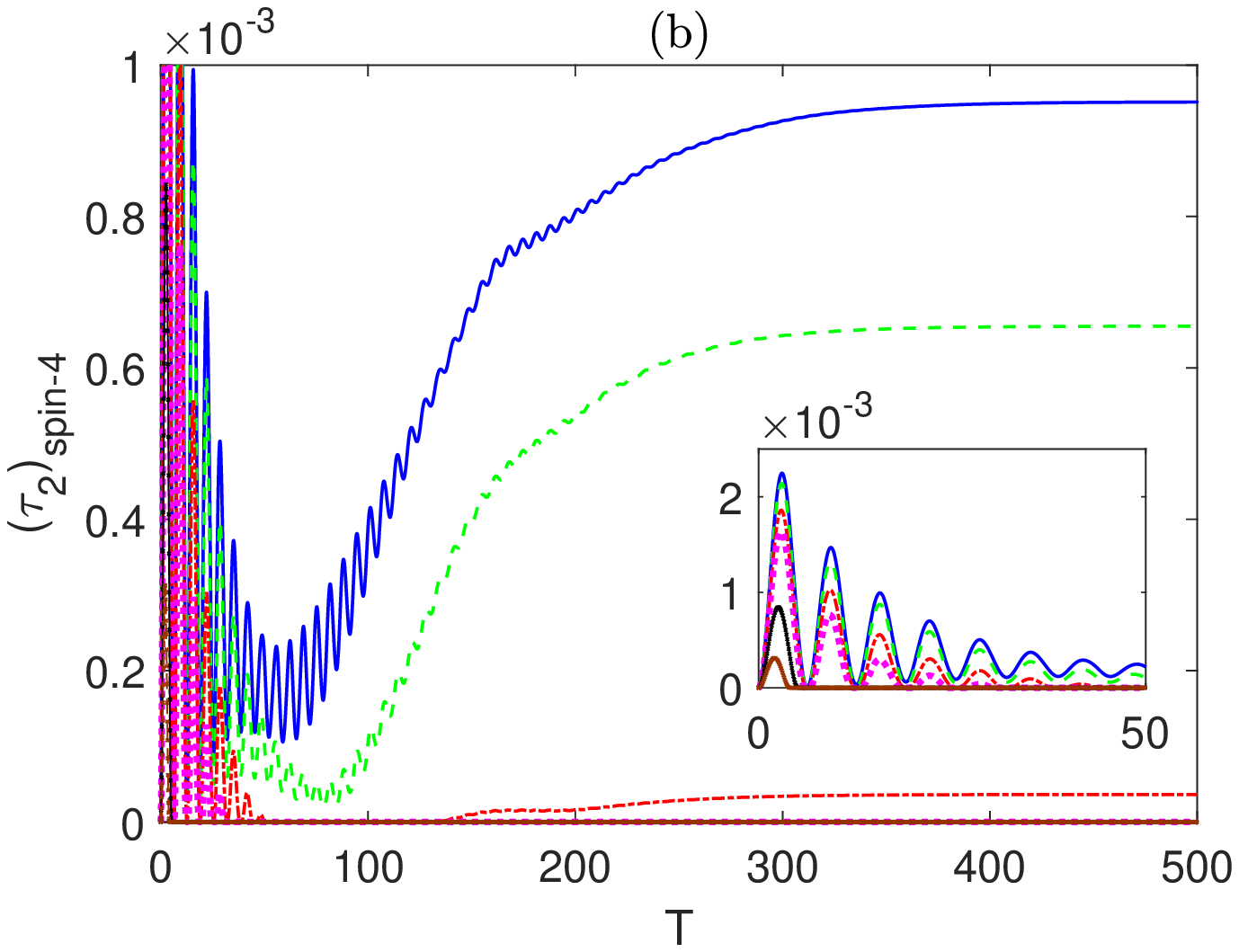}}\\
 \subfigure{\includegraphics[width=8 cm]{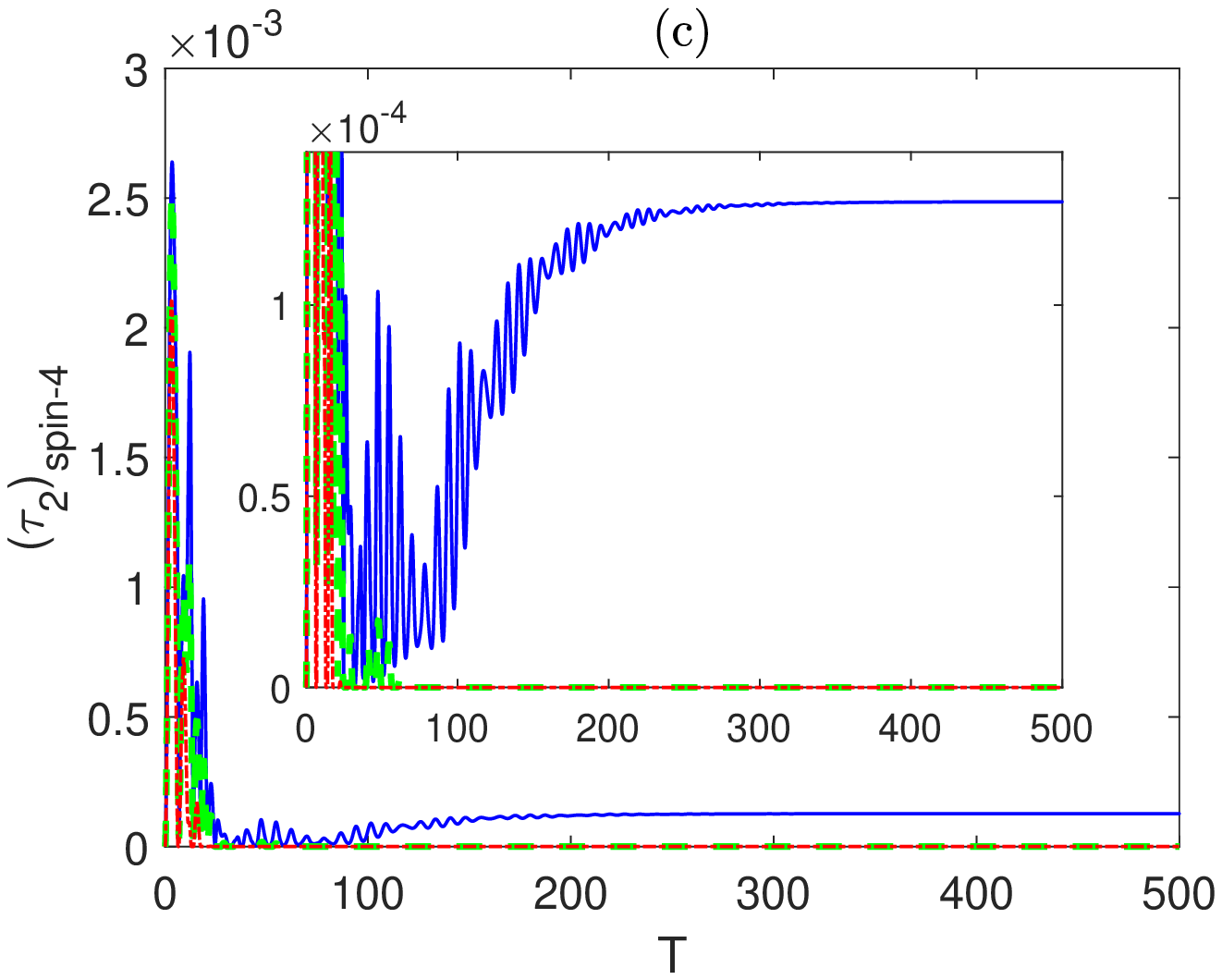}}\quad
 \caption{{\protect\footnotesize Time evolution of $\tau_2$ for spin 4 in the XYZ system in the presence of the environment $(\Gamma=0.05)$ starting from an initial maximally entangled state at different temperatures ($0 \leq \bar{n} \leq 0.1$), and different magnetic field strengths (a) $B_1=1$ and $B_2=1$, (b) $B_1=1$ and $B_2=0.1$, and (c) $B_1=0.1$ and $B_2=1$. The legend for all panels is as shown in panel (a).}}
\label{fig14}
\end{figure} 
 The dynamics of the global bipartite entanglement $\tau_2$ for spin 1, in the XYZ system, starting from a maximally entangled state is presented in Fig.~\ref{fig13}. The overall behavior of $\tau_2$ is very close to what has been observed in the Ising case, shown in Fig.~\ref{fig5}, where the entanglement decays from a maximum value of one, reaching a minimum value or zero, depending on the temperature, before reviving again and rising to reach asymptotically a steady state. The steady state values presented in Fig.~\ref{fig13}(a) and (b) are reduced to about one third of the corresponding values in the Ising case with less thermal robustness. The anisotropic magnetic field, where $B_1<B_2$ case presented in Fig.~\ref{fig13}(c) shows a very slight decrease in the asymptotic values and thermal robustness compared with the Ising case illustrated in Fig.~\ref{fig5}(c). Testing the time evolution $\tau_2$ for spin 4, presented in Fig.~\ref{fig14} reveals that its asymptotic values doubles that of $\tau_2$ for spin 1 for the homogeneous magnetic field case shown in panel Fig.~\ref{fig14}(a), while it increases multiple times compared with that of spin 1 for the inhomogeneous case presented in Fig.~\ref{fig14}(b), but there is no noticeable difference in thermal robustness in both cases. However, In the inhomogeneous field case, illustrated in panel (c), where $B_1<B_2$, $\tau_2$ for spin 4 has a much lower steady state values compared with that of spin 1 with much lower thermal robustness. Comparing the behavior of $\tau_2$ for spin 4 in both of the Ising system, illustrated in Fig.~\ref{fig6} and the XYZ system, illustrated in Fig.~\ref{fig14}, indicates that lower anisotropy reduces the steady state values significantly and weakens thermal robustness in the system.

\begin{figure}[htbp]
 \centering
  \subfigure{\includegraphics[width=8cm]{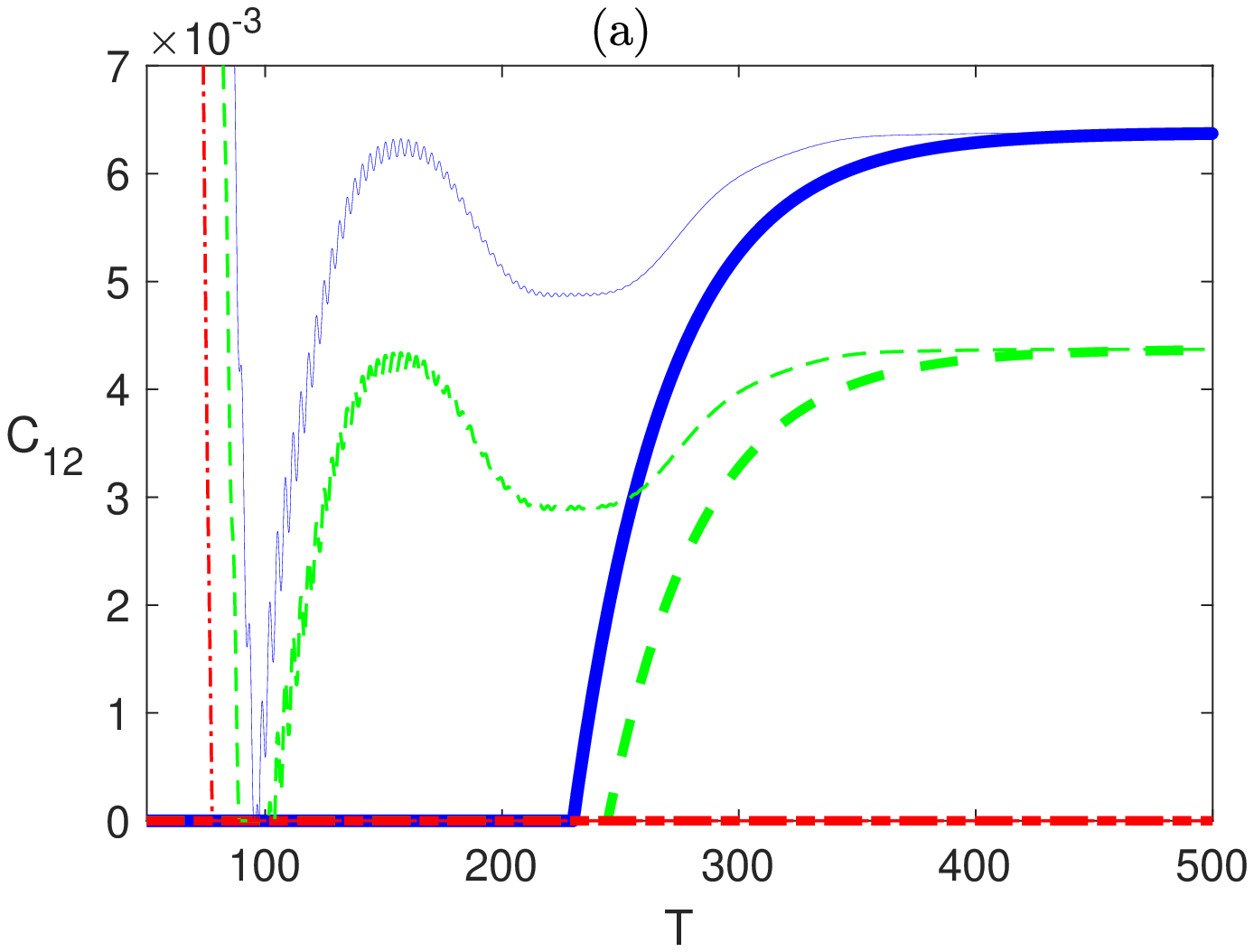}}\quad 
 \subfigure{\includegraphics[width=8cm]{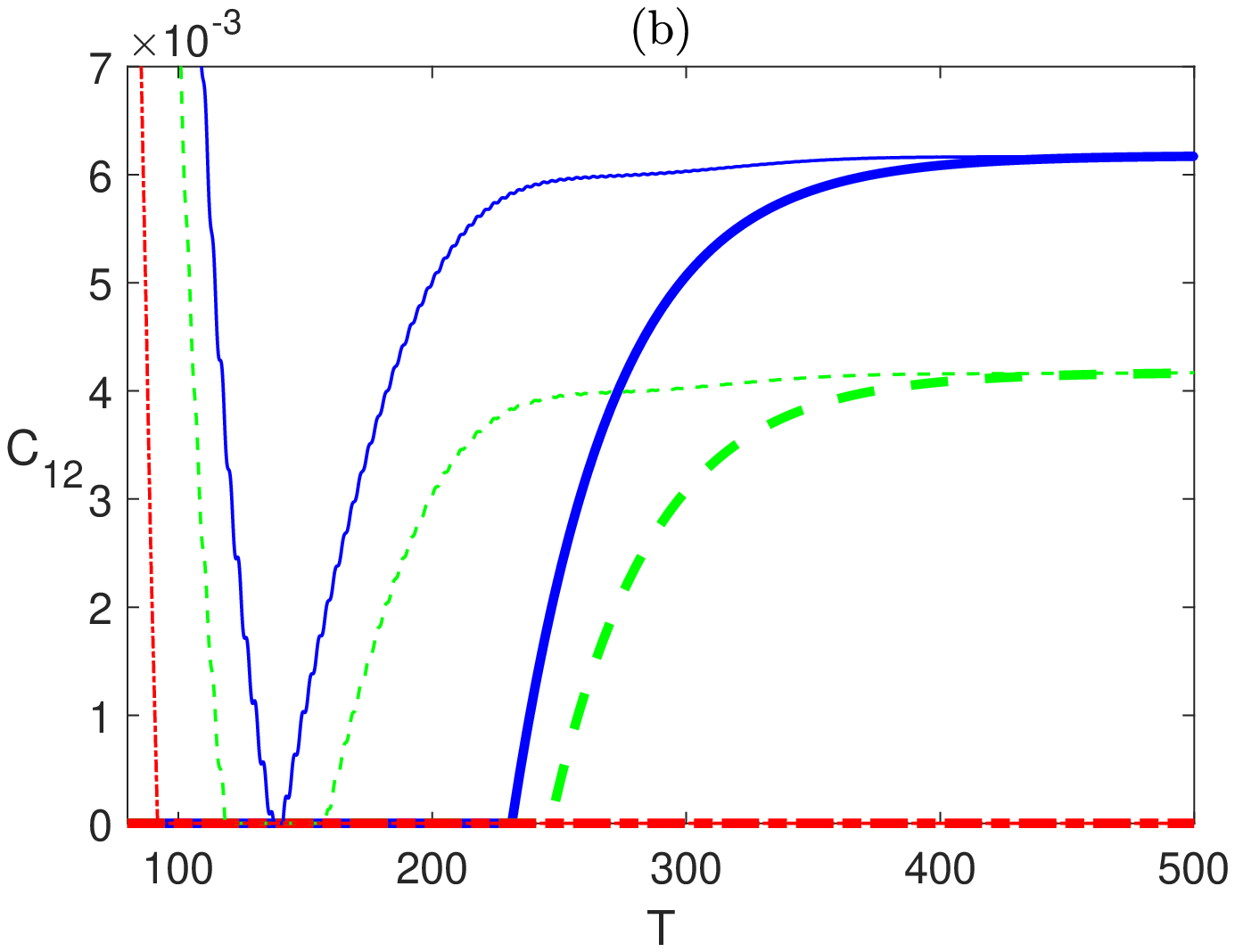}}\\
 \subfigure{\includegraphics[width=8cm]{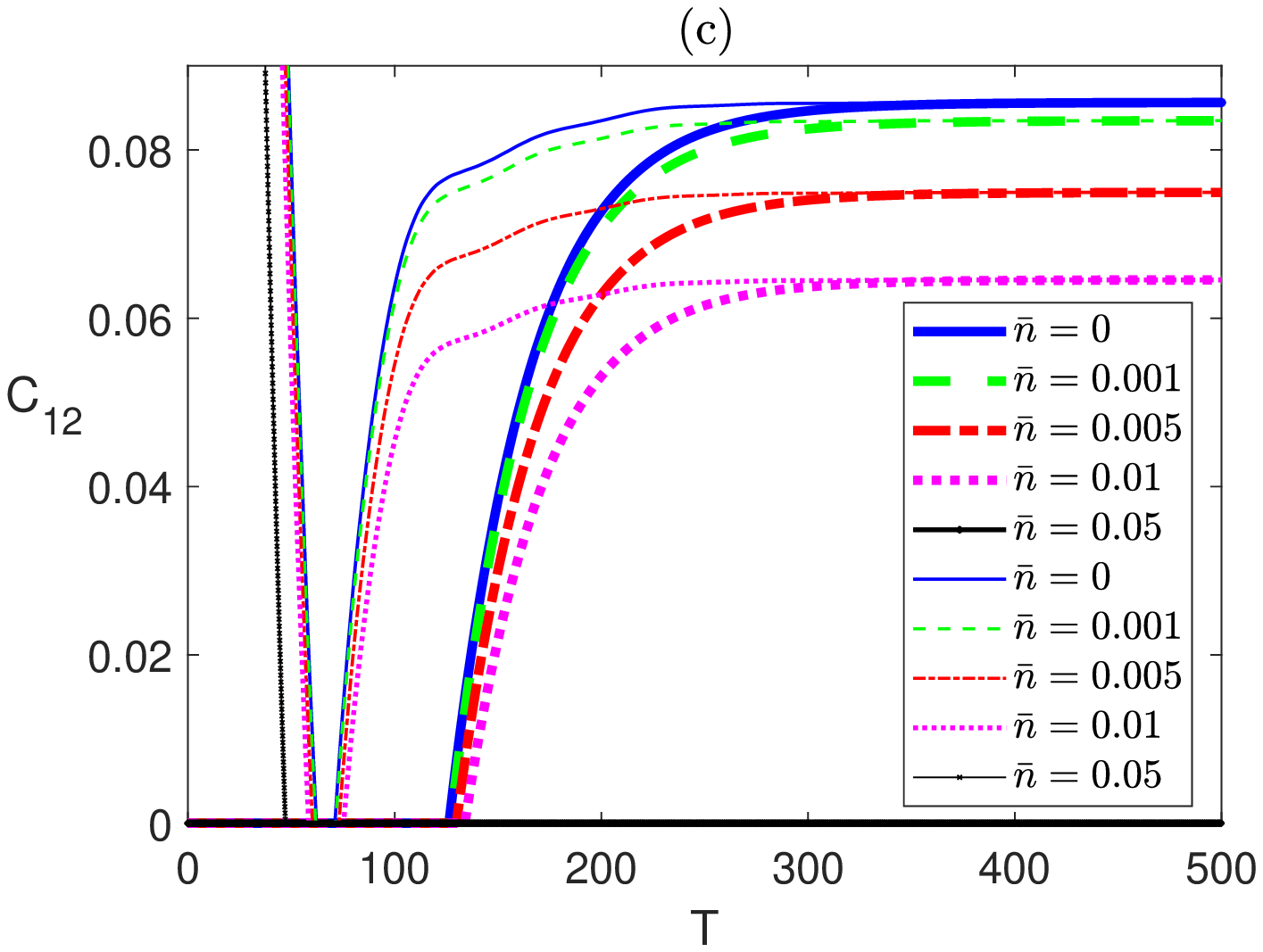}}\quad
 \caption{{\protect\footnotesize Time evolution of $C_{12}$ in the XYZ system in the presence of the environment $(\Gamma=0.05)$ starting from an initial disentangled state (bold lines) and a maximally entangled state (thin lines) at different temperatures ($0 \leq \bar{n} \leq 0.05$), and different magnetic field strengths (a) $B_1=1$ and $B_2=1$, (b) $B_1=1$ and $B_2=0.1$, and (c) $B_1=0.1$ and $B_2=1$. The legend for all panels is as shown in panel (c).}}
\label{fig15}
\end{figure}
We explored the time evolution of the dissipative XYZ system under the effect of different magnetic field configurations starting from an initial disentangled state. The different entanglements, nn, nnn and nnnn bipartite and global, start from a zero value before reviving and increasing monotonically to asymptotically reach steady state values that coincide with the corresponding ones in the case of an initial maximally entangled state, as was discussed in Figs.~\ref{fig9}-\ref{fig14}, in a very similar fashion to what was observed in the Ising system. In Fig.~\ref{fig16}, as an illustration, we depict the time evolution of $C12$ starting from the disentangled initial state, using the bold lines, and from the maximally entangled initial state, using the thin lines. It shows that, regardless of the initial state, $C_{12}$ evolves to reach asymptotically the same steady state value, which depends on the anisotropy of the system, the temperature and the inhomogeneity of the magnetic field. We did not insert the graphs of the comparison of the other cases to save space and avoid redundancy.
\subsection{Isotropic system (XXX Model)}
\begin{figure}[htbp]
 \centering
  \subfigure{\includegraphics[width=8cm]{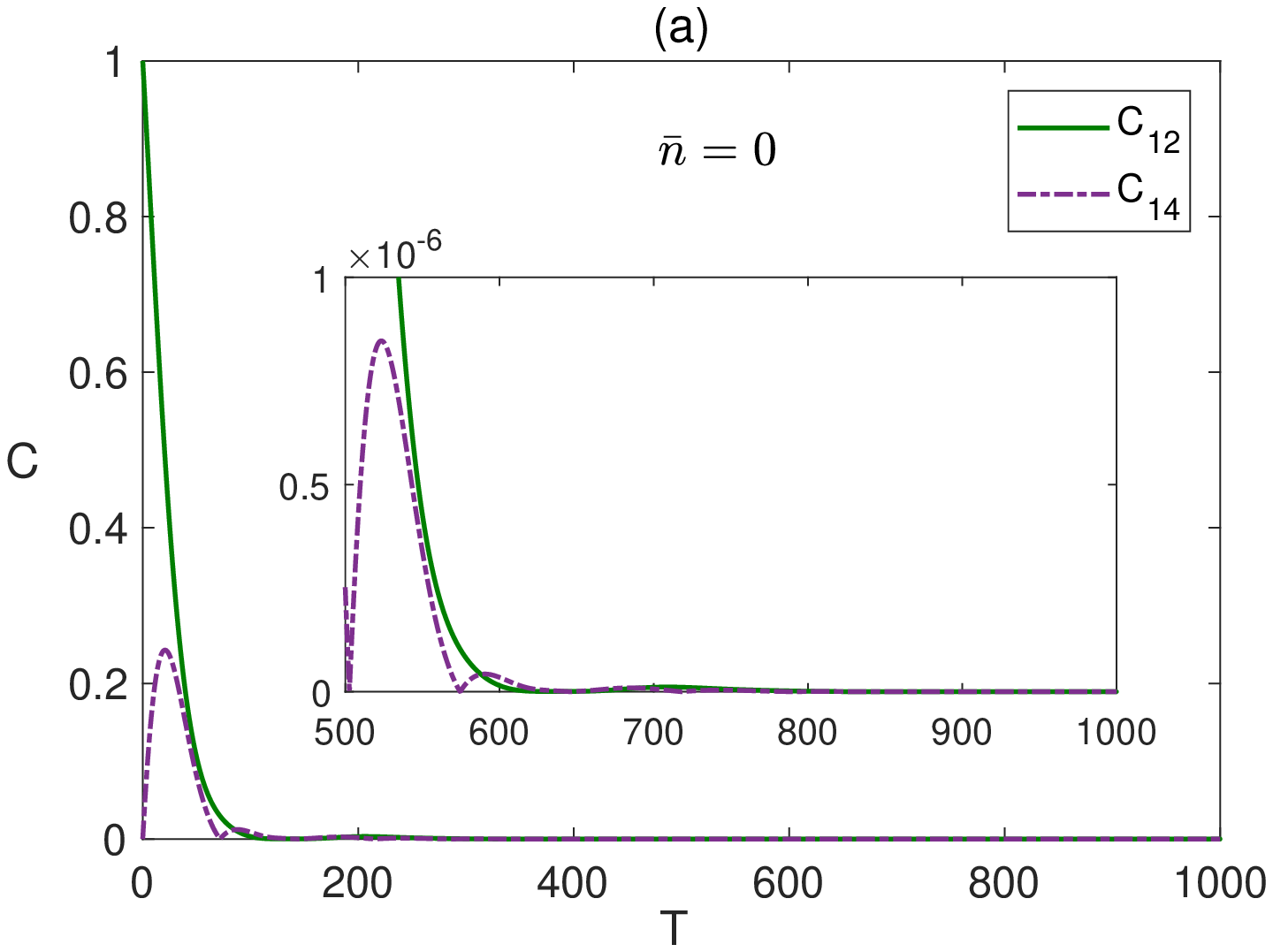}}\quad 
 \subfigure{\includegraphics[width=8cm]{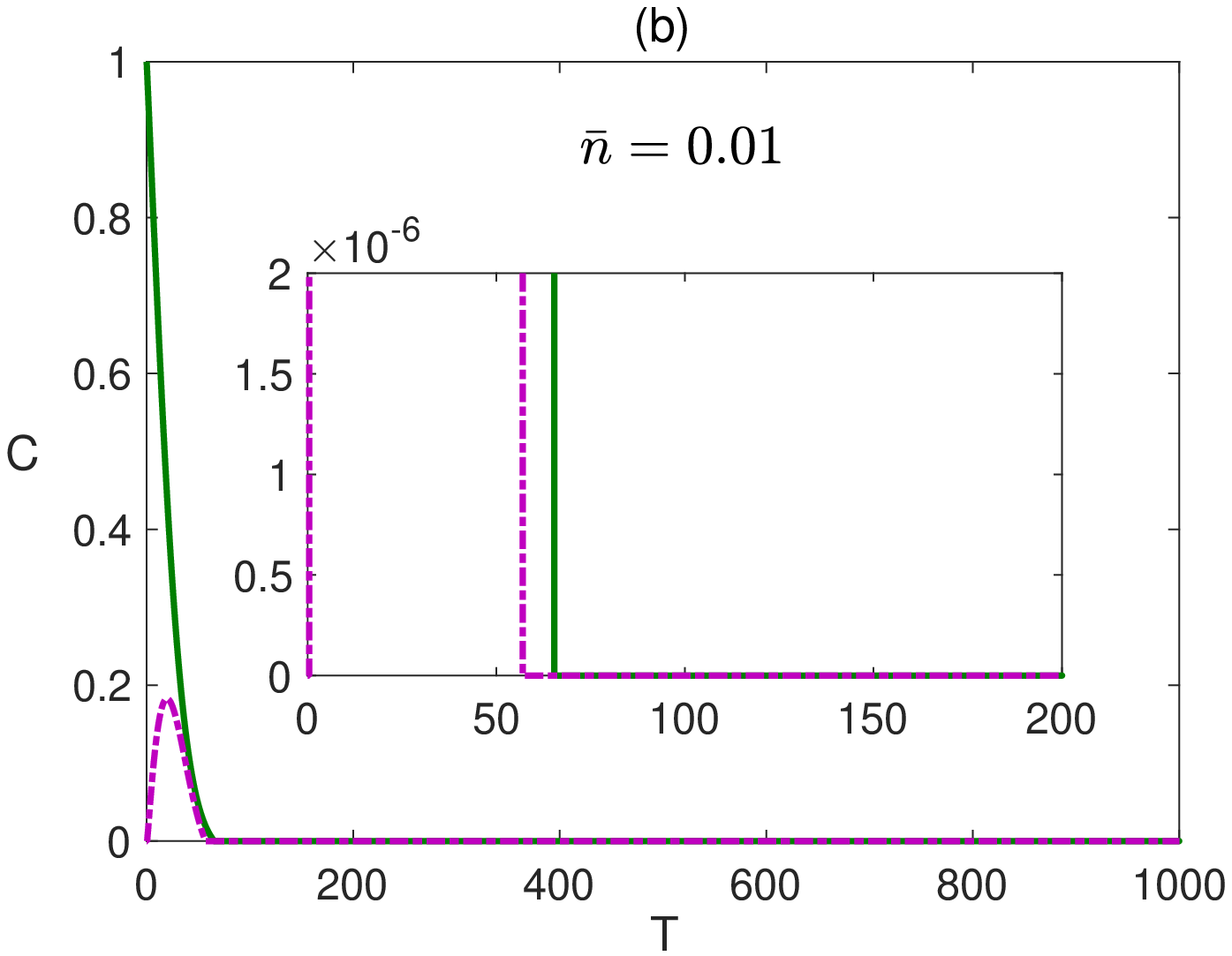}}\\
 \subfigure{\includegraphics[width=8cm]{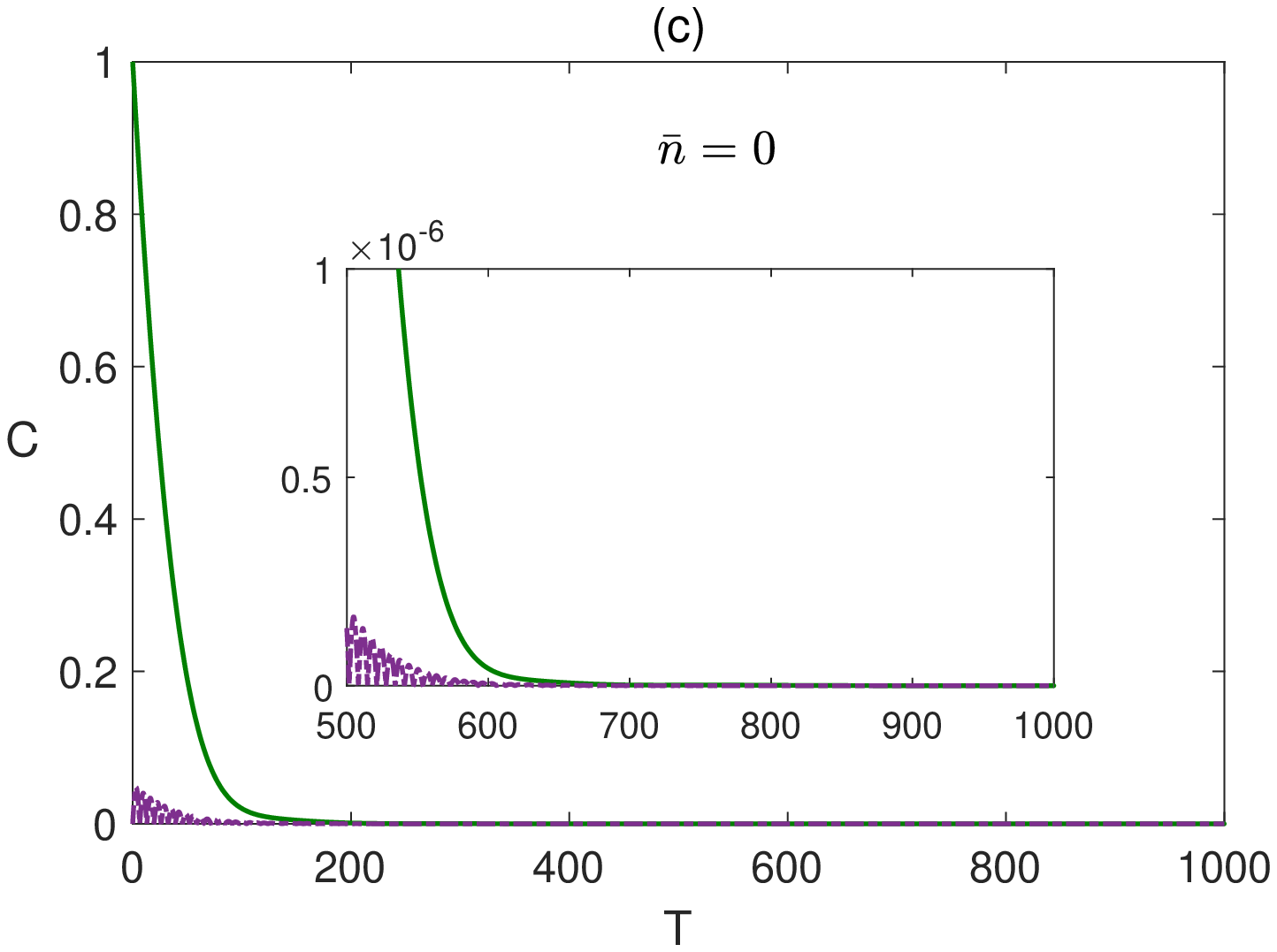}}\quad
  \subfigure{\includegraphics[width=8cm]{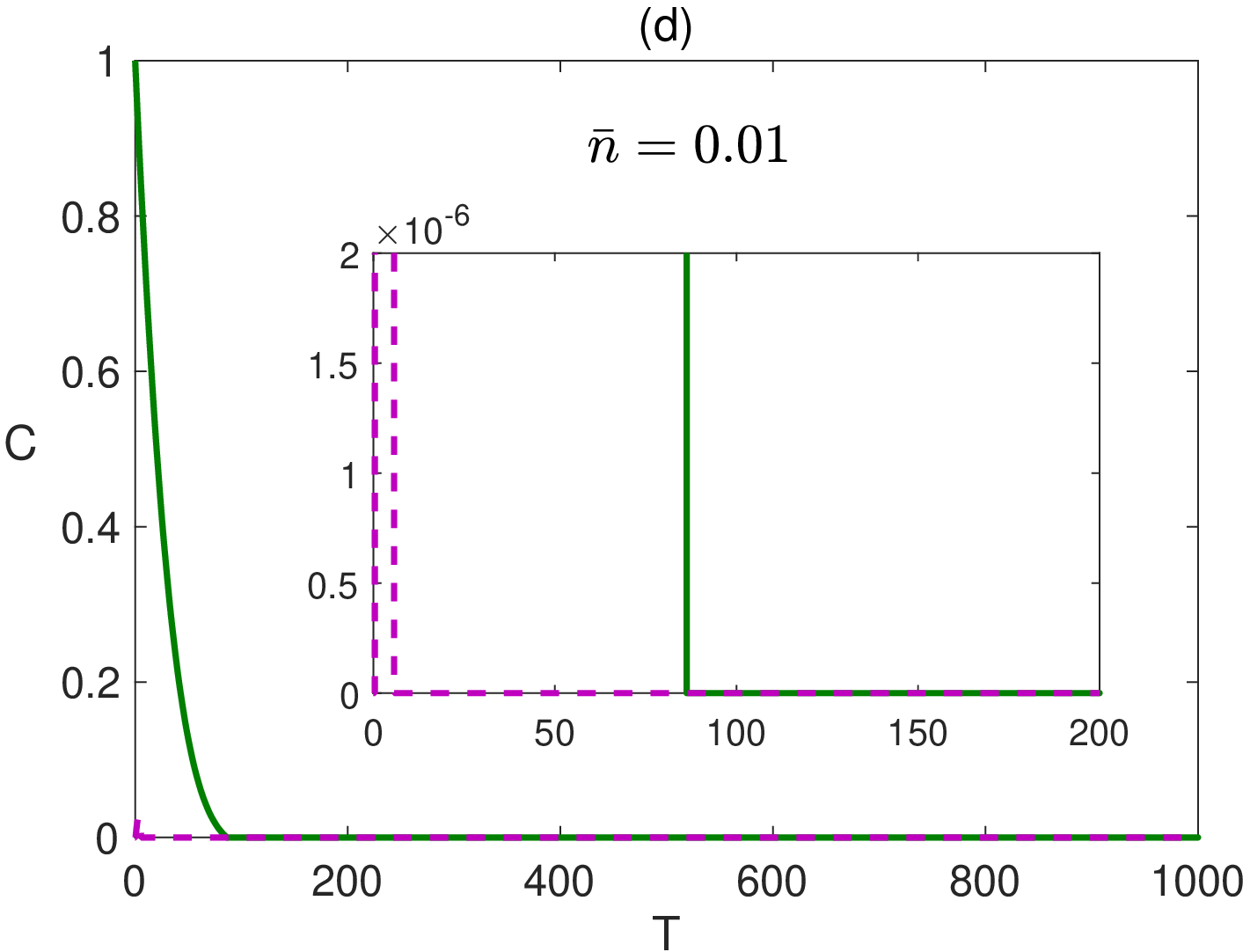}}\quad
 \caption{{\protect\footnotesize Time evolution of $C_{12}$ and $C_{14}$ in the $XXX$ system in presence of the environment $(\Gamma=0.05)$ starting from an initial maximally entangled state, at different temperatures and magnetic fields, where in (a) $\bar{n}=0$, $B_1=1$ and $B_2=1$, (b) $\bar{n}=0$, $B_1=1$ and $B_2=0.1$, (c) $\bar{n}=0.01$, $B_1=1$ and $B_2=1$ and (c) $\bar{n}=0.01$, $B_1=0.1$ and $B_2=1$. The legend for all panels is as shown in panel (a).}}
\label{fig16}
\end{figure}

The time evolution of the bipartite entanglements in a completely isotropic (XXX) system is explored in Fig.~\ref{fig16}, starting from a maximally entangled state, where the legend in this figure is different from the default one in this paper. In Fig.~\ref{fig16}(a), the dynamics of $C_{12}$ and $C_{14}$ is depicted under the effect of a homogeneous magnetic field at zero temperature. The entanglement $C_{14}$, after displaying an oscillatory behavior vanishes within a finite period of time, while $C_{12}$ decays from a maximum value to zero monotonically within a smaller period of time. Raising the temperature, in the presence of a homogeneous magnetic field, causes a sudden death of entanglement at a much earlier time as illustrated in Fig.~\ref{fig16}(b) 

The effect of an inhomogeneous magnetic field at different temperatures is shown Fig.~\ref{fig16}(c) and (d). Applying any inhomogeneous magnetic field, $B_1>B_2$, at zero temperature, reduces the entanglement oscillation, and the entanglement vanishes within a finite period of time close to that of the homogeneous field case, as illustrated in Fig.~\ref{fig16}(c). Considering the other inhomogeneous filed case, at zero temperature, does not show a significant change from what is shown in Fig.~\ref{fig16}(c), but as the temperature is raised, a sudden death behavior is observed where the death times, for $C_{12}$ and $C_{14}$, are quite distinguished from each other, as illustrated in Fig.~\ref{fig16}(d), compared with the homogeneous field case shown in Fig.~\ref{fig16}(b). 
But in all cases, the entanglement in the XXX system vanishes asymptotically regardless of the system setup, where the environment dissipative decay effect dominates over the net magnetic field acting on the spins, aligning all spins down, at zero temperature, or close to down at finite temperature, to a separable steady state state, which will be discussed further in the next section. 

\section{Spin relaxation}
\subsection{Ising system}

\begin{figure}[htbp]
 \centering
 \subfigure{\includegraphics[width=8cm]{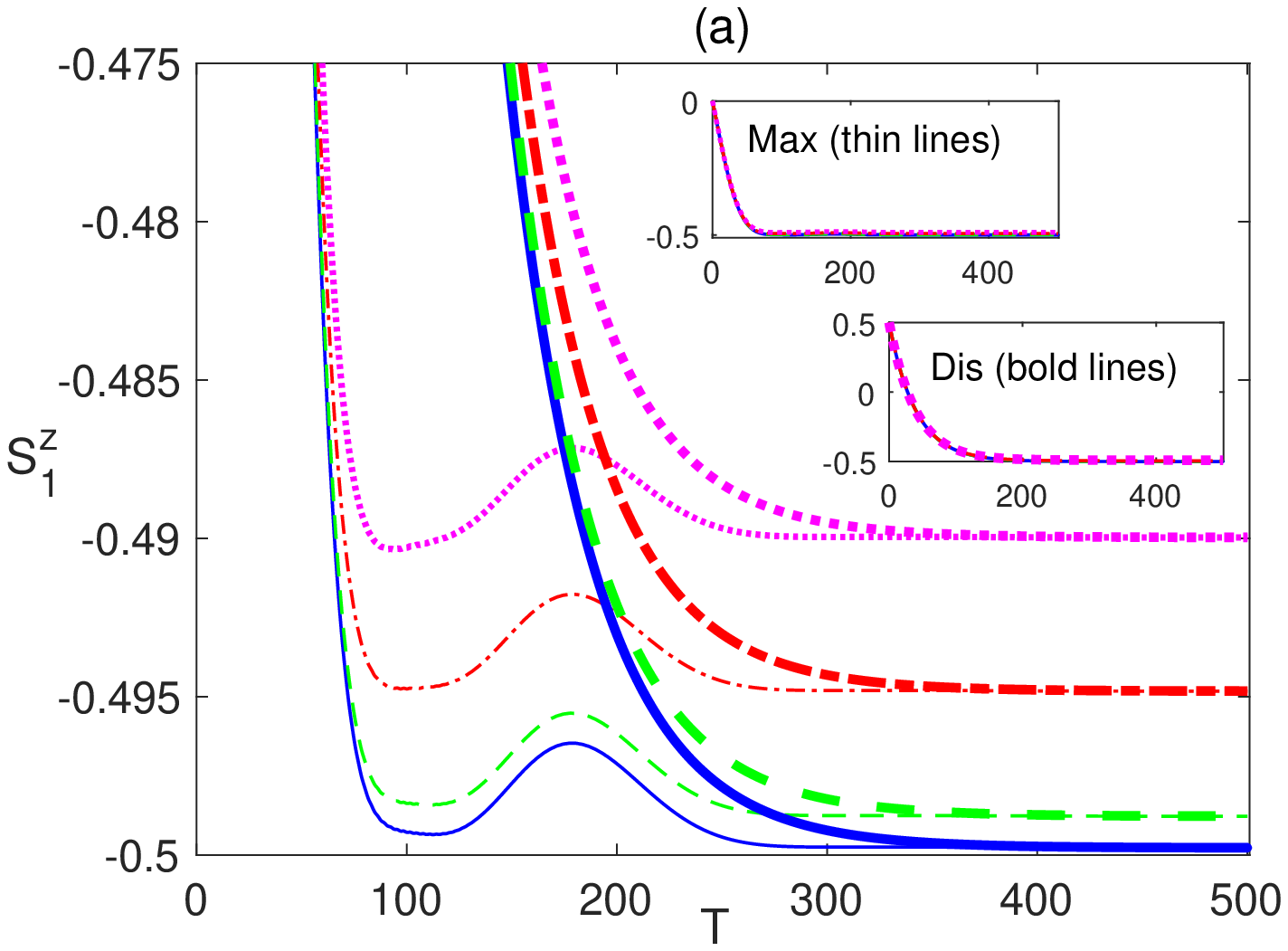}}\quad 
 \subfigure{\includegraphics[width=8cm]{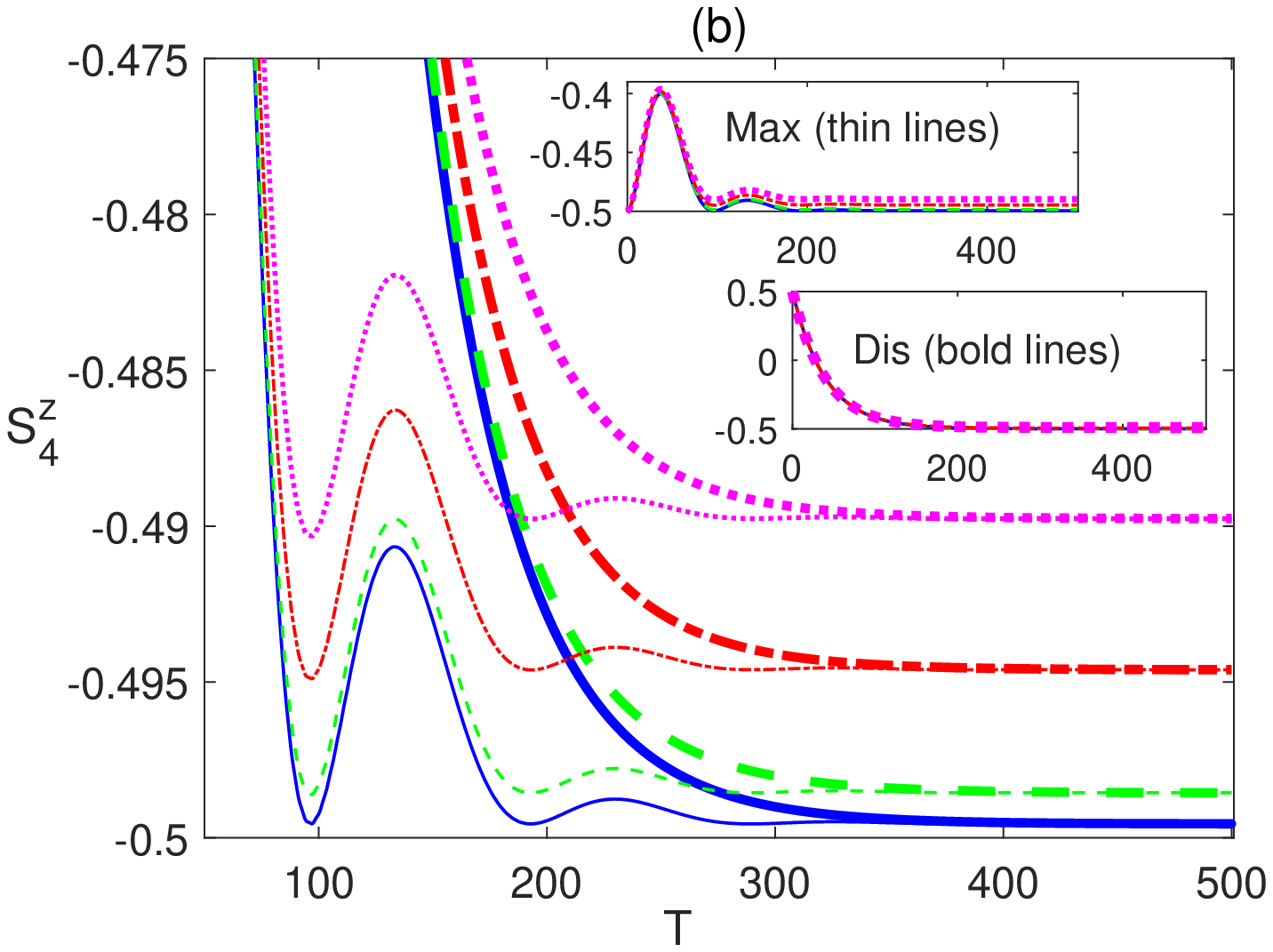}}\\
 \subfigure{\includegraphics[width=8cm]{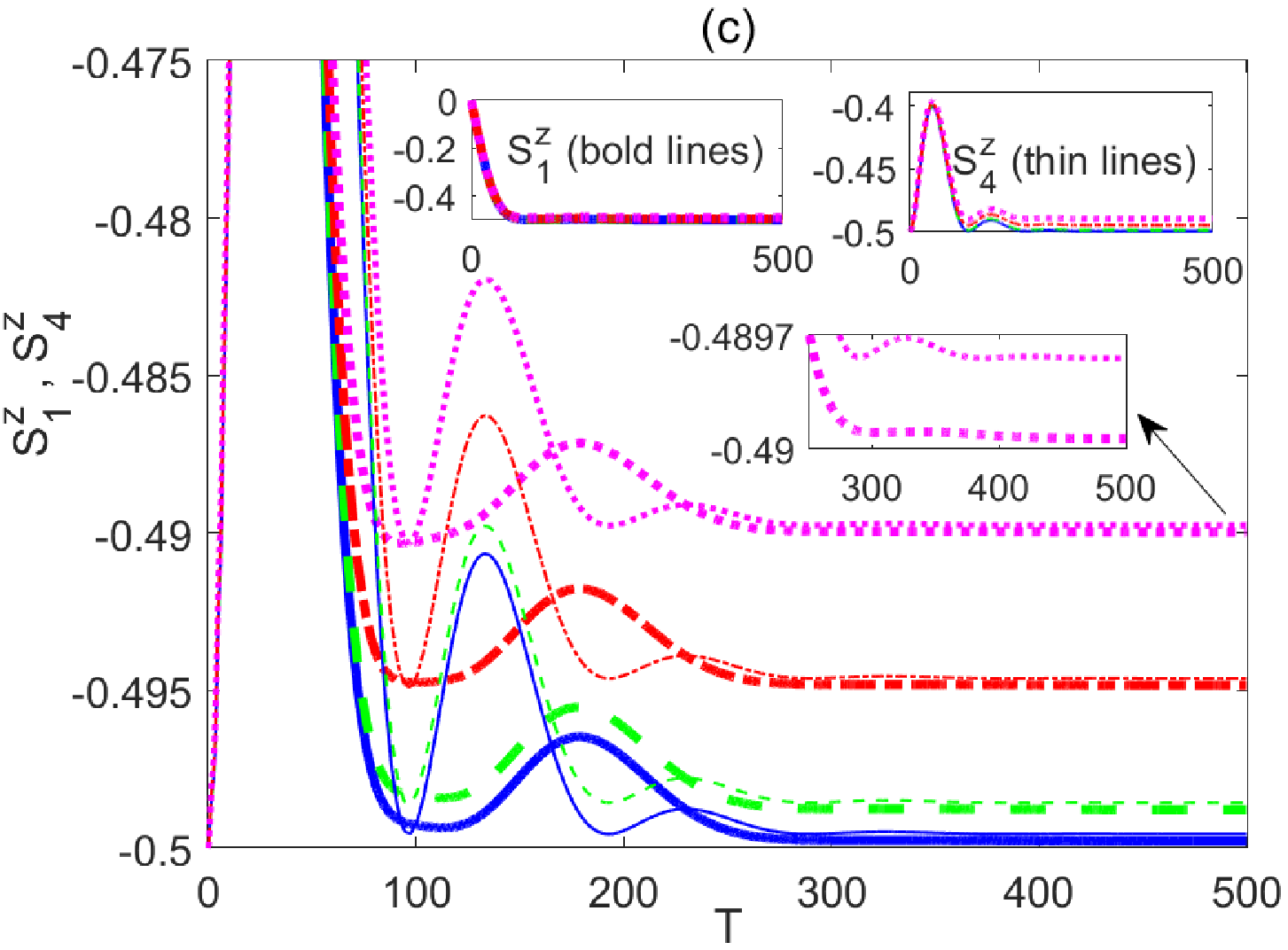}}\quad
 \caption{{\protect\footnotesize Time evolution of the spin state $\left\langle S^z_1\right\rangle $ and $\left\langle S^z_4\right\rangle $ in the Ising system in the presence of the environment $(\Gamma=0.05)$ at different temperatures ($0 \leq \bar{n} \leq 0.01$), and magnetic field strengths $B_1=1$ and $B_2=1$, where in (a) $\left\langle S^z_1\right\rangle $ starts from a maximally entangled state (thin lines) and a disentangled state (bold lines), (b) $\left\langle S^z_4\right\rangle $ starts from a maximally entangled state (thin lines) and a disentangled state (bold lines), and (c) $\left\langle S^z_1\right\rangle$ (bold lines) and $\left\langle S^z_4\right\rangle $ (thin lines), each starts from a maximally entangled state. The legend for all panels is as shown in Fig.~\ref{fig15}(c).}}
\label{fig17}
\end{figure}

It is very important and enlightening to explore how the spin state evolves in time under the different system configurations, compare and correlate it with the time evolution of the corresponding entanglements reported above, particularly their asymptotic behavior. 
In Fig.~\ref{fig17}, we study the time evolution of the spin state at the border site 1 and the central site 4 in the dissipative Ising system in the presence of a homogeneous magnetic field, $B_1=B_2=1$, at different temperatures starting from different initial states.
In Fig.~\ref{fig17}(a), we compare the time evolution of the spin 1 state $\left\langle S^z_1\right\rangle $ starting from two different initial states, maximally entangled and disentangled (separable), represented by  $\left|\psi_m\right\rangle$ and $\left|\psi_s\right\rangle$ respectively.
In the maximally entangled state case, represented by the thin plots, at all temperatures, $\left\langle S^z_1\right\rangle $ starts from zero, decays, reaches a minimum value, then shows a brief oscillation before reaching a steady state value that is very close to $-0.5$ at zero temperature but increases as the temperature is raised. Obviously, the spin state starts at zero value but asymptotically under the decay effect of the environment it is pushed downward, but due to the impact of the precession motion induced by the net magnetic field, it ends up at a steady state value that is slightly higher than $-0.5$ at zero temperature. It deviates further up, away from $-0.5$, as the temperature increases due to the thermal excitation as expected. Starting from a disentangled state value, $0.5$, depicted in Fig.~\ref{fig17}(a) as bold lines, it decays monotonically reaching a steady state value that coincides with that of the maximally entangled initial state case at all temperatures. The inner inset plots in this figure and all coming figures represent the overall dynamics of the concerned spin state.
\begin{figure}[htbp]
 \centering
 \subfigure{\includegraphics[width=8cm]{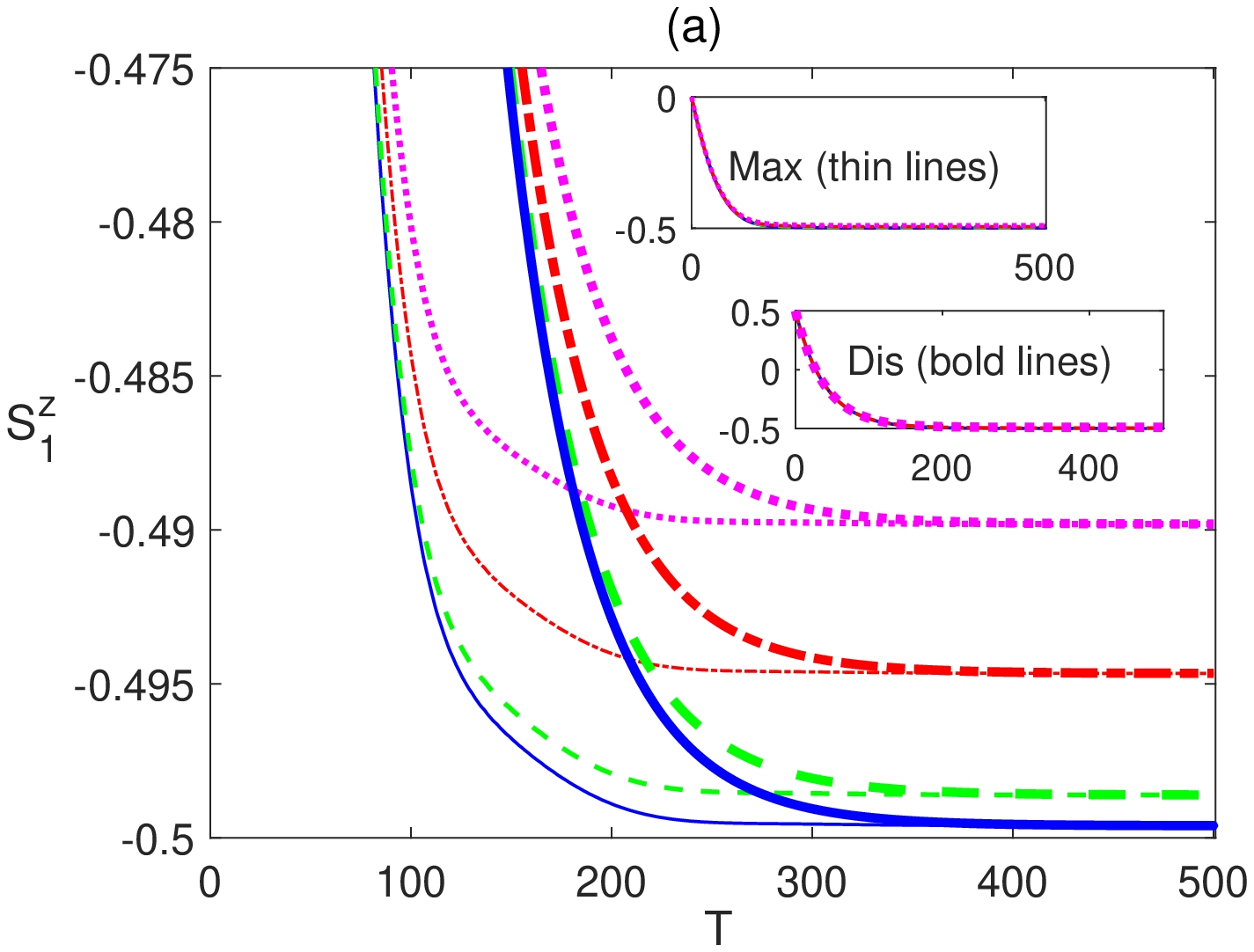}}\quad 
 \subfigure{\includegraphics[width=8cm]{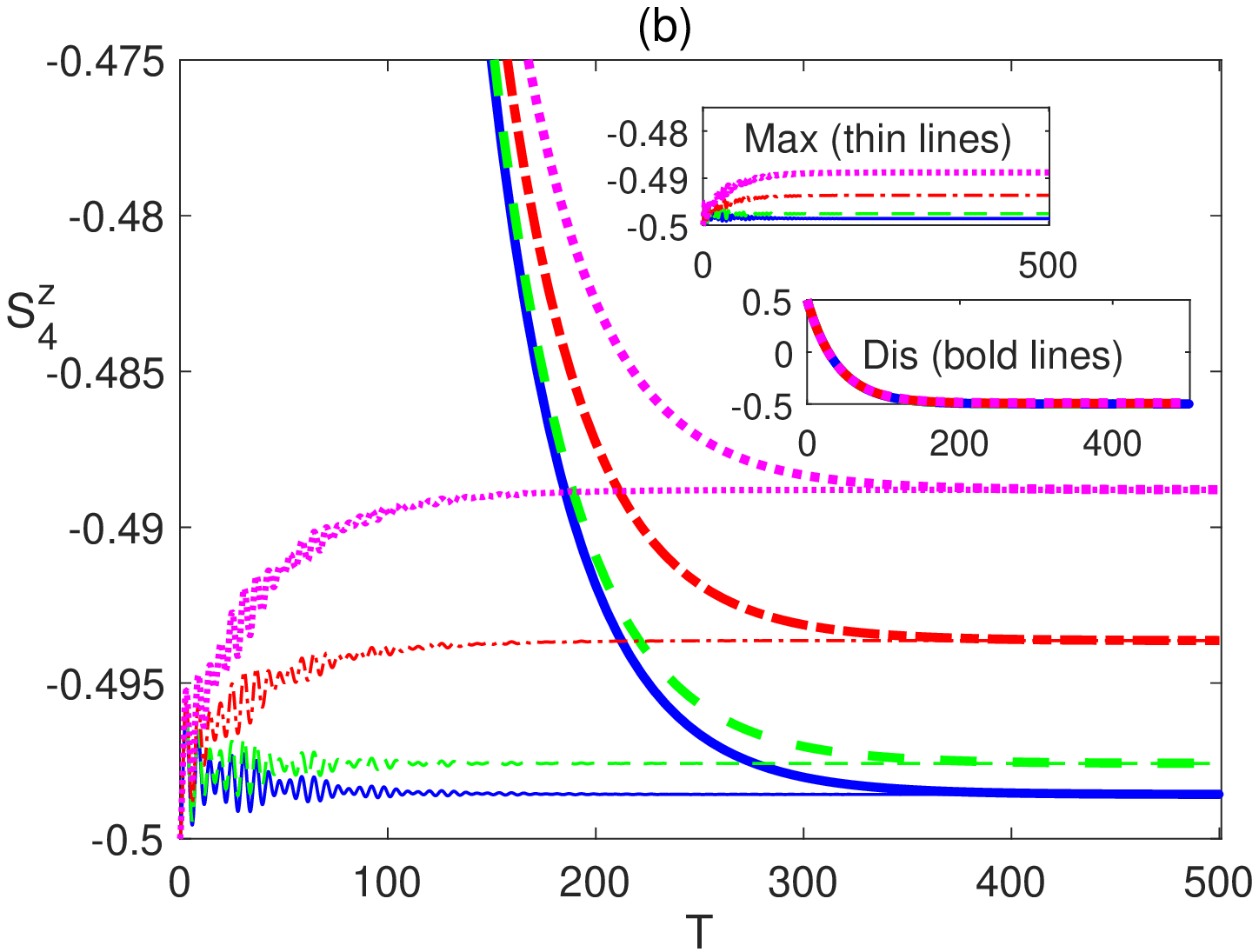}}\\
 \subfigure{\includegraphics[width=8cm]{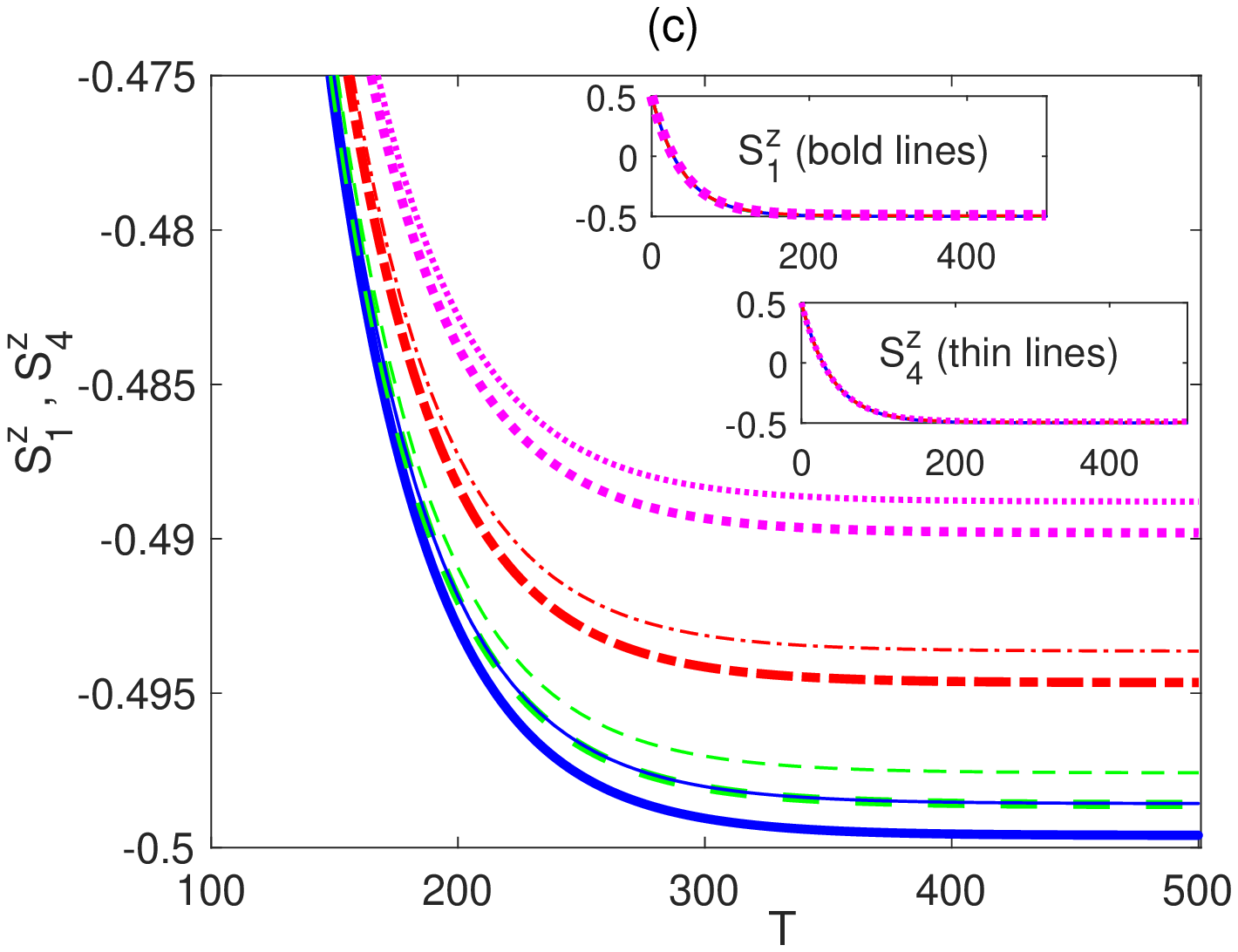}}\quad
 \caption{{\protect\footnotesize Time evolution of the spin state $\left\langle S^z_1\right\rangle $ and $\left\langle S^z_4\right\rangle $ in the Ising system in the presence of the environment $(\Gamma=0.05)$ at different temperatures ($0 \leq \bar{n} \leq 0.01$), and magnetic field strengths $B_1=1$ and $B_2=0.1$, where in (a) $\left\langle S^z_1\right\rangle $ starts from a maximally entangled state (thin lines) and a disentangled state (bold lines), (b) $\left\langle S^z_4\right\rangle $ starts from a maximally entangled state (thin lines) and a disentangled state (bold lines), and (c) $\left\langle S^z_1\right\rangle$ (bold lines) and $\left\langle S^z_4\right\rangle $ (thin lines), each starts from a disentangled state. The legend for all panels is as shown in Fig.~\ref{fig15}(c).}}
\label{fig18}
\end{figure}
In fact, comparing the dynamics of the spin state, in the current set up, with the corresponding bipartite entanglements (as well as the global bipartite entanglement) that we reported in Figs.~\ref{fig2}-\ref{fig8}, one can notice a strong resemblance. The entanglements corresponding to the maximally entangled state show oscillatory behavior before reaching a steady state value, while that starting from a disentangled state increases from zero monotonically before reaching the same steady state asymptotically. In addition, the time rate to reach the steady state is very much the same for the spin state and the entanglements.
\begin{figure}[htbp]
 \centering
 \subfigure{\includegraphics[width=8cm]{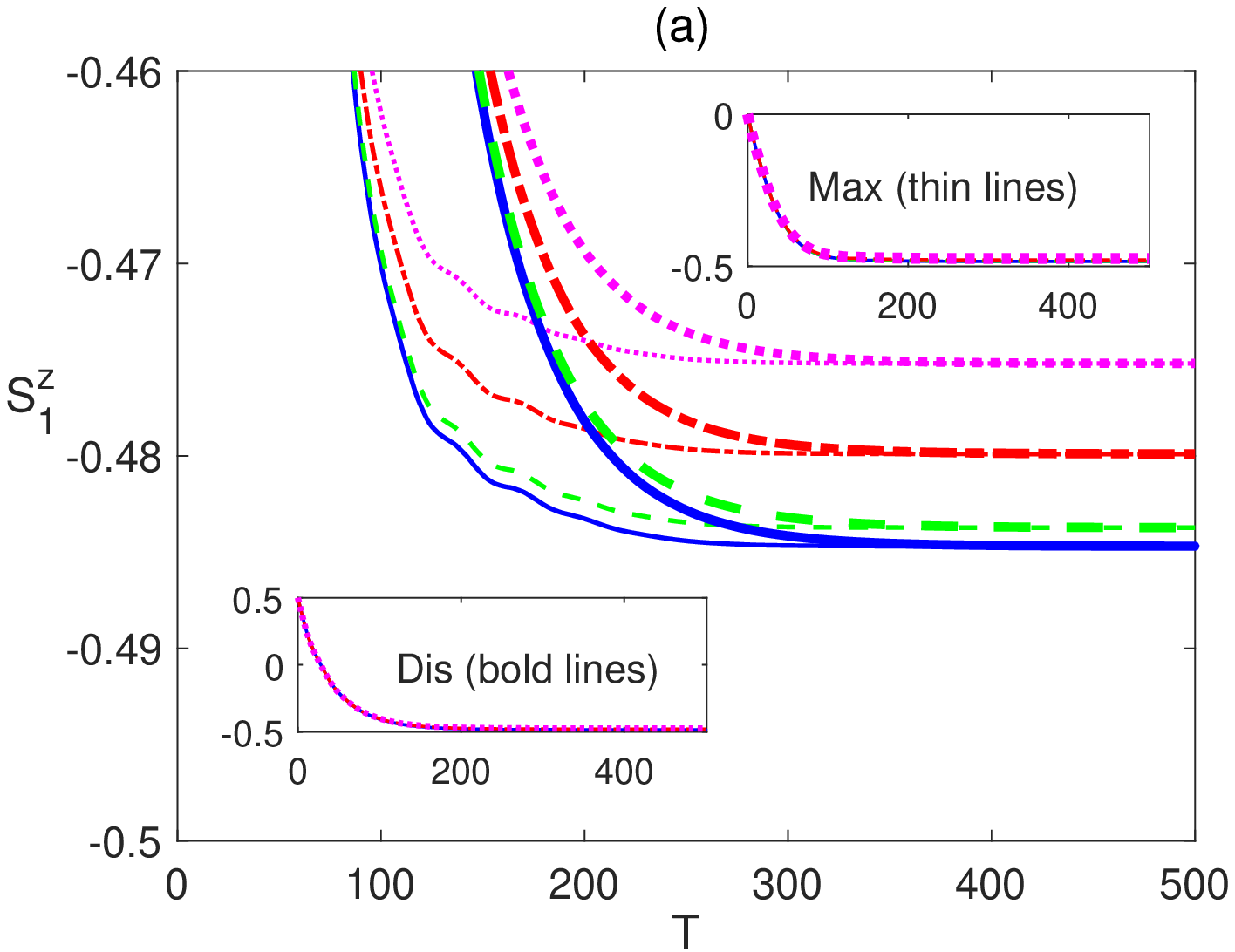}}\quad 
 \subfigure{\includegraphics[width=8cm]{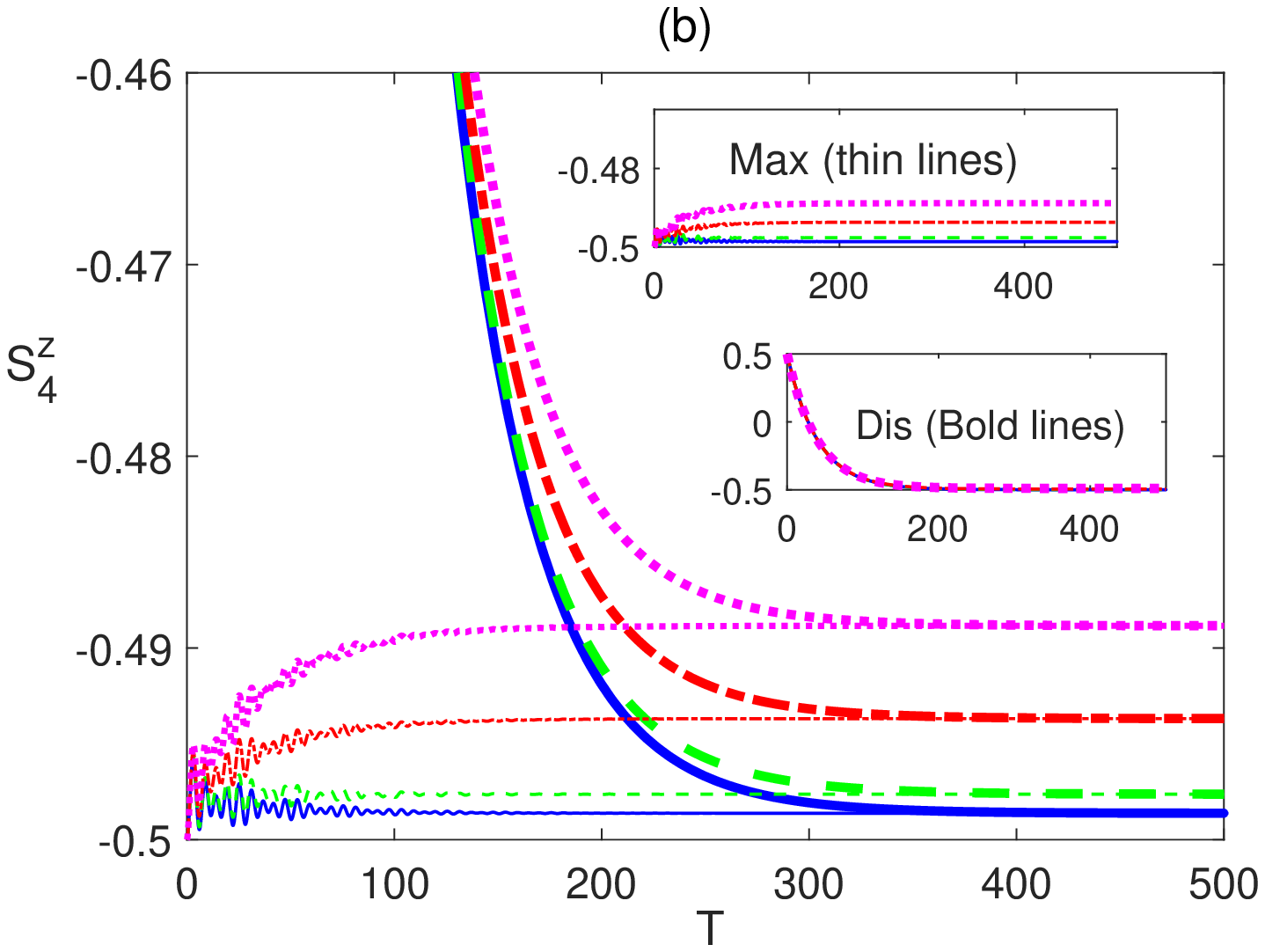}}\\
 \caption{{\protect\footnotesize Time evolution of the spin state $\left\langle S^z_1\right\rangle $ and $\left\langle S^z_4\right\rangle $ in the Ising system in the presence of the environment $(\Gamma=0.05)$ at different temperatures ($0 \leq \bar{n} \leq 0.01$), and magnetic field strengths $B_1=0.1$ and $B_2=1$, where in (a) $\left\langle S^z_1\right\rangle $ starts from a maximally entangled state (thin lines) and a disentangled state (bold lines), and (b) $\left\langle S^z_4\right\rangle $ starts from a maximally entangled state (thin lines) and a disentangled state (bold lines). The legend for all panels is as shown in Fig.~\ref{fig15}(c).}}
\label{fig19}
\end{figure}
In Fig.~\ref{fig17}(b), we explore the time evolution of the state of the central spin, 4, in the Ising system starting from the maximally entangled state (thin lines) and the disentangled one (bold lines). As can be noticed, the dynamics of $\left\langle S^z_4\right\rangle$ shows a very similar behavior to that of $\left\langle S^z_1\right\rangle$, which depends on the initial state but it reaches asymptotically a common steady state value regardless of the initial state at all temperatures. In Fig.~\ref{fig17}(c), we compare the behavior of $\left\langle S^z_1\right\rangle$ (bold lines) and $\left\langle S^z_4\right\rangle$ (thin lines), starting from an initial maximally entangled state. They reach their steady state values at around the same time but these values are not coinciding as can be seen, the value of $\left\langle S^z_4\right\rangle$ is slightly higher than that of $\left\langle S^z_1\right\rangle$. This behavior should be expected as the central spin is interacting with more nearest neighbor spins compared with the border spin, which provides a stronger precession and consequently a stronger resistance to the environment decay effect.
\begin{figure}[htbp]
 \centering
 \subfigure{\includegraphics[width=8cm]{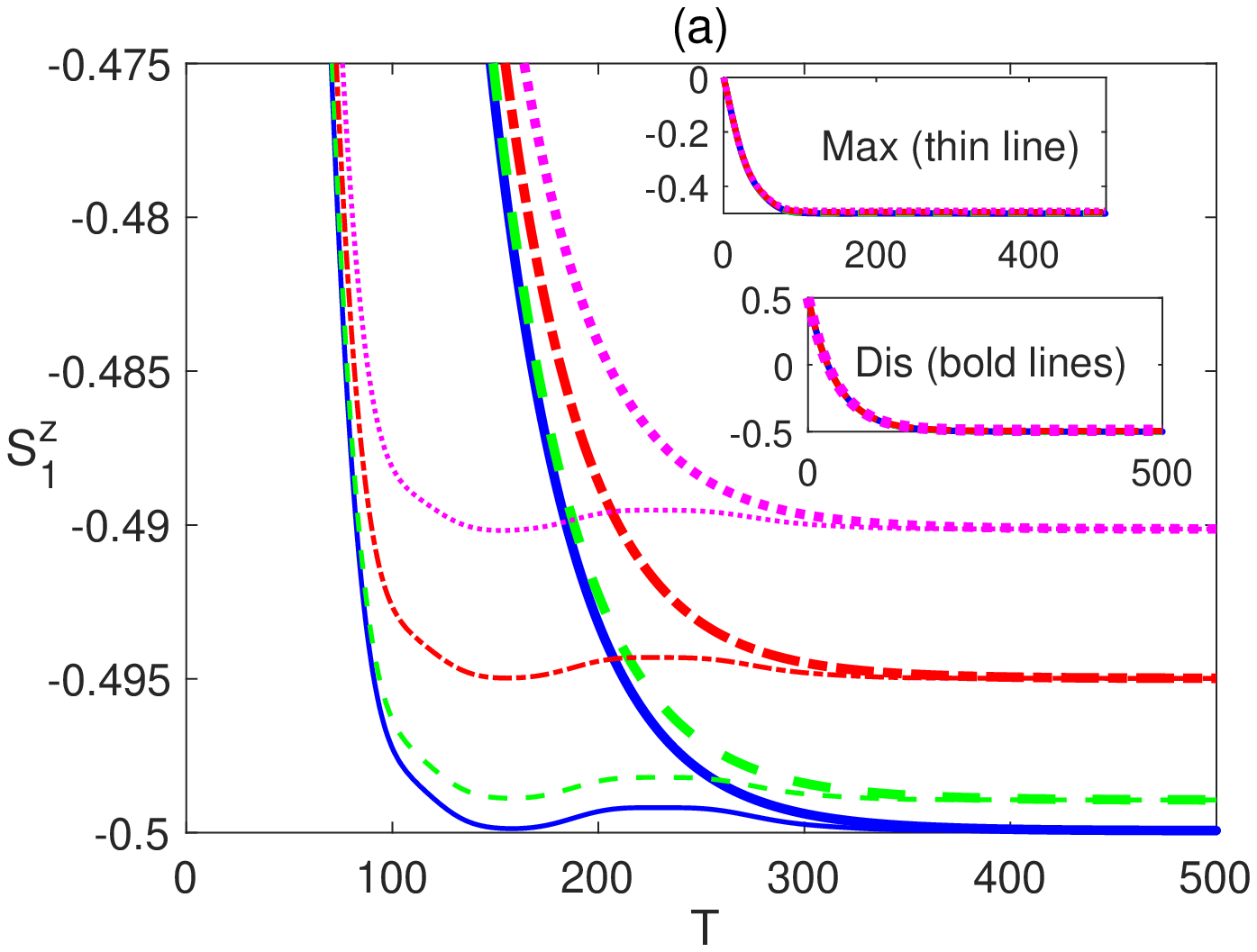}}\quad 
 \subfigure{\includegraphics[width=8cm]{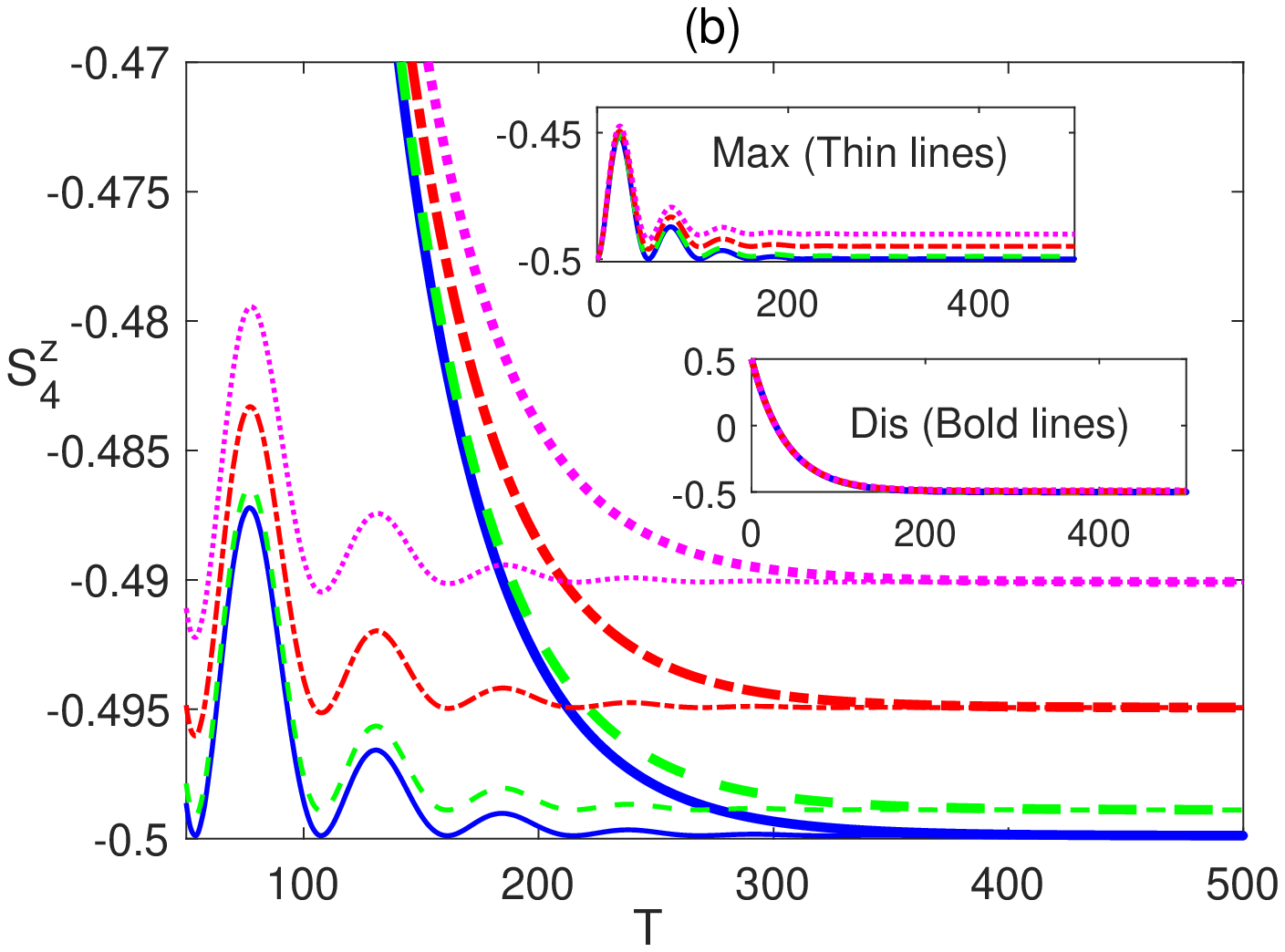}}\\
 \subfigure{\includegraphics[width=8cm]{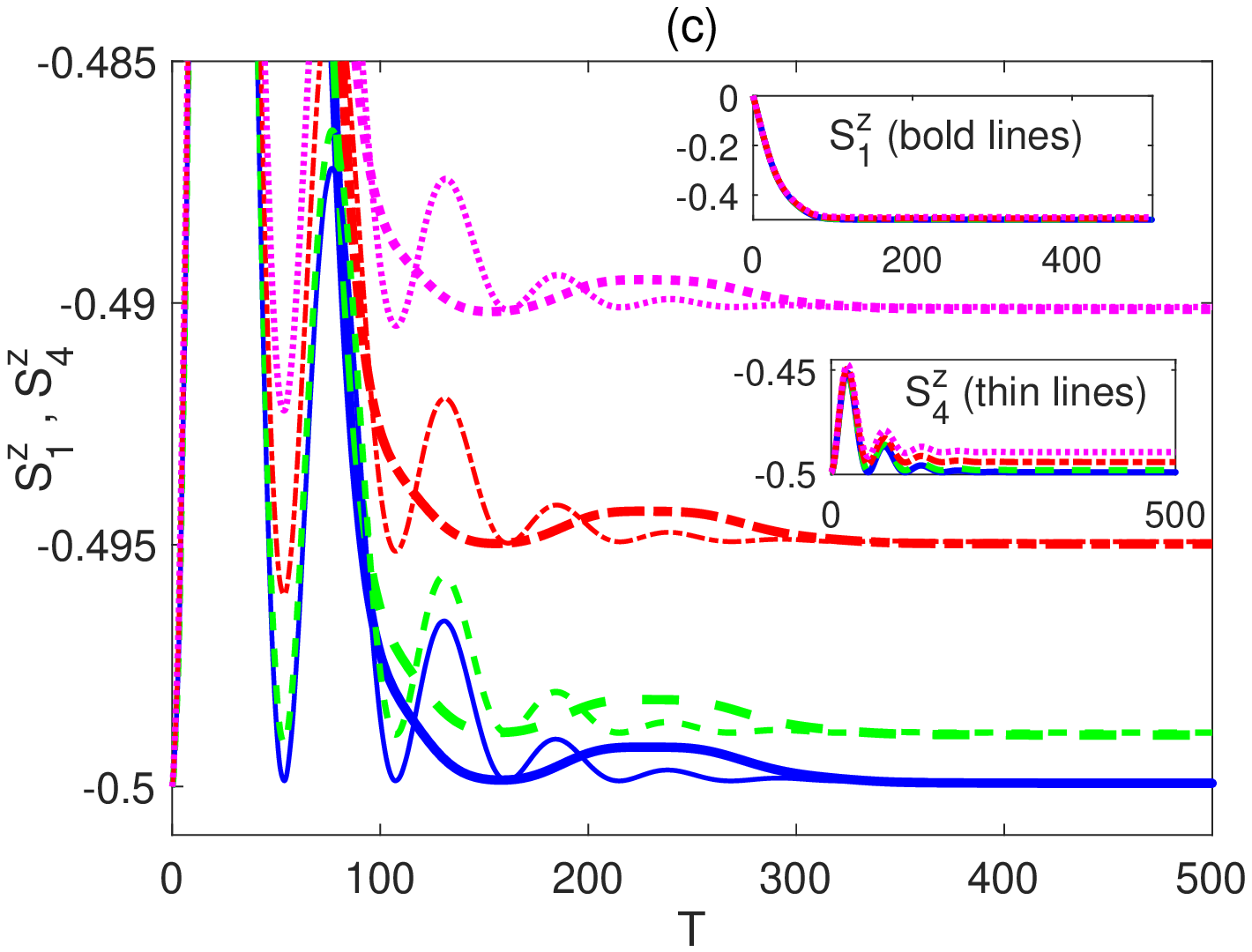}}\quad
 \caption{{\protect\footnotesize Time evolution of the spin state $\left\langle S^z_1\right\rangle $ and $\left\langle S^z_4\right\rangle $ in the XYZ system in the presence of the environment $(\Gamma=0.05)$ at different temperatures ($0 \leq \bar{n} \leq 0.01$), and magnetic field strengths $B_1=1$ and $B_2=1$, where in (a) $\left\langle S^z_1\right\rangle $ starts from a maximally entangled state (thin lines) and a disentangled state (bold lines), (b) $\left\langle S^z_4\right\rangle $ starts from a maximally entangled state (thin lines) and a disentangled state (bold lines), and (c) $\left\langle S^z_1\right\rangle$ (bold lines) and $\left\langle S^z_4\right\rangle $ (thin lines), each starts from a maximally entangled state. The legend for all panels is as shown in Fig.~\ref{fig15}(c).}}
\label{fig20}
\end{figure}
The effect of an inhomogeneous magnetic field on the time evolution of the spin states of the Ising system is explored in Fig.~\ref{fig18}, where $B_1>B_2$. The dynamics of $\left\langle S^z_1\right\rangle$ is depicted in Fig.~\ref{fig18}(a) starting from a maximally entanglement state (thin lines) and a disentangled state (bold lines), which shows a very similar behavior to what was observed in the homogeneous case except that for both of the initial states the spin state decays monotonically without any oscillation, reaching a common steady state value at each temperature, which is very slightly higher than that of the homogeneous field case. Therefore, although the magnetic field at the boarder sites is still of the the same strength, applying a weaker field at the center deviates the asymptotic value further away from $-0.5$.
\begin{figure}[htbp]
 \centering
 \subfigure{\includegraphics[width=8cm]{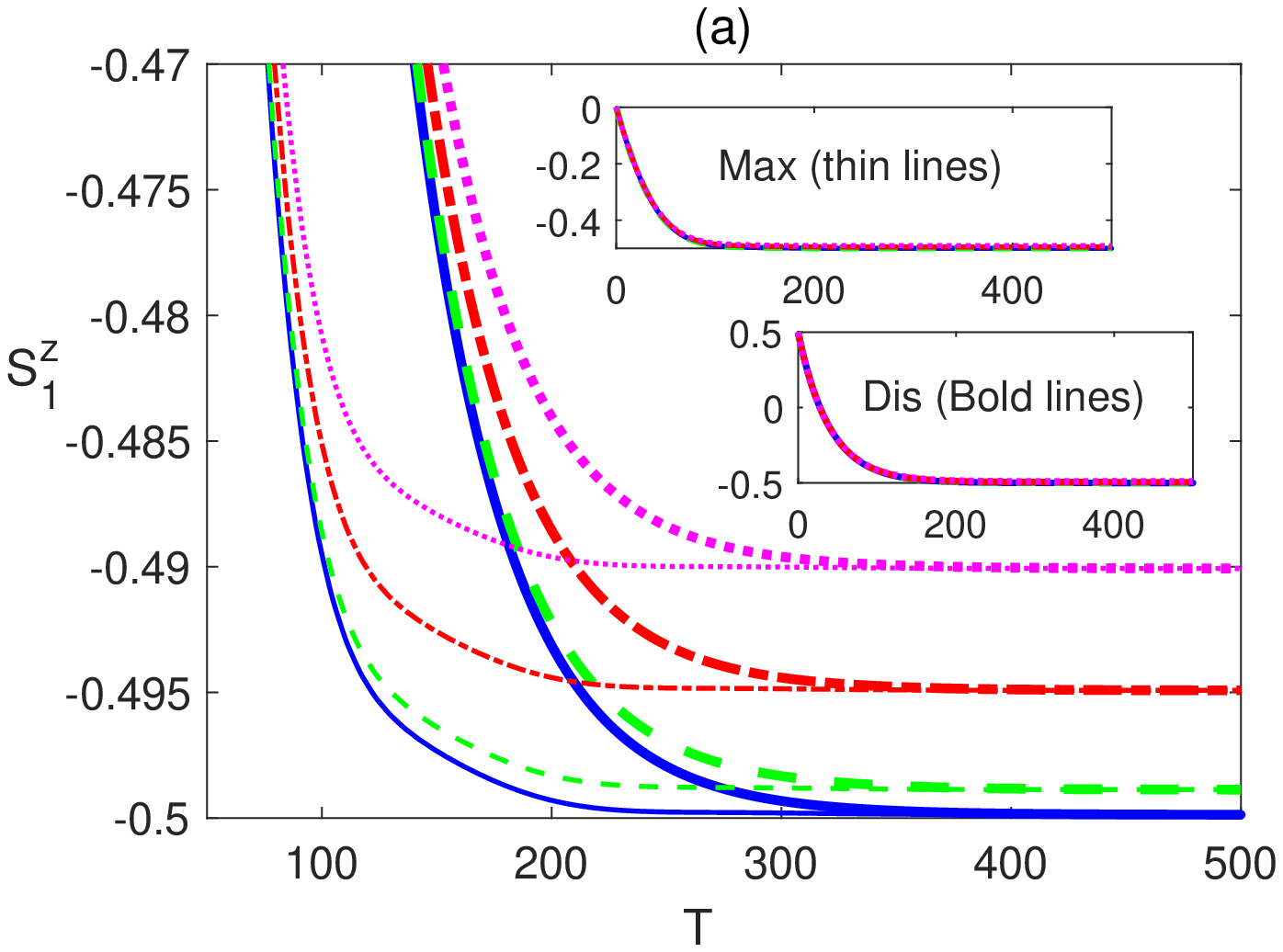}}\quad 
 \subfigure{\includegraphics[width=8cm]{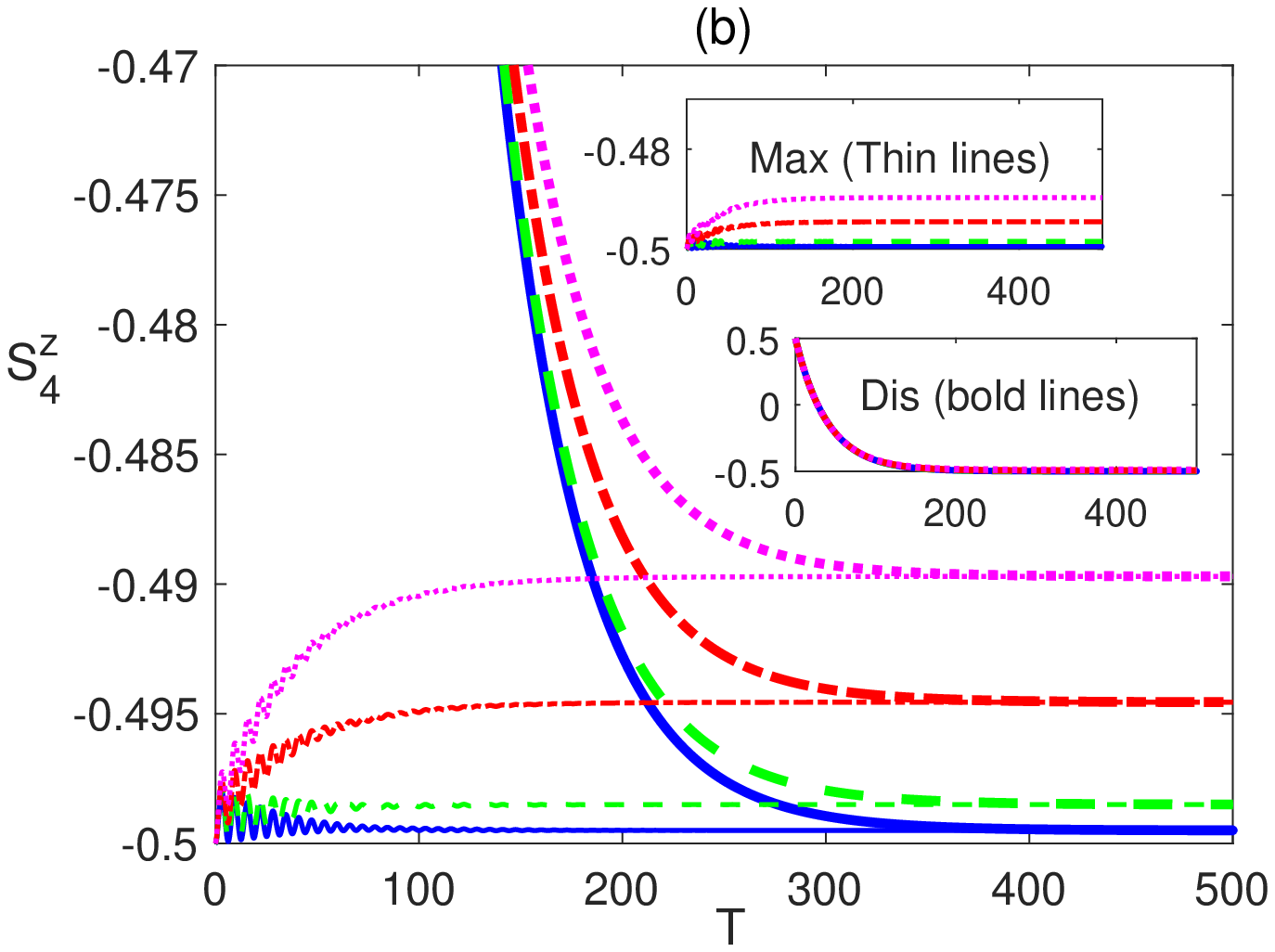}}\\
 \subfigure{\includegraphics[width=8cm]{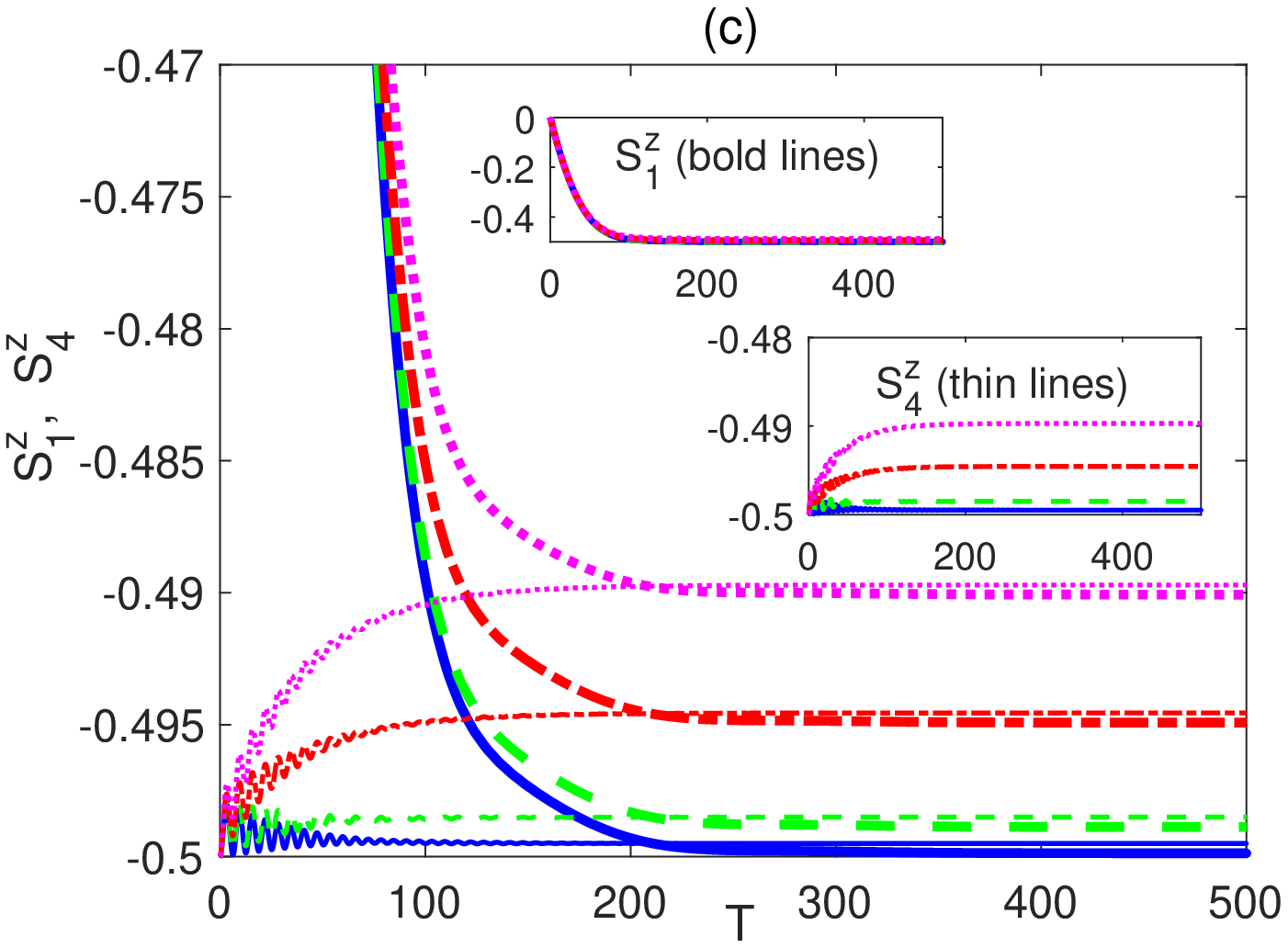}}\quad
 \caption{{\protect\footnotesize Time evolution of the spin state $\left\langle S^z_1\right\rangle $ and $\left\langle S^z_4\right\rangle $ in the XYZ system in the presence of the environment $(\Gamma=0.05)$ at different temperatures ($0 \leq \bar{n} \leq 0.01$), and magnetic field strengths $B_1=1$ and $B_2=0.1$, where in (a) $\left\langle S^z_1\right\rangle $ starts from a maximally entangled state (thin lines) and a disentangled state (bold lines), (b) $\left\langle S^z_4\right\rangle $ starts from a maximally entangled state (thin lines) and a disentangled state (bold lines), and (c) $\left\langle S^z_1\right\rangle$ (bold lines) and $\left\langle S^z_4\right\rangle $ (thin lines), each starts from a maximally entangled state. The legend for all panels is as shown in Fig.~\ref{fig15}(c).}}
\label{fig21}
\end{figure}
The dynamics of $\left\langle S^z_4\right\rangle$ is plotted in Fig.~\ref{fig18}(b), where it starts form $-0.5$ for the maximally entangled state (thin lines) and $0.5$ for the disentangled state (bold lines) but in both cases it evolves to a common steady state value. Comparing Fig.~\ref{fig18}(b) with Fig.~\ref{fig18}(a), one can notice an increase in the steady state values of $\left\langle S^z_4\right\rangle$ at all temperatures as a result of applying a weaker magnetic field at the central spin. 
In Fig.~\ref{fig18}(c), we compare the dynamics of $\left\langle S^z_1\right\rangle$ (bold lines) and $\left\langle S^z_4\right\rangle$ (thin lines) and their asymptotic steady state values starting form a  disentangled state, where both spins are initially pointing upward with the same value $0.5$. The steady state value of $\left\langle S^z_4\right\rangle$ is clearly higher than that of $\left\langle S^z_1\right\rangle$, at all temperatures, as a results of the applied inhomogeneous magnetic field. The difference in the steady state values is larger than what was seen in the homogeneous magnetic field case reported in Fig.~\ref{fig17}(c).
The other inhomogeneous magnetic field case, where $B_1=0.1$ and $B_2=1$, is presented in Fig.~\ref{fig19}. The common steady state values of $\left\langle S^z_1\right\rangle$, as illustrated in Fig.~\ref{fig19}(a), is much higher than that was reported in the previous cases in Fig.~\ref{fig18}(a) and (b), as a result of the weak magnetic field strength applied at the border sites in the current case. On the other hand, the common steady state values of $\left\langle S^z_4\right\rangle$, shown in Fig.~\ref{fig19}(b), is slightly lower than what was observed in Fig.~\ref{fig17}(b) and Fig.~\ref{fig18}(b), which emphasizes the very small impact of varying the magnetic field strength on the central spin. Comparing Fig.~\ref{fig19}(a) and Fig.~\ref{fig19}(b) shows a huge difference between the corresponding steady state values of $\left\langle S^z_1\right\rangle$ and $\left\langle S^z_4\right\rangle$, compared with what has been observed in the two previous magnetic fields configurations reported in Figs.~\ref{fig17} and \ref{fig18}.
This shows that, in this particular configuration, $\left\langle S^z_1\right\rangle $ and $\left\langle S^z_4\right\rangle $ end up in two completely different states and the steady state of $\left\langle S^z_1\right\rangle $ deviates the most away from the downward state, in contrary to $\left\langle S^z_4\right\rangle$, which an indication that the net magnetic field in this case is significantly enhancing the spin precession motion against the dissipative decay effect, causing the bipartite entanglement and the global entanglement of spin 1, as well as the robustness against thermal effects, to be considerably boosted, as was illustrated Figs.~\ref{fig2}(c), \ref{fig5}(c) and \ref{fig7}(c).

\begin{figure}[htbp]
 \centering
 \subfigure{\includegraphics[width=8cm]{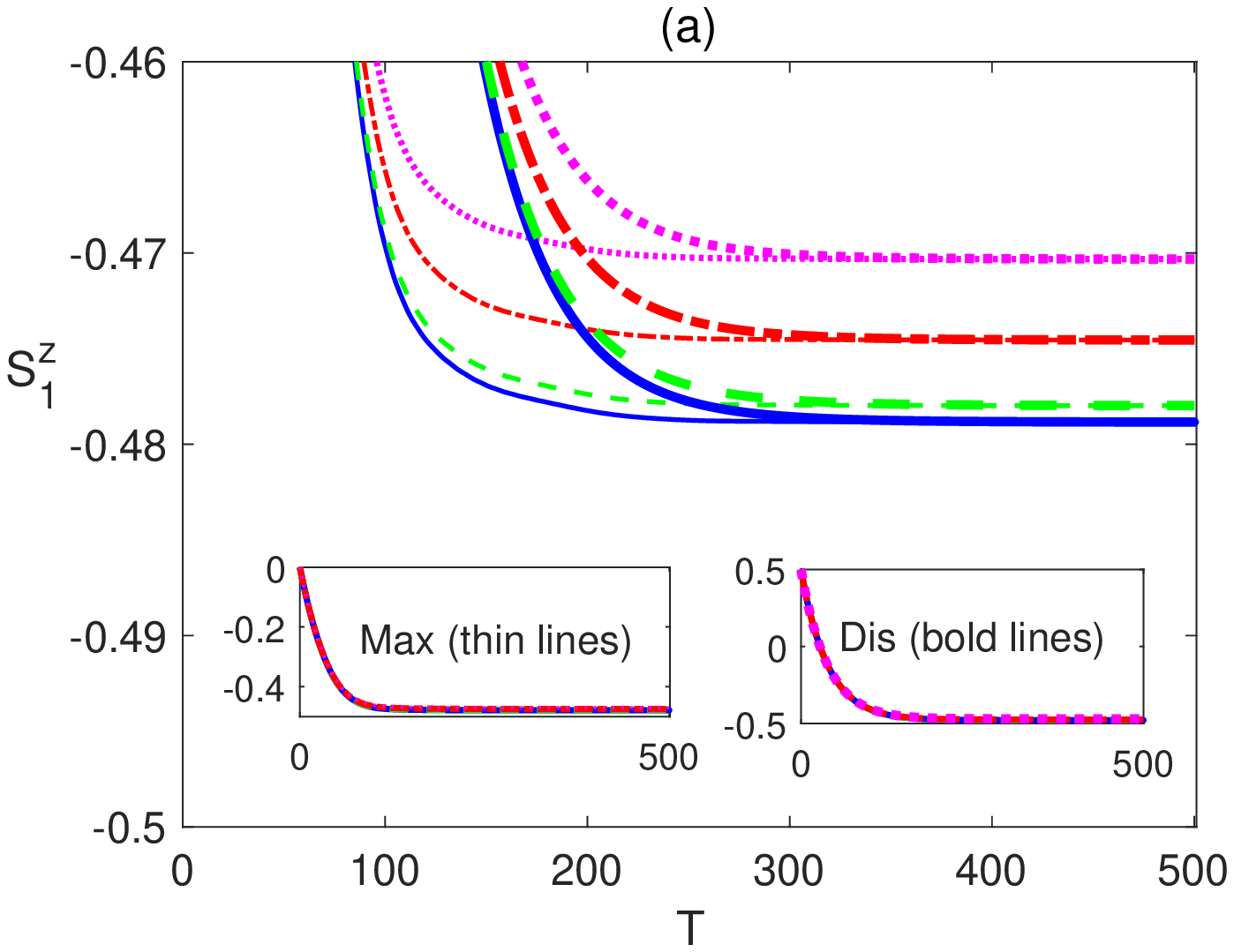}}\quad 
 \subfigure{\includegraphics[width=8cm]{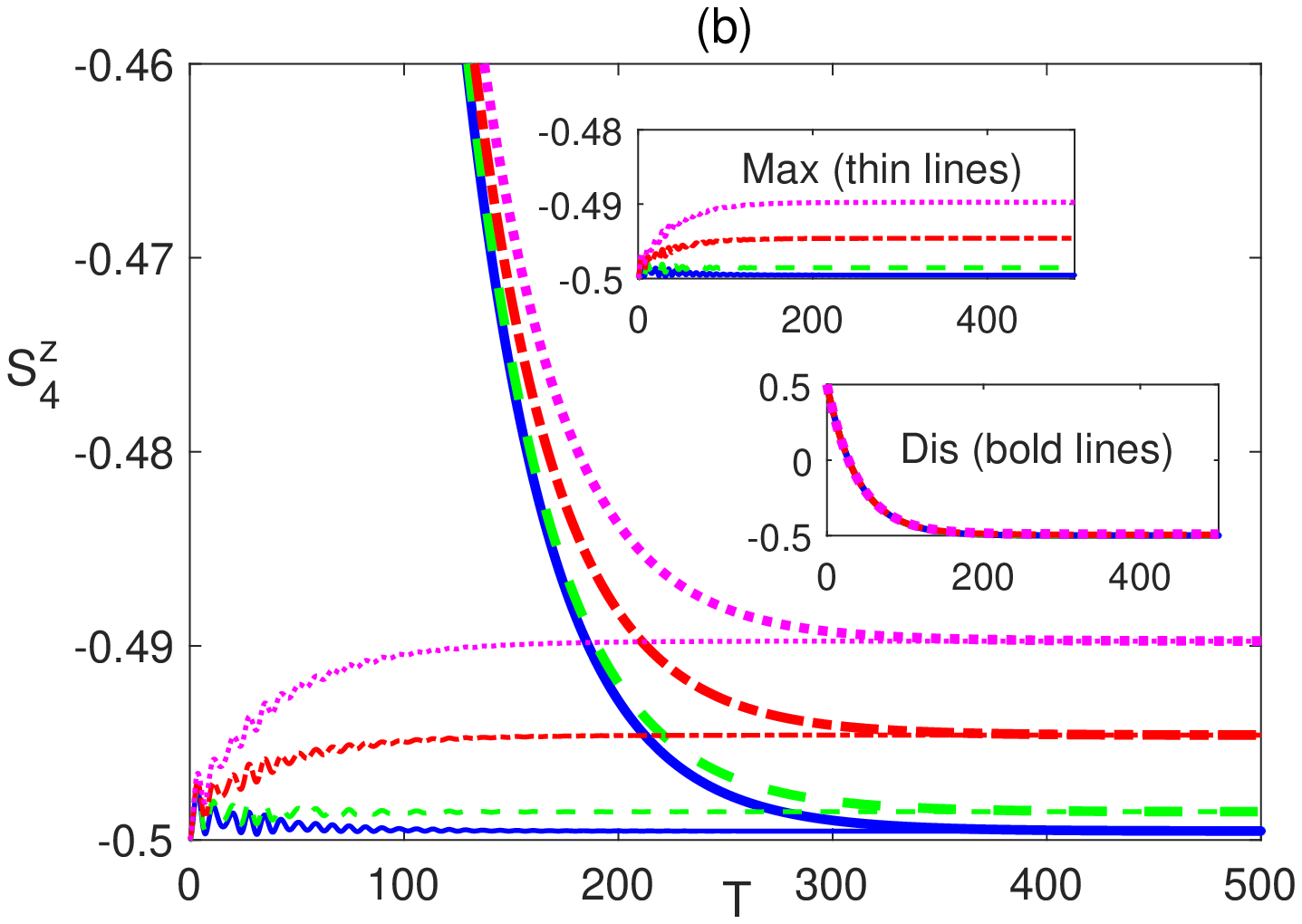}}\\
 \caption{{\protect\footnotesize Time evolution of the spin state $\left\langle S^z_1\right\rangle $ and $\left\langle S^z_4\right\rangle $ in the XYZ system in the presence of the environment $(\Gamma=0.05)$ at different temperatures ($0 \leq \bar{n} \leq 0.01$), and magnetic field strengths $B_1=0.1$ and $B_2=1$, where in (a) $\left\langle S^z_1\right\rangle $ starts from a maximally entangled state (thin lines) and a disentangled state (bold lines), and (b) $\left\langle S^z_4\right\rangle $ starts from a maximally entangled state (thin lines) and a disentangled state (bold lines). The legend for all panels is as shown in Fig.~\ref{fig15}(c).}}
\label{fig22}
\end{figure}
\subsection{XYZ system}

The spin state dynamics of the dissipative partially anisotropic (XYZ) spin system under different magnetic field configurations is explored in Figs.~\ref{fig20}-\ref{fig22}.
Comparing Fig.~\ref{fig20} with Fig.~\ref{fig17}, where the magnetic field is homogeneous, one can see that $\left\langle S^z_1\right\rangle $ and $\left\langle S^z_4\right\rangle $ in the partially anisotropic system behave in a very similar way to the Ising system case except that the oscillatory behavior is reduced in the XYZ system and the asymptotic values, at all temperatures, are slightly lower than that of the Ising case, which is expected as the effect of the spin-spin interaction in the current case is lower than the anisotropic case.
Turning to the inhomogeneous magnetic field case, where where $B_1 > B_2$, comparing Fig.~\ref{fig21} and Fig.~\ref{fig18}, it shows that, again, the dynamic behavior of $\left\langle S^z_1\right\rangle$ and $\left\langle S^z_4\right\rangle$ is very similar to the Ising case, but the asymptotic equilibrium values are lower even compared with the homogeneous field case. 

The interesting change takes place in the other inhomogeneous field case, $B_1 < B_2$, presented in Fig~\ref{fig22}, where the magnetic field strength at the border sites is much smaller than the one at the central site. While there is no notable change in the $\left\langle S^z_4\right\rangle$ behavior compared with that depicted in Fig~\ref{fig21}(b), the steady state values of $\left\langle S^z_1\right\rangle$ are much higher than the corresponding ones in the Ising case, illustrated in Fig~\ref{fig19}(a). One may have expected a different behavior for $\left\langle S^z_1\right\rangle$, and all the border spins, where they should have relaxed asymptotically to a steady state value that is lower, closer to the downward state, than the corresponding one of the Ising system, similar to the two previous magnetic field configurations, illustrated in Figs~\ref{fig20} and \ref{fig21}. However, the observed behavior means that the precession motion of the border spins is enhanced and persistent against the dissipative decay effect and thermal excitation, which may explain the strong long range, beyond nearest neighbors, entanglement observed in this system configuration, as illustrated in Figs~\ref{fig11}(c) and \ref{fig12}(c). This also might be a sign of a critical behavior of the system at this particular combination of the system parameters that needs further investigation.

\subsection{XXX system}
\begin{figure}[htbp]
 \centering
  \subfigure{\includegraphics[width=8cm]{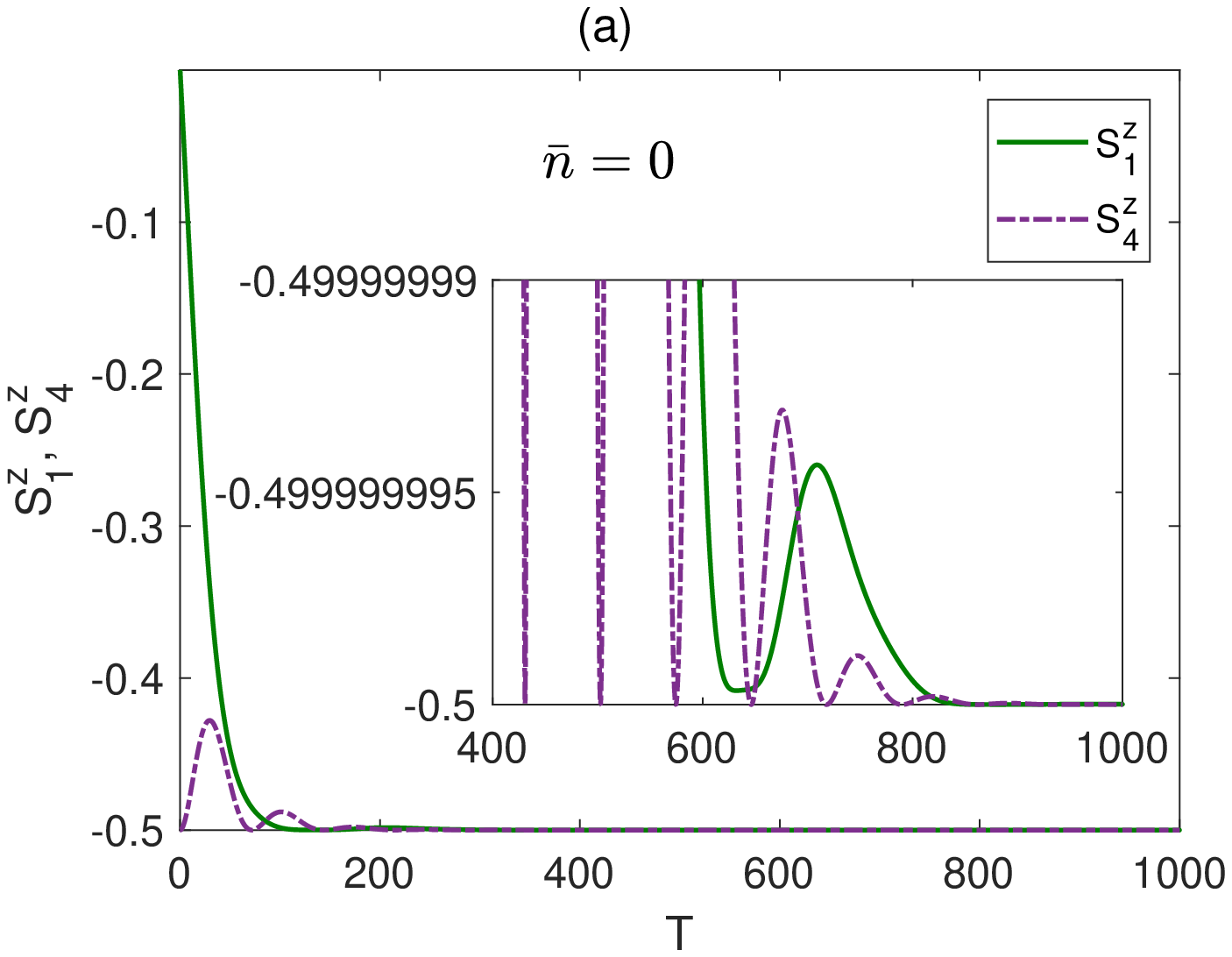}}\quad 
 \subfigure{\includegraphics[width=8cm]{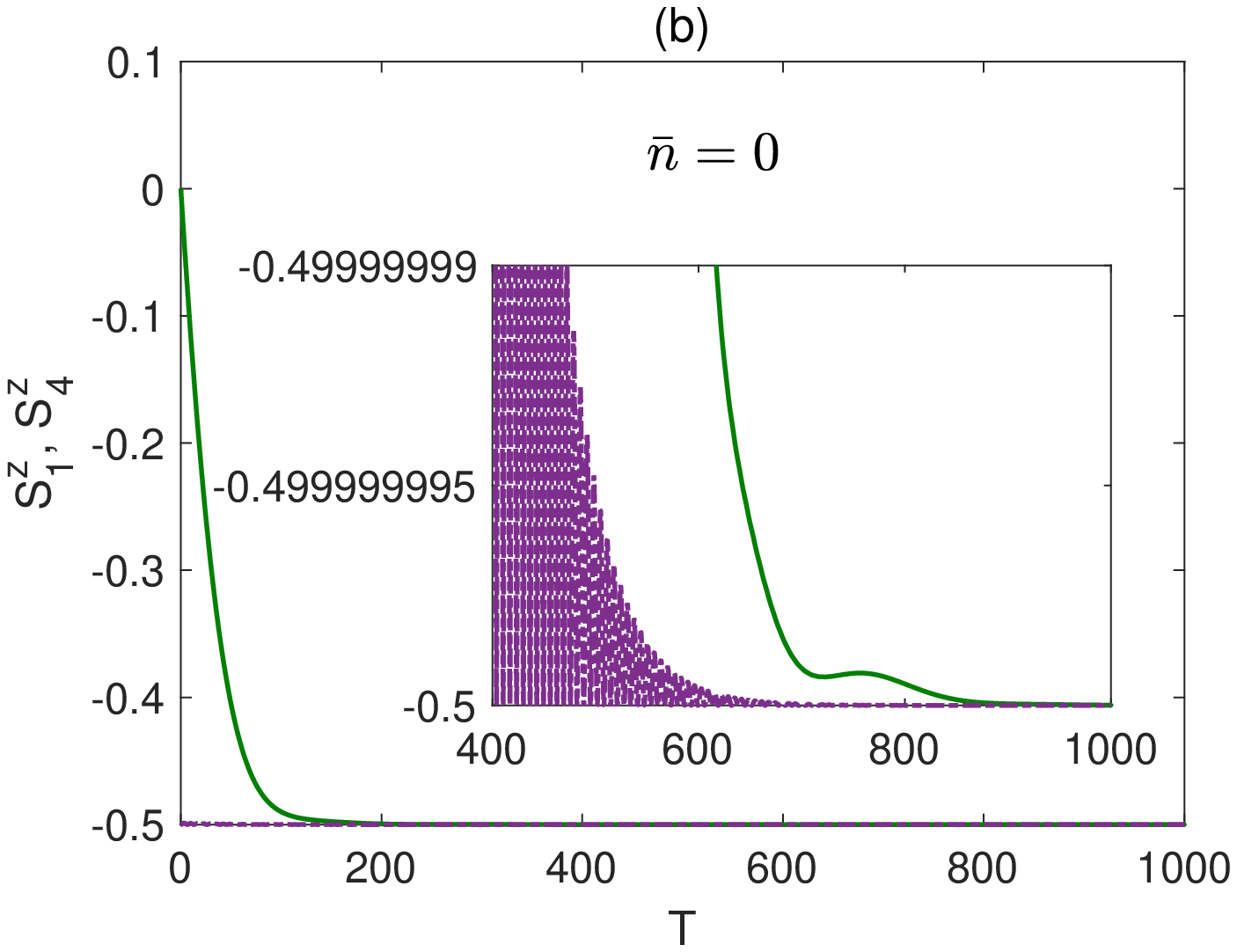}}\\
 \subfigure{\includegraphics[width=8cm]{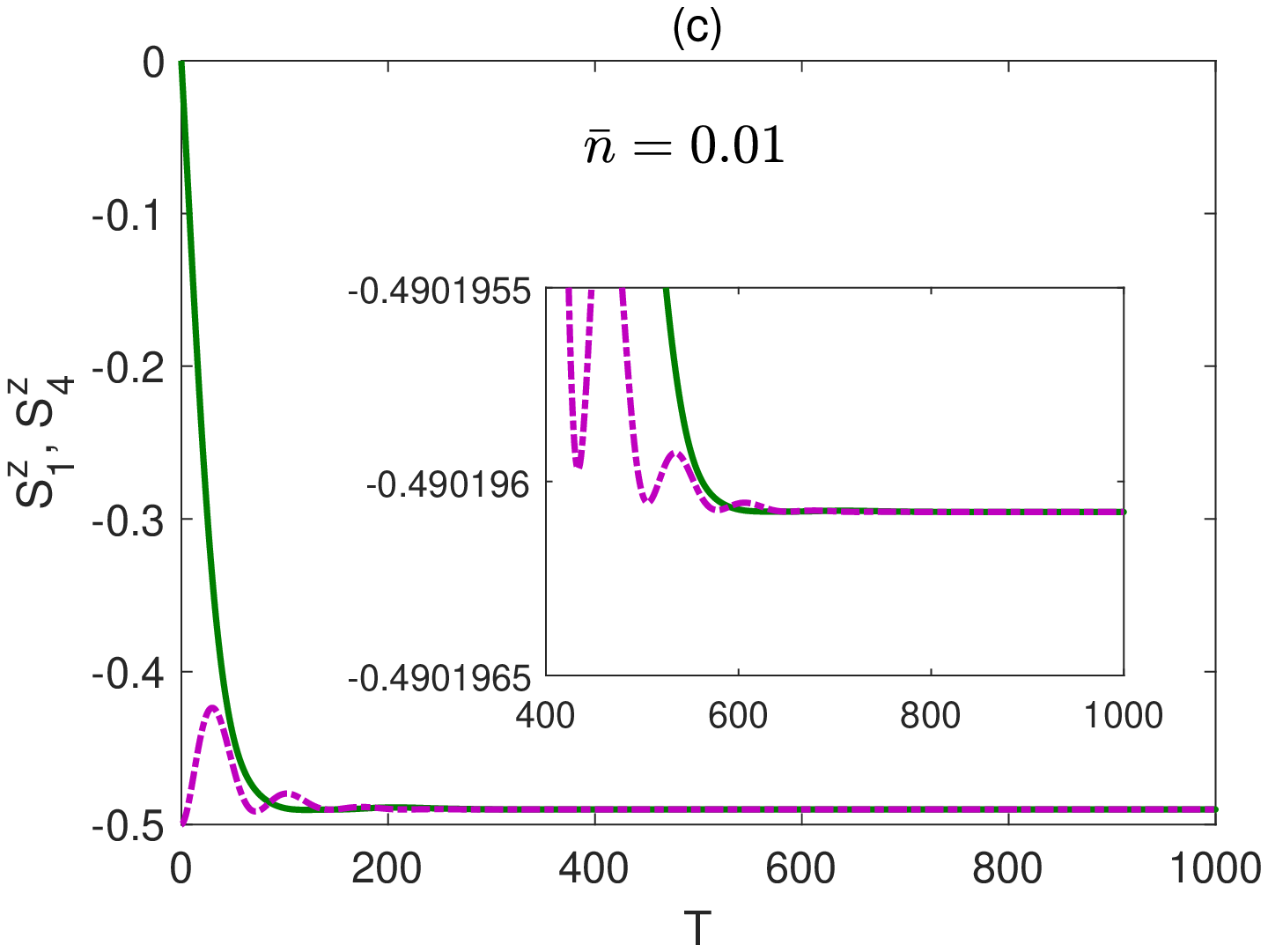}}\quad
  \subfigure{\includegraphics[width=8cm]{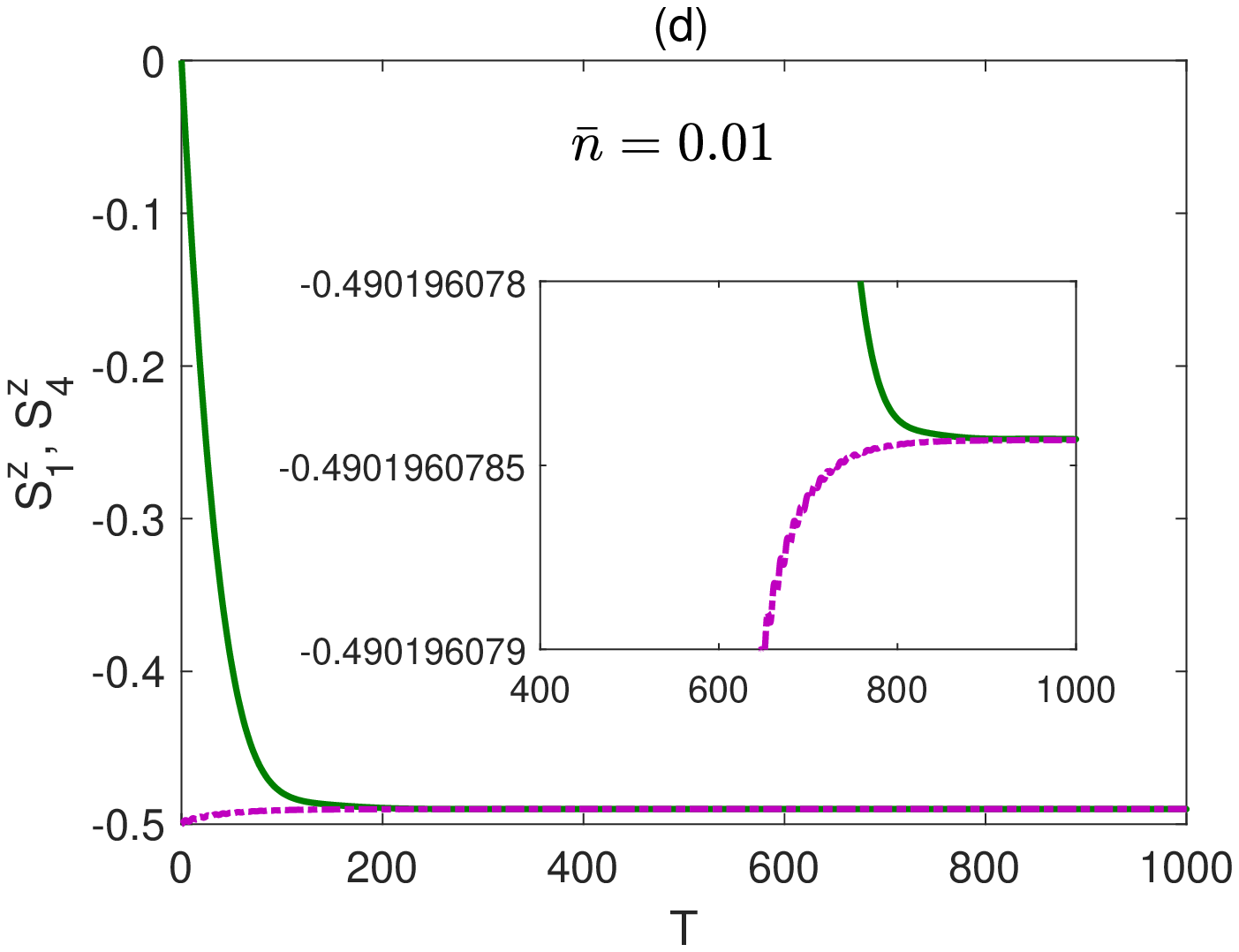}}\quad
 \caption{{\protect\footnotesize Time evolution of $\left\langle S^z_1\right\rangle$ and $\left\langle S^z_4\right\rangle $ in the $XXX$ system in the presence of the environment $(\Gamma=0.05)$ starting from an initial maximally entangled state, at different temperatures and magnetic fields, where in (a) $\bar{n}=0$, $B_1=1$ and $B_2=1$, (b) $\bar{n}=0$, $B_1=1$ and $B_2=0.1$, (d) $\bar{n}=0.01$, $B_1=1$ and $B_2=1$ and (c) $\bar{n}=0.01$, $B_1=0.1$ and $B_2=1$. The legend for all panels is as shown in panel (a).}}
\label{fig23}
\end{figure}

We study here the time evolution of the spin state in the completely isotropic dissipative (XXX) system, where a representative sample of the results are depicted in Fig~\ref{fig23}.
We found that, at zero temperature regardless of the magnetic field configuration or the initial state, the steady state of both of $\left\langle S^z_1\right\rangle$ and $\left\langle S^z_4\right\rangle$ takes exactly the value $-0.5$, which as we discussed before is due to the fact that the spins are pointing along the net magnetic field direction with no precession motion at all and as a result the disspiative decay effect dominates and force all the spins to, eventually, align downward, parallel to each other reaching a final separable (disentangled) state, as shown in Fig.~\ref{fig23}(a) and (b). At a non-zero temperature, the spin states $\left\langle S^z_1\right\rangle$ and $\left\langle S^z_4\right\rangle$, due to thermal excitation, relax to a common final steady state value that is higher than $-0.5$, where the higher the temperature is, the larger is the deviation, as shown in Fig.~\ref{fig23}(c). Applying an inhomogeneous magnetic field doesn't split the common steady state value of $\left\langle S^z_1\right\rangle$ and $\left\langle S^z_4\right\rangle$, even at non-zero temperature as can be noticed in Fig.~\ref{fig23}(d). 
This means, in the XXX system case, all the spins end up pointing in the same direction reaching a separable disentangled steady state, as was pointed out in Fig.~\ref{fig16}, regardless of the initial state or the system parameters.
\section{Conclusions}
We studied a finite two-dimensional Heisenberg spin lattice with nearest-neighbor spin interaction coupled to a dissipative Lindblad environment in the presence of an external inhomogeneous magnetic field at finite temperature. The spin lattice consists of a central spin surrounded by 6 border spins equally distant form it in a triangular symmetric structure.
We developed an exact numerical solution for the Lindblad master equation of the system, under the Brorn-Markovian constrain, in Liouville space.
We have shown that applying an inhomogeneous magnetic field, compared with the homogeneous one, has a  great impact on the entanglement distribution among the spins in the lattice and can be utilized to significantly enhance the bipartite and global bipartite entanglement among the spins in the system, even beyond nearest neighbors, and boost their thermal robustness at different degrees of anisotropy.
Particularly, applying an inhomogeneous magnetic field with a gradient directed inward, where the central spin is exposed to higher magnetic field strength compared with the border spins, has the most significant impact on the entanglement enhancement and robustness against the thermal dissipative environment at all degrees of anisotropy, compared with the other magnetic field configurations.
Applying such a field to a completely anisotropic (Ising) system, enhanced the nearest neighbor entanglement, as well as the global one, among the border spins in the steady state considerably, though the bipartite entanglements involving the central spin were slightly reduced. All the beyond nearest neighbor bipartite entanglements of all spins vanish asymptotically in this setup. 
However, when the same inhomogeneous field was applied to a partially anisotropic (XYZ) system, not only it has significantly enhanced the steady state entanglement among the nearest neighbor border spins, and its thermal robustness, but also among all the beyond nearest neighbor spins in a remarkable way, which indicates that combining an inhomogeneous external magnetic field with anisotropic spin-spin interaction creates a long range quantum correlation across the lattice. 
The entanglement in the isotropic (XXX) system was found to asymptotically vanish regardless of the initial state of the system, the temperature or the degree of inhomogeneity of the magnetic field, where the last affects only the length of the time period the system spends before completely losing its entanglement as well as the loss rate.
Furthermore, we investigated the spin state dynamics in the system and its correlation to the entanglement behavior under the different system configurations. We have demonstrated that in the isotropic (XXX) system, the dissipative decay effect dominates entirely, over the other system parameters influences, forcing all the spins to align parallel to each other, downward at zero temperature or very slightly away from the downward at finite temperature, into a separable (disentangled) steady state, regardless of the system initial state or the inhomogeneity of the magnetic field. On the other hand, for anisotropic system, the inhomogeneous magnetic field, with again an inward gradient, was found to have the greatest impact on the spins dynamics and steady state. The complete anisotropy, in the Ising system, enhanced the system robustness against the environment thermal dissipative decay effect, protecting the mutual entanglement among the spins and causing them to asymptotically relax to different steady states that depend on their locations in the lattice, deviating from the downward state induced by the decay. 
Interestingly, the XYZ system, where the degree of anisotropy is lower than that of the Ising system,  exhibited a stronger robustness to the environment decay effect, which was reflected in a more distinguished steady states of the spins, from each other, and further away from the downward state. These states, as we have already pointed out, were accompanied by a long range quantum entanglement among the spins across the lattice, which is a sign of a critical behavior of the system taking place at this configuration that may need further investigation in a future work.
\section*{Acknowledgment}
This work was supported by the University of Sharjah, Vice Chancellor of research office, grant No.
1602143030-P.



\begin{thebibliography}{26}
\expandafter\ifx\csname natexlab\endcsname\relax\def\natexlab#1{#1}\fi
\expandafter\ifx\csname bibnamefont\endcsname\relax
  \def\bibnamefont#1{#1}\fi
\expandafter\ifx\csname bibfnamefont\endcsname\relax
  \def\bibfnamefont#1{#1}\fi
\expandafter\ifx\csname citenamefont\endcsname\relax
  \def\citenamefont#1{#1}\fi
\expandafter\ifx\csname url\endcsname\relax
  \def\url#1{\texttt{#1}}\fi
\expandafter\ifx\csname urlprefix\endcsname\relax\def\urlprefix{URL }\fi
\providecommand{\bibinfo}[2]{#2}
\providecommand{\eprint}[2][]{\url{#2}}

  \bibitem[{\citenamefont{Peres}(1993)}]{Peres1993}
  \bibinfo{author}{\bibfnamefont{A.}~\bibnamefont{Peres}},
  \emph{\bibinfo{title}{Quantum Theory: Concepts and Methods}},
 (\bibinfo{publisher}{Kluwer, Dordrecht, The Netherlands},
  \bibinfo{year}{1993}).


  \bibitem[{\citenamefont{Sachdev}(2001)}]{Sachdev2001}
  \bibinfo{author}{\bibfnamefont{S.}~\bibnamefont{Sachdev}},
  \emph{\bibinfo{title}{Quantum Phase Transitions}},
  (\bibinfo{publisher}{Cambridge Univ. Press, Cambridge},
  \bibinfo{year}{2001}).

  \bibitem[{\citenamefont{Nielsen and Chuang}(2000)}]{Nielsen2000}
  \bibinfo{author}{\bibfnamefont{M.}~\bibnamefont{Nielsen}} \bibnamefont{and}
  \bibinfo{author}{\bibfnamefont{I.}~\bibnamefont{Chuang}},
  \emph{\bibinfo{title}{Quantum Computation and Quantum Communication}},
  (\bibinfo{publisher}{Cambridge Univ. Press, Cambridge},
  \bibinfo{year}{2000}).

  \bibitem[{\citenamefont{Zurek et~al.}(1991)\citenamefont{Zurek}}]{Zurek1991}
  \bibinfo{author}{\bibfnamefont{W.}~\bibnamefont{Zurek}},
  \bibinfo{journal}{Phys. Today} \textbf{\bibinfo{volume}{44}},
  \bibinfo{pages}{36-44} (\bibinfo{year}{1991}).

  \bibitem[{\citenamefont{Bacon et~al.}(2000)\citenamefont{Bacon, Kempe, Lidar, and Whaley}}]{Bacon2000}
  \bibinfo{author}{\bibfnamefont{D.}~\bibnamefont{Bacon}},
  \bibinfo{author}{\bibfnamefont{J.}~\bibnamefont{Kempe}},
  \bibinfo{author}{\bibfnamefont{D.~A.} \bibnamefont{Lidar}}, \bibnamefont{and}
  \bibinfo{author}{\bibfnamefont{K.~B.} \bibnamefont{Whaley}},
  \bibinfo{journal}{Phys. Rev. Lett.} \textbf{\bibinfo{volume}{85}},
  \bibinfo{pages}{1758} (\bibinfo{year}{2000}).

  \bibitem[{\citenamefont{Diehl et~al.}(2010)\citenamefont{Diehl, Tomadin, Micheli, Fazio, and Zoller}}]{Diehl2010}
  \bibinfo{author}{\bibfnamefont{S.}~\bibnamefont{Diehl}},
  \bibinfo{author}{\bibfnamefont{A.}~\bibnamefont{Tomadin}},
  \bibinfo{author}{\bibfnamefont{A.}~\bibnamefont{Micheli}},
  \bibinfo{author}{\bibfnamefont{R.} \bibnamefont{Fazio}}, \bibnamefont{and}
  \bibinfo{author}{\bibfnamefont{P.} \bibnamefont{Zoller}},
  \bibinfo{journal}{Phys. Rev. Lett.} \textbf{\bibinfo{volume}{105}},
  \bibinfo{pages}{015702} (\bibinfo{year}{2010}).


  \bibitem[{\citenamefont{Torre et~al.}(2010)\citenamefont{Torre, Demler, Giamarchi, and Altman}}]{Torre2010}
  \bibinfo{author}{\bibfnamefont{E.~G.~D.}~\bibnamefont{Torre}},
  \bibinfo{author}{\bibfnamefont{E.}~\bibnamefont{Demler}},
  \bibinfo{author}{\bibfnamefont{T.} \bibnamefont{Giamarchi}}, \bibnamefont{and}
  \bibinfo{author}{\bibfnamefont{E.} \bibnamefont{Altman}},
  \bibinfo{journal}{Nat. Phys.} \textbf{\bibinfo{volume}{6}},
  \bibinfo{pages}{806-810} (\bibinfo{year}{2010}).


  \bibitem[{\citenamefont{Ritsch et~al.}(2013)\citenamefont{Ritsch, Domokos, Brennecke, and Esslinger}}]{Ritsch2013}
  \bibinfo{author}{\bibfnamefont{H.}~\bibnamefont{Ritsch}},
  \bibinfo{author}{\bibfnamefont{P.}~\bibnamefont{Domokos}},
  \bibinfo{author}{\bibfnamefont{F.} \bibnamefont{Brennecke}}, \bibnamefont{and}
  \bibinfo{author}{\bibfnamefont{T.} \bibnamefont{Esslinger}},
  \bibinfo{journal}{Rev. Mod. Phys.} \textbf{\bibinfo{volume}{85}},
  \bibinfo{pages}{553--601} (\bibinfo{year}{2013}).


  \bibitem[{\citenamefont{Carusotto et~al.}(2013)\citenamefont{Carusotto, and Ciuti}}]{Carusotto2013}
  \bibinfo{author}{\bibfnamefont{I.}~\bibnamefont{Carusotto}}, \bibnamefont{and}
  \bibinfo{author}{\bibfnamefont{C.} \bibnamefont{Ciuti}},
  \bibinfo{journal}{Rev. Mod. Phys.} \textbf{\bibinfo{volume}{85}},
  \bibinfo{pages}{299--366} (\bibinfo{year}{2013}).


  \bibitem[{\citenamefont{Markus et~al.}(2012)\citenamefont{Markus, Sebastian, Guido, and Peter}}]{Markus2012}
  \bibinfo{author}{\bibfnamefont{M.}~\bibnamefont{Markus}},
  \bibinfo{author}{\bibfnamefont{D.}~\bibnamefont{Sebastian}},
  \bibinfo{author}{\bibfnamefont{P.} \bibnamefont{Guido}}, \bibnamefont{and}
  \bibinfo{author}{\bibfnamefont{Z.} \bibnamefont{Peter}},
  \bibinfo{journal}{Adv. At. Mol. Opt. Phys.} \textbf{\bibinfo{volume}{61}},
  \bibinfo{pages}{1-80} (\bibinfo{year}{2012}).

  \bibitem[{\citenamefont{Heule et~al.}(2011)\citenamefont{Heule, Bruder, Burgarth, and VMS}}]{Heule2011}
  \bibinfo{author}{\bibfnamefont{R.}~\bibnamefont{Heule}},
  \bibinfo{author}{\bibfnamefont{C.}~\bibnamefont{Bruder}},
  \bibinfo{author}{\bibfnamefont{D.}~\bibnamefont{Burgarth}},
  \bibnamefont{and}
  \bibinfo{author}{\bibnamefont{VMS}},
  \bibinfo{journal}{Eur. Phys. J. D} \textbf{\bibinfo{volume}{63}},
  \bibinfo{pages}{41-46} (\bibinfo{year}{2011}).

  \bibitem[{\citenamefont{Houck, et~al.}(2012)\citenamefont{Houck,, Tureci, and Koch,}}]{Houck2012}
  \bibinfo{author}{\bibfnamefont{A.~A.}~\bibnamefont{Houck,}},
  \bibinfo{author}{\bibfnamefont{H.~E.}~\bibnamefont{Tureci,}}, \bibnamefont{and}
  \bibinfo{author}{\bibfnamefont{J.} \bibnamefont{Koch,}},
  \bibinfo{journal}{Nat. Phys.} \textbf{\bibinfo{volume}{8}},
  \bibinfo{pages}{292} (\bibinfo{year}{2012}).

  \bibitem[{\citenamefont{Fitzpatrick et~al.}(2017)\citenamefont{Fitzpatrick, Sundaresan, Li, Koch, and Houck }}]{Fitzpatrick2017}
  \bibinfo{author}{\bibfnamefont{M.}~\bibnamefont{Fitzpatrick}},
  \bibinfo{author}{\bibfnamefont{N.~M.}~\bibnamefont{Sundaresan}},
  \bibinfo{author}{\bibfnamefont{A.~C.~Y.}~\bibnamefont{Li}},
  \bibinfo{author}{\bibfnamefont{J.}~\bibnamefont{Koch}}, \bibnamefont{and}
  \bibinfo{author}{\bibfnamefont{A.~A.}~\bibnamefont{Houck}},
  \bibinfo{journal}{Phys. Rev. X} \textbf{\bibinfo{volume}{7}},
  \bibinfo{pages}{011016} (\bibinfo{year}{2017}).





  

\bibitem[{\citenamefont{Sadiek et~al.}(2010)\citenamefont{Sadiek, Alkurtass, and Aldossary}}]{Sadiek2010}
  \bibinfo{author}{\bibfnamefont{G.}~\bibnamefont{Sadiek}},
  \bibinfo{author}{\bibfnamefont{B.}~\bibnamefont{Alkurtass}},
  \bibnamefont{and}
  \bibinfo{author}{\bibfnamefont{O.}~\bibnamefont{Aldossary}},
  \bibinfo{journal}{Phys. Rev. A} \textbf{\bibinfo{volume}{82}},
  \bibinfo{pages}{052337} (\bibinfo{year}{2010}).

 \bibitem[{\citenamefont{Barouch et~al.}(1970)\citenamefont{Barouch, McCoy, and Dresden}}]{Barouch1970}
   \bibinfo{author}{\bibfnamefont{E.}~\bibnamefont{Barouch}},
   \bibinfo{author}{\bibfnamefont{B.M.}~\bibnamefont{McCoy}},
   \bibnamefont{and}
   \bibinfo{author}{\bibfnamefont{M.}~\bibnamefont{Dresden}},
   \bibinfo{journal}{Phys. Rev. A} \textbf{\bibinfo{volume}{2}},
   \bibinfo{pages}{1075--1092} (\bibinfo{year}{1970}).

\bibitem[{\citenamefont{Sadiek et~al.}(2010)\citenamefont{Sadiek, Huang, Aldossary and Kais}}]{Sadiek2008}
  \bibinfo{author}{\bibfnamefont{G.}~\bibnamefont{Sadiek}},
  \bibinfo{author}{\bibfnamefont{Z.}~\bibnamefont{Huang}},
  \bibinfo{author}{\bibfnamefont{O.}~\bibnamefont{Aldossary}},
  \bibnamefont{and}
  \bibinfo{author}{\bibfnamefont{S.}~\bibnamefont{Kais}},
  \bibinfo{journal}{Mol. Phys.} \textbf{\bibinfo{volume}{106}},
  \bibinfo{pages}{1777} (\bibinfo{year}{2008}).

\bibitem[{\citenamefont{Sen(De) et~al.}(2004)\citenamefont{Sen(De), Sen, and
  Lewenstein}}]{Sen(De)2004}
  \bibinfo{author}{\bibfnamefont{A.}~\bibnamefont{Sen(De)}},
  \bibinfo{author}{\bibfnamefont{U.}~\bibnamefont{Sen}}, \bibnamefont{and}
  \bibinfo{author}{\bibfnamefont{M.}~\bibnamefont{Lewenstein}},
  \bibinfo{journal}{Phys Rev. A} \textbf{\bibinfo{volume}{70}},
  \bibinfo{pages}{060304} (\bibinfo{year}{2004}).

\bibitem[{\citenamefont{Huang and Kais}(2006)}]{HuangZ2006}
\bibinfo{author}{\bibfnamefont{Z.}~\bibnamefont{Huang}} \bibnamefont{and}
  \bibinfo{author}{\bibfnamefont{S.}~\bibnamefont{Kais}},
  \bibinfo{journal}{Phys. Rev. A} \textbf{\bibinfo{volume}{73}},
  \bibinfo{pages}{022339} (\bibinfo{year}{2006}).       

   \bibitem[{\citenamefont{Lieb et~al.}(1961)\citenamefont{Lieb, Schultz, and Mattis}}]
   {Lieb1961}
   \bibinfo{author}{\bibfnamefont{E.}~\bibnamefont{Lieb}},
   \bibinfo{author}{\bibfnamefont{T.}~\bibnamefont{Schultz}}, \bibnamefont{and}
   \bibinfo{author}{\bibnamefont{D.}~\bibnamefont{Mattis}},
   \bibinfo{journal}{Annals of Physics} \textbf{\bibinfo{volume}{16}},
   \bibinfo{pages}{407-466} (\bibinfo{year}{1961}).     

\bibitem[{\citenamefont{Lashin et~al.}(2014)\citenamefont{Lashin, Sadiek, Abdalla and Aldufeery}}]{Lashin2014}
  \bibinfo{author}{\bibfnamefont{E.}~\bibnamefont{Lashin}},
  \bibinfo{author}{\bibfnamefont{G.}~\bibnamefont{Sadiek}},
  \bibinfo{author}{\bibfnamefont{M. S.}~\bibnamefont{Abdalla}},
  \bibnamefont{and}
  \bibinfo{author}{\bibfnamefont{E.}~\bibnamefont{Aldufeery}},
  \bibinfo{journal}{Appl. Math. Inf. Sci.} \textbf{\bibinfo{volume}{8}},
  \bibinfo{pages}{1071} (\bibinfo{year}{2014}).


\bibitem[{\citenamefont{Xu et~al.}(2010)\citenamefont{Xu, Kais, Naumov, and
  Sameh}}]{XuQ2010}
\bibinfo{author}{\bibfnamefont{Q.}~\bibnamefont{Xu}},
  \bibinfo{author}{\bibfnamefont{S.}~\bibnamefont{Kais}},
  \bibinfo{author}{\bibfnamefont{M.}~\bibnamefont{Naumov}}, \bibnamefont{and}
  \bibinfo{author}{\bibfnamefont{A.}~\bibnamefont{Sameh}},
  \bibinfo{journal}{Phys. Rev. A} \textbf{\bibinfo{volume}{81}},
  \bibinfo{pages}{022324} (\bibinfo{year}{2010}).   

\bibitem[{\citenamefont{Xu et~al.}(2011)\citenamefont{Xu, Sadiek, and
  Kais}}]{XuQ2011}
\bibinfo{author}{\bibfnamefont{Q.}~\bibnamefont{Xu}},
  \bibinfo{author}{\bibfnamefont{G.}~\bibnamefont{Sadiek}}, \bibnamefont{and}
  \bibinfo{author}{\bibfnamefont{S.}~\bibnamefont{Kais}},
  \bibinfo{journal}{Phys Rev. A} \textbf{\bibinfo{volume}{83}},
  \bibinfo{pages}{062312} (\bibinfo{year}{2011}).     


\bibitem[{\citenamefont{Sadiek and Kais}(2013)\citenamefont{Xu, Sadiek, and Kais}}]{Sadiek2013}
\bibinfo{author}{\bibfnamefont{G.}~\bibnamefont{Sadiek}}, \bibnamefont{and}
  \bibinfo{author}{\bibfnamefont{S.}~\bibnamefont{Kais}},
  \bibinfo{journal}{J. Phys. B} \textbf{\bibinfo{volume}{46}},
  \bibinfo{pages}{245501} (\bibinfo{year}{2013}).  
 
   \bibitem[{\citenamefont{Sadiek et~al.}(2019)\citenamefont{Sadiek, Al-Drees and Sebaweh}}]{Sadiek2019}
   \bibinfo{author}{\bibfnamefont{G.}~\bibnamefont{Sadiek}},
   \bibinfo{author}{\bibfnamefont{W.}~\bibnamefont{Al-Drees}}, \bibnamefont{and}
   \bibinfo{author}{\bibfnamefont{A.~S.}~\bibnamefont{Sebaweh}},
   \bibinfo{journal}{Optics Express} \textbf{\bibinfo{volume}{27}},
   \bibinfo{pages}{33799} (\bibinfo{year}{2019}).     


  \bibitem[{\citenamefont{Wang et~al.}(2006)\citenamefont{Wang, Batelaan, Podany, and Starace}}]{Wang2006}
  \bibinfo{author}{\bibfnamefont{J.}~\bibnamefont{Wang}},
  \bibinfo{author}{\bibfnamefont{H.}~\bibnamefont{Batelaan}},
  \bibinfo{author}{\bibfnamefont{J.}~\bibnamefont{Podany}},
  \bibnamefont{and}
  \bibinfo{author}{\bibfnamefont{A.~F.}~\bibnamefont{Starace}},
  \bibinfo{journal}{J. Phys. B: Atomic, Molecular and Optical Physics} \textbf{\bibinfo{volume}{39}},
  \bibinfo{pages}{4343} (\bibinfo{year}{2006}).



  \bibitem[{\citenamefont{Abliz et~al.}(2006)\citenamefont{Abliz, Gao, Xie, Wu, and Liu}}]{ABLIZ2006}
  \bibinfo{author}{\bibfnamefont{A.}},~\bibinfo{author}{\bibfnamefont{Abliz}},
  \bibinfo{author}{\bibfnamefont{H.~J.}~ \bibnamefont{Gao}},
  \bibinfo{author}{\bibfnamefont{X.~C.}~\bibnamefont{Xie}},
  \bibinfo{author}{\bibfnamefont{Y.~S.}~\bibnamefont{Wu}}, \bibnamefont{and}
  \bibinfo{author}{\bibfnamefont{W.~M.}~\bibnamefont{Liu}},
  \bibinfo{journal}{Phys. Rev. A.} \textbf{\bibinfo{volume}{74}},
  \bibinfo{pages}{052105} (\bibinfo{year}{2006}).

  \bibitem[{\citenamefont{Dubi et~al.}(2009)\citenamefont{Dubi, and Di}}]{Dubi2009}
  \bibinfo{author}{\bibfnamefont{Y.}~\bibnamefont{Dubi}},
  \bibnamefont{and}
  \bibinfo{author}{\bibfnamefont{V.~M.} \bibnamefont{Di}},
  \bibinfo{journal}{Phys. Rev. A} \textbf{\bibinfo{volume}{79}},
  \bibinfo{pages}{012328} (\bibinfo{year}{2009}).


  \bibitem[{\citenamefont{Hein et~al.}(2005)\citenamefont{Hein, D\"ur, and Briegel}}]{Hein2005}
  \bibinfo{author}{\bibfnamefont{M.}~\bibnamefont{Hein}},
  \bibinfo{author}{\bibfnamefont{W.}~\bibnamefont{D\"ur}},
  \bibnamefont{and}
  \bibinfo{author}{\bibfnamefont{H.~J.} \bibnamefont{Briegel}},
  \bibinfo{journal}{Phys. Rev. A} \textbf{\bibinfo{volume}{71}},
  \bibinfo{pages}{032350} (\bibinfo{year}{2005}).

  \bibitem[{\citenamefont{Tsomokos et~al.}(2007)\citenamefont{Tsomokos, Hartmann, Huelga, and Plenio}}]{Tsomokos2007}
  \bibinfo{author}{\bibfnamefont{D.~I.} \bibnamefont{Tsomokos}},
  \bibinfo{author}{\bibfnamefont{M.~J.} \bibnamefont{Hartmann}},
  \bibinfo{author}{\bibfnamefont{S.~F.} \bibnamefont{Huelga}},
  \bibnamefont{and}
  \bibinfo{author}{\bibfnamefont{M.~B.} \bibnamefont{Plenio}},
  \bibinfo{journal}{New J. Phys.} \textbf{\bibinfo{volume}{9}},
  \bibinfo{pages}{79} (\bibinfo{year}{2007}).

  \bibitem[{\citenamefont{Buric et~al.}(2008)\citenamefont{Buric}}]{Buric2008}
  \bibinfo{author}{\bibfnamefont{N.}~\bibnamefont{Buric}},
  \bibinfo{journal}{Phys. Rev. A} \textbf{\bibinfo{volume}{77}},
  \bibinfo{pages}{012321} (\bibinfo{year}{2008}).

  \bibitem[{\citenamefont{Hu et~al.}(2009)\citenamefont{Hu and Xi}}]{Hu2009}
  \bibinfo{author}{\bibfnamefont{M.~L.}~\bibnamefont{Hu}},
  \bibnamefont{and}
  \bibinfo{author}{\bibfnamefont{X.~Q.}~\bibnamefont{Xi}},
  \bibinfo{journal}{Opt. Commun.} \textbf{\bibinfo{volume}{282}},
  \bibinfo{pages}{4819} (\bibinfo{year}{2009}).

  \bibitem[{\citenamefont{Hu et~al.}(2009)\citenamefont{Hu, Xi, and Lian}}]{Hu2009a}
  \bibinfo{author}{\bibfnamefont{M.~L.} \bibnamefont{Hu}},
  \bibinfo{author}{\bibfnamefont{X.~Q.} \bibnamefont{Xi}},
  \bibnamefont{and}
  \bibinfo{author}{\bibfnamefont{H.~L.} \bibnamefont{Lian}},
  \bibinfo{journal}{Physica B: Condensed Matter} \textbf{\bibinfo{volume}{404}},
  \bibinfo{pages}{3499} (\bibinfo{year}{2009}).

  \bibitem[{\citenamefont{Buric et~al.}(2009)\citenamefont{Buric and Lindén}}]{Buric2009}
  \bibinfo{author}{\bibfnamefont{N.}~\bibnamefont{Buric}},
  \bibnamefont{and}
  \bibinfo{author}{\bibfnamefont{B.~L.} \bibnamefont{Lindén}},
  \bibinfo{journal}{Phys. Lett. A } \textbf{\bibinfo{volume}{373}},
  \bibinfo{pages}{1531} (\bibinfo{year}{2009}).        

  \bibitem[{\citenamefont{Pumulo et~al.}(2011)\citenamefont{Pumulo, Sinayskiy, and Petruccione}}]{Pumulo2011}
  \bibinfo{author}{\bibfnamefont{N.}~\bibnamefont{Pumulo}},
  \bibinfo{author}{\bibfnamefont{I.}~\bibnamefont{Sinayskiy}},
  \bibnamefont{and}
  \bibinfo{author}{\bibfnamefont{F.}~\bibnamefont{Petruccione}},
  \bibinfo{journal}{Phys. Lett. A} \textbf{\bibinfo{volume}{375}},
  \bibinfo{pages}{3157-3166} (\bibinfo{year}{2011}).

  \bibitem[{\citenamefont{Zhang et~al.}(2013)\citenamefont{Zhang, Zhang, Zhang, and Wang}}]{Zhang2013}
  \bibinfo{author}{\bibfnamefont{Xiu-Xing.}~\bibnamefont{Zhang}},
  \bibinfo{author}{\bibfnamefont{Al-Ping}~\bibnamefont{Zhang}},
  \bibinfo{author}{\bibfnamefont{Jia}~\bibnamefont{Zhang}},
  \bibnamefont{and}
  \bibinfo{author}{\bibfnamefont{Ju-Xia}~\bibnamefont{Wang}},
  \bibinfo{journal}{Mod. Phys. Lett. B} \textbf{\bibinfo{volume}{27}},
  \bibinfo{pages}{1350078} (\bibinfo{year}{2013}).

  \bibitem[{\citenamefont{Lakshminarayan et~al.}(2001)\citenamefont{Lakshminarayan}}]{Lakshminarayan2001}
  \bibinfo{author}{\bibfnamefont{A.}~\bibnamefont{Lakshminarayan}},
  \bibinfo{journal}{Phys. Rev. E} \textbf{\bibinfo{volume}{64}},
  \bibinfo{pages}{036207} (\bibinfo{year}{2001}).

  \bibitem[{\citenamefont{Petrosyan et~al.}(2010)\citenamefont{Petrosyan, Nikolopoulos, and Lambropoulos}}]{Petrosyan2010}
  \bibinfo{author}{\bibfnamefont{D.}~\bibnamefont{Petrosyan}},
  \bibinfo{author}{\bibfnamefont{G.~M.} \bibnamefont{Nikolopoulos}},
  \bibnamefont{and}
  \bibinfo{author}{\bibfnamefont{P.}~\bibnamefont{Lambropoulos}},
  \bibinfo{journal}{Phys. Rev. A} \textbf{\bibinfo{volume}{81}},
  \bibinfo{pages}{042307} (\bibinfo{year}{2010}).

  \bibitem[{\citenamefont{Ronke et~al.}(2011)\citenamefont{Ronke, Spiller, and D'Amico}}]{Ronke2011}
  \bibinfo{author}{\bibfnamefont{R.}~\bibnamefont{Ronke}},
  \bibinfo{author}{\bibfnamefont{T.~P.} \bibnamefont{Spiller}},
  \bibnamefont{and}
  \bibinfo{author}{\bibfnamefont{I.}~\bibnamefont{D'Amico}},
  \bibinfo{journal}{Phys. Rev. A} \textbf{\bibinfo{volume}{83}},
  \bibinfo{pages}{012325} (\bibinfo{year}{2011}).      

 \bibitem[{\citenamefont{Alkurtass and Wichterich}(2013)\citenamefont{Alkurtass, Wichterich, and Sougato}}]{Alkurtass2013}
  \bibinfo{author}{\bibfnamefont{B.}~\bibnamefont{Alkurtass}},
  \bibinfo{author}{\bibfnamefont{H.}~\bibnamefont{Wichterich}},	
  \bibnamefont{and}
  \bibinfo{author}{\bibfnamefont{Bose,}~\bibnamefont{Sougato}},
  \bibinfo{journal}{Phys. Rev. A} \textbf{\bibinfo{volume}{88}},
  \bibinfo{pages}{062325} (\bibinfo{year}{2013}).           

  \bibitem[{\citenamefont{Wu et~al.}(2014)\citenamefont{Wu, Nanduri, and Rabitz}}]{Wu2014}
  \bibinfo{author}{\bibfnamefont{N.}~\bibnamefont{Wu}},
  \bibinfo{author}{\bibfnamefont{A.}~\bibnamefont{Nanduri}},
  \bibnamefont{and}
  \bibinfo{author}{\bibfnamefont{H.}~\bibnamefont{Rabitz}},
  \bibinfo{journal}{Phys. Rev. A} \textbf{\bibinfo{volume}{89}},
  \bibinfo{pages}{062105} (\bibinfo{year}{2014}).          

 \bibitem[{\citenamefont{Ciliberti et~al.}(2013)\citenamefont{Ciliberti, Canosa, and Rossignoli}}]{Ciliberti2013}
  \bibinfo{author}{\bibfnamefont{L.}~\bibnamefont{Ciliberti}},
  \bibinfo{author}{\bibfnamefont{N.}~\bibnamefont{Canosa}}, 	
  \bibnamefont{and}
  \bibinfo{author}{\bibfnamefont{R.}~\bibnamefont{Rossignoli}},
  \bibinfo{journal}{Phys. Rev. A} \textbf{\bibinfo{volume}{88}},
  \bibinfo{pages}{012119} (\bibinfo{year}{2013}).          

 \bibitem[{\citenamefont{Sadiek et~al.}(2016)\citenamefont{Sadiek and Almalki}}]{Sadiek2016}
  \bibinfo{author}{\bibfnamefont{G.}~\bibnamefont{Sadiek}},	
  \bibnamefont{and}
  \bibinfo{author}{\bibfnamefont{S.}~\bibnamefont{Almalki}},  
  \bibinfo{journal}{Phys. Rev. A} \textbf{\bibinfo{volume}{94}},
  \bibinfo{pages}{012341} (\bibinfo{year}{2016}).            

 \bibitem[{\citenamefont{Lee et~al.}(2013)\citenamefont{Lee, Gopalakrishnan, and Lukin}}]{Lee2013}
  \bibinfo{author}{\bibfnamefont{T.~E.}~\bibnamefont{Lee}},
  \bibinfo{author}{\bibfnamefont{S.}~\bibnamefont{Gopalakrishnan}}, 	
  \bibnamefont{and}
  \bibinfo{author}{\bibfnamefont{M.~D.}~\bibnamefont{Lukin}},
  \bibinfo{journal}{Phys. Rev. Lett.} \textbf{\bibinfo{volume}{110}},
  \bibinfo{pages}{257204} (\bibinfo{year}{2013}).          

  \bibitem[{\citenamefont{Rios et~al.}(2017)\citenamefont{Rios, Rossignoli, and Canosa}}]{Rios2017}
  \bibinfo{author}{\bibfnamefont{E.}~\bibnamefont{Rios}},
  \bibinfo{author}{\bibfnamefont{R.}~\bibnamefont{Rossignoli}}, \bibnamefont{and}
  \bibinfo{author}{\bibfnamefont{N.}~\bibnamefont{Canosa}},
  \bibinfo{journal}{J. Phys. B: At. Mol. Opt. Phys.} \textbf{\bibinfo{volume}{50}},
  \bibinfo{pages}{095501} (\bibinfo{year}{2017}).


 \bibitem[{\citenamefont{Sun et~al.}(2003)\citenamefont{Sun, Chen, and Chen}}]{Sun2003}
  \bibinfo{author}{\bibfnamefont{Y.}~\bibnamefont{Sun}},
  \bibinfo{author}{\bibfnamefont{Y.}~\bibnamefont{Chen}},	
  \bibnamefont{and}
  \bibinfo{author}{\bibfnamefont{H.}~\bibnamefont{Chen}},
  \bibinfo{journal}{Phys. Rev. A.} \textbf{\bibinfo{volume}{68}},
  \bibinfo{pages}{044301} (\bibinfo{year}{2003}).          

  \bibitem[{\citenamefont{Asoudeh et~al.}(2005)\citenamefont{Asoudeh and Karimipour}}]{Asoudeh2005}
   \bibinfo{author}{\bibfnamefont{M.}~\bibnamefont{Asoudeh}}, 	
   \bibnamefont{and}
   \bibinfo{author}{\bibfnamefont{V.}~\bibnamefont{Karimipour}},  
   \bibinfo{journal}{Phys. Rev. A.} \textbf{\bibinfo{volume}{71}},
   \bibinfo{pages}{022308} (\bibinfo{year}{2005}).         

  \bibitem[{\citenamefont{Zhang et~al.}(2005)\citenamefont{Zhang and Li}}]{Zhang2005}
  \bibinfo{author}{\bibfnamefont{G.~F.}~\bibnamefont{Zhang}},	
  \bibnamefont{and}
  \bibinfo{author}{\bibfnamefont{S.~S.}~\bibnamefont{Li}},  
  \bibinfo{journal}{Phys. Rev. A.} \textbf{\bibinfo{volume}{72}},
  \bibinfo{pages}{034302} (\bibinfo{year}{2005}).

  \bibitem[{\citenamefont{Hu et~al.}(2007)\citenamefont{Hu, Yi and Park}}]{Hu2007}
  \bibinfo{author}{\bibfnamefont{Z-N}~\bibnamefont{Hu}},
  \bibinfo{author}{\bibfnamefont{K.~S.}~\bibnamefont{Yi}},   	
  \bibnamefont{and}
  \bibinfo{author}{\bibfnamefont{K-S}~\bibnamefont{Park}},  
  \bibinfo{journal}{Phys. A: Math. Theor.} \textbf{\bibinfo{volume}{40}},
  \bibinfo{pages}{7283} (\bibinfo{year}{2007}).

  \bibitem[{\citenamefont{Hassan et~al.}(2010)\citenamefont{Hassan, Lari and Joag}}]{Hassan2010}
  \bibinfo{author}{\bibfnamefont{A.~S.~M.}~\bibnamefont{Hassan}},
  \bibinfo{author}{\bibfnamefont{B.}~\bibnamefont{Lari}}, 	
  \bibnamefont{and}
  \bibinfo{author}{\bibfnamefont{P.~S.}~\bibnamefont{Joag}},    
  \bibinfo{journal}{Phys. A: Math. Theor.} \textbf{\bibinfo{volume}{43}},
  \bibinfo{pages}{485302} (\bibinfo{year}{2010}).

 \bibitem[{\citenamefont{Guo et~al.}(2011)\citenamefont{Guo, Mi, Zhang and Song}}]{Guo2011}
  \bibinfo{author}{\bibfnamefont{J.-L.}~\bibnamefont{Guo}},
  \bibinfo{author}{\bibfnamefont{Y.-J.}~\bibnamefont{Mi}},
  \bibinfo{author}{\bibfnamefont{J.}~\bibnamefont{Zhang}},	
  \bibnamefont{and}
  \bibinfo{author}{\bibfnamefont{H.-S.}~\bibnamefont{Song}},
  \bibinfo{journal}{Phys. B: At. Mol. Opt. Phys.} \textbf{\bibinfo{volume}{44}},
  \bibinfo{pages}{065504} (\bibinfo{year}{2011}).

 \bibitem[{\citenamefont{Zhang et~al.}(2011)\citenamefont{Zhang, Fan, Ji, Jiang, Abliz and Liu}}]{Zhang2011}
  \bibinfo{author}{\bibfnamefont{G.~F.}~\bibnamefont{Zhang}},
  \bibinfo{author}{\bibfnamefont{H.}~\bibnamefont{Fan}},
  \bibinfo{author}{\bibfnamefont{A.~L.}~\bibnamefont{Ji}},
  \bibinfo{author}{\bibfnamefont{Z.~T.}~\bibnamefont{Jiang}},
  \bibinfo{author}{\bibfnamefont{A.}~\bibnamefont{Abliz}},	
  \bibnamefont{and}
  \bibinfo{author}{\bibfnamefont{W.~M.}~\bibnamefont{Liu}},
  \bibinfo{journal}{Ann. Phys. NY} \textbf{\bibinfo{volume}{326}},
  \bibinfo{pages}{2694-2701} (\bibinfo{year}{2011}).

 \bibitem[{\citenamefont{Albayrak}(2010)\citenamefont{Albayrak}}]{Albayrak2010}
  \bibinfo{author}{\bibfnamefont{E.}~\bibnamefont{Albayrak}},
  \bibinfo{journal}{Chin. Phys. B} \textbf{\bibinfo{volume}{19}},
  \bibinfo{pages}{090319} (\bibinfo{year}{2010}).

 \bibitem[{\citenamefont{Guo et~al.}(2010)\citenamefont{Guo, Liang, Xu and Zhu}}]{Guo2010}
  \bibinfo{author}{\bibfnamefont{K.~T.}~\bibnamefont{Guo}},
  \bibinfo{author}{\bibfnamefont{M.~C.}~\bibnamefont{Liang}},
  \bibinfo{author}{\bibfnamefont{H.~Y.}~\bibnamefont{Xu}},	
  \bibnamefont{and}
  \bibinfo{author}{\bibfnamefont{C.~B.}~\bibnamefont{Zhu}},
  \bibinfo{journal}{Phys. A: Math. Theor.} \textbf{\bibinfo{volume}{43}},
  \bibinfo{pages}{505301} (\bibinfo{year}{2010}).

 \bibitem[{\citenamefont{Hu et~al.}(2014)\citenamefont{Hu, Da-Chuang, Xian-Ping, Ming and Zhuo-Liang}}]{Hu2014}
  \bibinfo{author}{\bibfnamefont{L.}~\bibnamefont{Hu}},
  \bibinfo{author}{\bibfnamefont{L.}~\bibnamefont{Da-Chuang}},
  \bibinfo{author}{\bibfnamefont{W.}~\bibnamefont{Xian-Ping}},
  \bibinfo{author}{\bibfnamefont{Y.}~\bibnamefont{Ming}},	
  \bibnamefont{and}
  \bibinfo{author}{\bibfnamefont{C.}~\bibnamefont{Zhuo-Liang}},
  \bibinfo{journal}{Chem. Phys. Lett} \textbf{\bibinfo{volume}{31}},
  \bibinfo{pages}{040301} (\bibinfo{year}{2014}).

 \bibitem[{\citenamefont{Cerezo et~al.}(2017)\citenamefont{Cerezo, Rossignoli, Canosa and R\'{\i}os}}]{Cerezo2017}
  \bibinfo{author}{\bibfnamefont{M.}~\bibnamefont{Cerezo}},
  \bibinfo{author}{\bibfnamefont{R.}~\bibnamefont{Rossignoli}},
  \bibinfo{author}{\bibfnamefont{N.}~\bibnamefont{Canosa}}, 	
  \bibnamefont{and}
  \bibinfo{author}{\bibfnamefont{E.}~\bibnamefont{R\'{\i}os}},
  \bibinfo{journal}{Phys. Rev. Lett.} \textbf{\bibinfo{volume}{119}},
  \bibinfo{pages}{220605} (\bibinfo{year}{2017}).

 \bibitem[{\citenamefont{Rota et~al.}(2017)\citenamefont{Rota, Storme, Bartolo, Fazio and Ciuti}}]{Rota2017}
  \bibinfo{author}{\bibfnamefont{R.}~\bibnamefont{Rota}},
  \bibinfo{author}{\bibfnamefont{F.}~\bibnamefont{Storme}},  
  \bibinfo{author}{\bibfnamefont{N.}~\bibnamefont{Bartolo}},
  \bibinfo{author}{\bibfnamefont{R.}~\bibnamefont{Fazio}},	
  \bibnamefont{and}
  \bibinfo{author}{\bibfnamefont{C.}~\bibnamefont{Ciuti}},
  \bibinfo{journal}{Phys. Rev. B} \textbf{\bibinfo{volume}{95}},
  \bibinfo{pages}{134431} (\bibinfo{year}{2017}).

 \bibitem[{\citenamefont{Cerezo et~al.}(2019)\citenamefont{Cerezo, Rossignoli, Canosa and Lamas}}]{Cerezo2019}
  \bibinfo{author}{\bibfnamefont{M.}~\bibnamefont{Cerezo}},
  \bibinfo{author}{\bibfnamefont{R.}~\bibnamefont{Rossignoli}},
  \bibinfo{author}{\bibfnamefont{N.}~\bibnamefont{Canosa}}, 	
  \bibnamefont{and}
  \bibinfo{author}{\bibfnamefont{C. A.}~\bibnamefont{Lamas}},
  \bibinfo{journal}{Phys. Rev. B} \textbf{\bibinfo{volume}{99}},
  \bibinfo{pages}{014409} (\bibinfo{year}{2019}).
  
   \bibitem[{\citenamefont{Lindblad}(1976)\citenamefont{Lindblad}}]{Lindblad1976}
   \bibinfo{author}{\bibfnamefont{G.}~\bibnamefont{Lindblad}},
   \bibinfo{journal}{Comm. Math. Phys.} \textbf{\bibinfo{volume}{48}},
   \bibinfo{pages}{119-130} (\bibinfo{year}{1976}).     

  \bibitem[{\citenamefont{Breuer}(2002)}]{Breuer2002}
  \bibinfo{author}{\bibfnamefont{H.~P.}~\bibnamefont{Breuer}},
  \bibnamefont{and}
  \bibinfo{author}{\bibfnamefont{F.}~\bibnamefont{Petruccione}},
  \emph{\bibinfo{title}{The Theory of Open Quantum Systems}}
  (\bibinfo{publisher}{Oxford University Press},
  \bibinfo{year}{2002}).

  \bibitem[{\citenamefont{Mintert et~al.}(2005)\citenamefont{Mintert, Carvalho, Kus, and Buchleitner}}]{Mintert2005}
  \bibinfo{author}{\bibfnamefont{F.}~\bibnamefont{Mintert}},
  \bibinfo{author}{\bibfnamefont{A.~R.~R.}~\bibnamefont{Carvalho}},
  \bibinfo{author}{\bibfnamefont{M.}~\bibnamefont{Kus}},
  \bibnamefont{and}
  \bibinfo{author}{\bibfnamefont{A.}~\bibnamefont{Buchleitner}},
  \bibinfo{journal}{Physics Reports} \textbf{\bibinfo{volume}{415}},
  \bibinfo{pages}{207-259} (\bibinfo{year}{2005}).
 
   \bibitem[{\citenamefont{Wootters}(1998)\citenamefont{Wootters}}]{Wootters1998}
   \bibinfo{author}{\bibfnamefont{William~K.}~\bibnamefont{Wootters}},
   \bibinfo{journal}{Phys. Rev. Lett.} \textbf{\bibinfo{volume}{80}},
   \bibinfo{pages}{2245-2248} (\bibinfo{year}{1998}).     

  \bibitem[{\citenamefont{Amico et~al.}(2004)\citenamefont{Amico, Osterloh, Plastina, Fazio, and Palma}}]{Amico2004}
  \bibinfo{author}{\bibfnamefont{L.}~\bibnamefont{Amico}},
  \bibinfo{author}{\bibfnamefont{A.}~\bibnamefont{Osterloh}},
  \bibinfo{author}{\bibfnamefont{F.}~\bibnamefont{Plastina}},
  \bibinfo{author}{\bibfnamefont{R.}~\bibnamefont{Fazio}},
  \bibnamefont{and}
  \bibinfo{author}{\bibfnamefont{G.~M.}~\bibnamefont{Palma}},
  \bibinfo{journal}{Phys. Rev. A.} \textbf{\bibinfo{volume}{69}},
  \bibinfo{pages}{022304} (\bibinfo{year}{2004}).
   
  \bibitem[{\citenamefont{Roscilde et~al.}(2004)\citenamefont{Roscilde, Verrucchi, Fubini, Haas and Tognetti}}]{Roscilde2004}
  \bibinfo{author}{\bibfnamefont{T.}~\bibnamefont{Roscilde}},
  \bibinfo{author}{\bibfnamefont{P.}~\bibnamefont{Verrucchi}},
  \bibinfo{author}{\bibfnamefont{A.}~\bibnamefont{Fubini}},
  \bibinfo{author}{\bibfnamefont{S.}~\bibnamefont{Haas}},
  \bibnamefont{and}
  \bibinfo{author}{\bibfnamefont{V.}~\bibnamefont{Tognetti}},
  \bibinfo{journal}{Phys. Rev. Lett.} \textbf{\bibinfo{volume}{93}},
  \bibinfo{pages}{167203} (\bibinfo{year}{2004}).
                       
\end{thebibliography}
\end{document}